\documentclass[
reprint,
bibnotes,
amsmath,amssymb,
aps,
prx,
]{revtex4-2}

\usepackage{graphicx}
\usepackage{dcolumn}
\usepackage{bm}
\usepackage{enumerate}
\usepackage{enumitem}

\begin{document}

\title{A Unified Theory for Chaotic Mixing in Porous Media: from Pore Networks to Granular Systems}

\author{Daniel R. Lester}
\email[]{daniel.lester@rmit.edu.au}
\affiliation{School of Engineering, RMIT University, 3000 Melbourne, Victoria, Australia.}
\author{Joris Heyman}
\affiliation{Universit\'{e} de Rennes, CNRS, G\'eosciences Rennes, UMR 6118, 35000 Rennes, France.}
\author{Yves M\'eheust} 
\affiliation{Universit\'{e} de Rennes, CNRS, G\'eosciences Rennes, UMR 6118, 35000 Rennes, France.}
\author{Tanguy Le Borgne}
\affiliation{Universit\'{e} de Rennes, CNRS, G\'eosciences Rennes, UMR 6118, 35000 Rennes, France.}


\date{\today}

\begin{abstract}
Recent studies have revealed the central role of chaotic stretching and folding at the pore scale in controlling mixing within porous media, whether the solid phase is discrete (as in granular and packed media) or continuous (as in vascular networks and open porous structures). Despite its widespread occurrence, a unified theory of chaotic mixing across these diverse systems remains  to be developed. Furthermore, previous studies have focused on fluid stretching mechanisms but the folding mechanisms are largely unknown. We address these shortcomings by presenting a unified theory of mixing in porous media. We thus show that fluid stretching and folding (SF) arise through the same fundamental kinematics driven by the topological complexity of the medium. We find that mixing in continuous porous media manifests as discontinuous mixing through a combination of SF and cutting and shuffling (CS) actions, but the rate of mixing is governed by SF only. Conversely, discrete porous media involves SF motions only. We unify these diverse systems and mechanisms by showing that continuous media represents an analog of discrete media with finite-sized grain contacts. This unified theory provides insights into the generation of pore-scale chaotic mixing and points to design of novel porous architectures with tuneable mixing and transport properties.
\end{abstract}

\maketitle

\section{Introduction}
\label{sec:intro}

Fluid flow in porous media plays host to a broad range of chemical, physical, biological and geological processes~\cite{Dentz:2011aa,Rolle:2019aa,Valocchi:2019aa}. These processes are chiefly controlled by the transport, mixing and dispersion of solutes, nutrients, colloids and microorganisms in the fluid phase. Therefore, detailed knowledge of the mixing dynamics of these flows is required to quantify, understand and predict these phenomena. For example, incomplete mixing of solute plumes has been recognised~\cite{Gramling:2002aa,Berkowitz:2016aa, Wright:2017aa,Valocchi:2019aa}
to significantly impact the propagation of chemical reactions. Hence, reactive transport models based on assumptions of complete mixing can lead to large errors~\cite{Dentz:2011aa}.

In recent years a new understanding of mixing in porous media has emerged through the characterization of the pore-scale kinematics driving mixing~\cite{Lester:2013aa,Lester:2016aa,Kree:2017aa,Turuban:2018aa,Turuban:2019aa,Souzy:2020aa,Heyman:2020aa,Heyman:2021aa}. These studies have established that chaotic mixing - where fluid particles undergo chaotic orbits and fluid elements are stretched exponentially in time - is inherent to steady flow in almost all 3D porous media. These ubiquitous kinematics have profound consequences for fluid-borne phenomena~\cite{Aref:2017aa}, including sustenance of chemical gradients at the pore-scale~\cite{Heyman:2020aa}, accelerated diffusive mixing~\cite{Lester:2014aa}, transport~\cite{Turuban:2019aa} and dispersion~\cite{Lester:2014ab} of diffusive solutes, augmented clustering and deposition of colloidal particles~\cite{Sapsis:2010aa,Ouellette:2008aa}, singular enhancement of autocatalytic and competitive reactions~\cite{Tel:2005aa,Karolyi:2000aa}, enhanced chemical signalling~\cite{Stocker:2012aa} metabolic pathways~\cite{Neufeld:2009aa,Karolyi:2002aa}, augmented alignment of particles~\cite{John:2007aa}. A complete understanding of the mechanisms of chaotic mixing in porous media is critical for characterisation and prediction of these phenomena. Such understanding also facilitates development of novel methods to control and optimise these processes directly in engineered systems or via interventions in natural systems.

\begin{figure}[h!]
\begin{centering}
\begin{tabular}{c c}
\includegraphics[width=0.48\columnwidth]{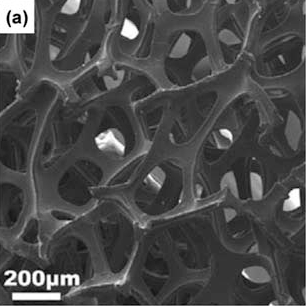}&
\includegraphics[width=0.48\columnwidth]{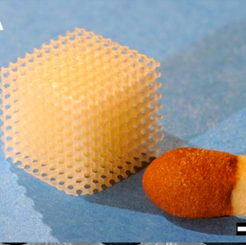}\\
(a) & (b) \\
\includegraphics[width=0.48\columnwidth]{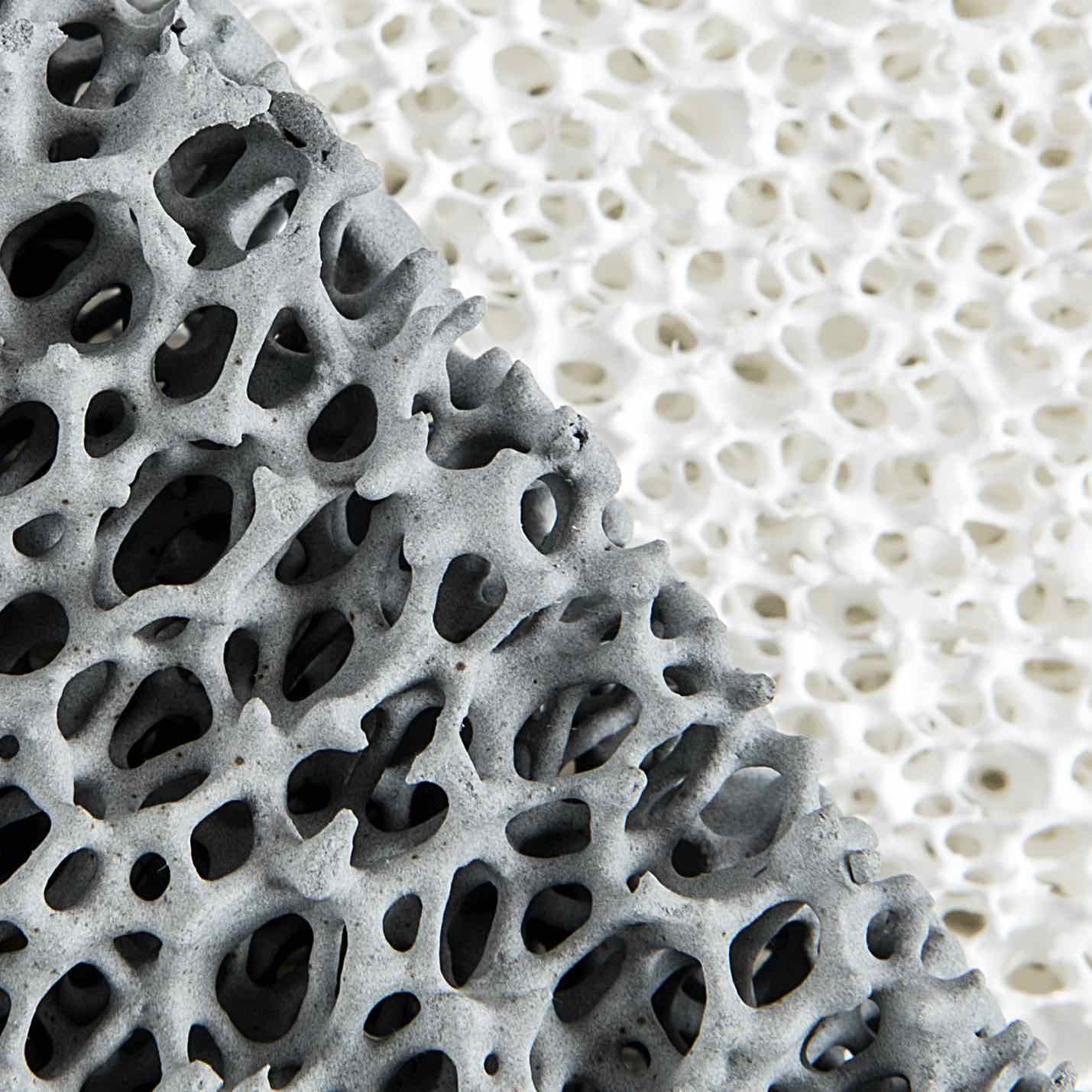}&
\includegraphics[width=0.48\columnwidth]{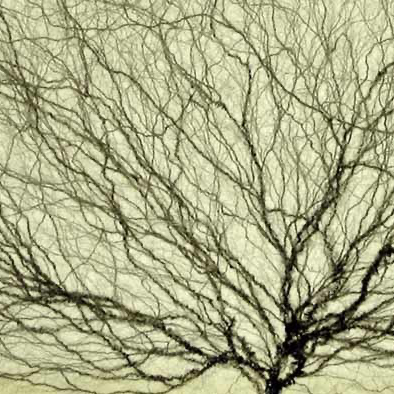}\\
 (c) & (d) \\
\multicolumn{2}{c}{\includegraphics[width=0.96\columnwidth]{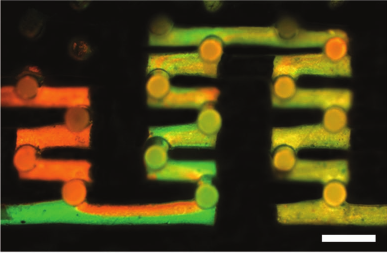}}\\
\multicolumn{2}{c}{(e)}
\end{tabular}
\end{centering}
\caption{Various porous networks as examples of \emph{continuous} porous media: (a) tofu microstructure~\cite{Huang:2018aa}, (b) gyroidal tissue scaffold~ \cite{Melchels:2009aa}, (c) ceramic foam (https://filterceramic.com/alu-ceramic-foam-filter), (d) vascular network of the heart ~\cite{Huang:2009aa}. (e) Mixing of dyes in a 3D micromixer~ \cite{Therriault:2003aa}.}\label{fig:continuous}
\end{figure}

\begin{figure}[h]
\begin{centering}
\begin{tabular}{c c}
\includegraphics[width=0.48\columnwidth]{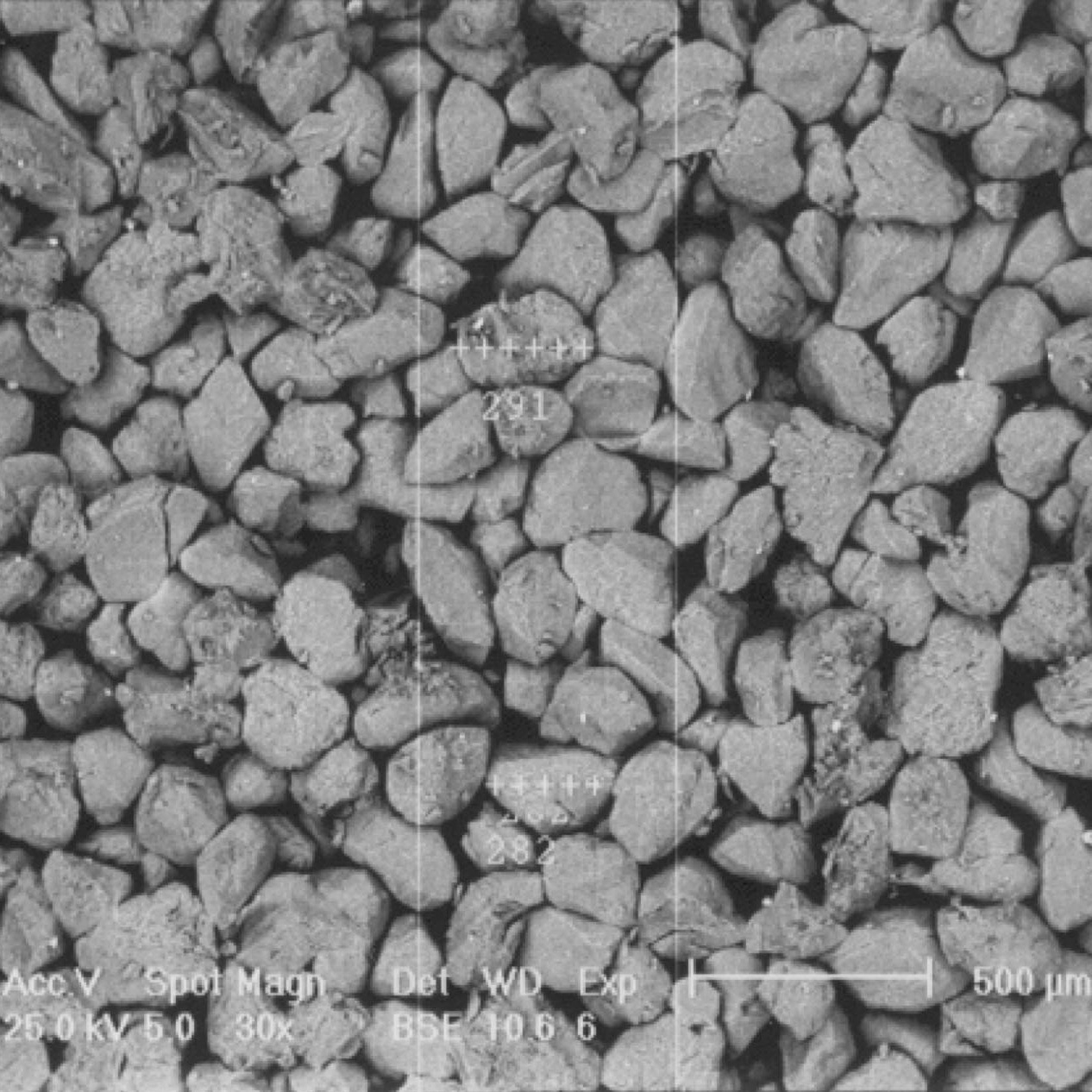}&
\includegraphics[width=0.48\columnwidth]{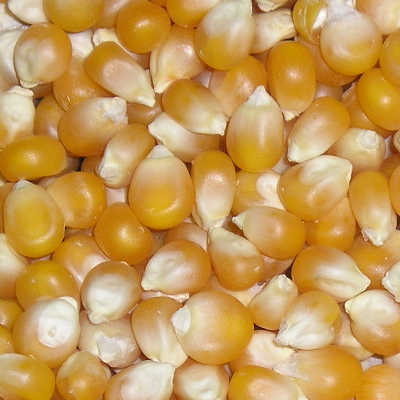}\\
(a) & (b) \\
\includegraphics[width=0.48\columnwidth]{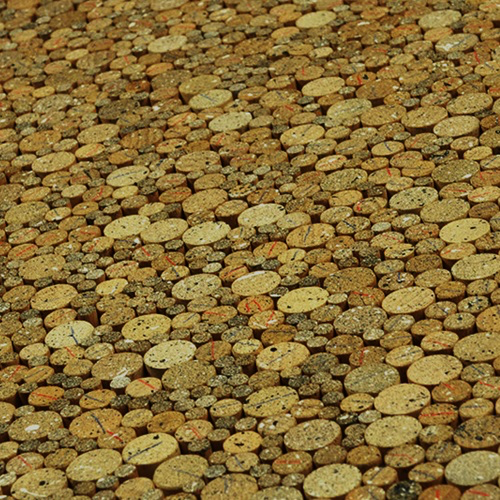}&
\includegraphics[width=0.48\columnwidth]{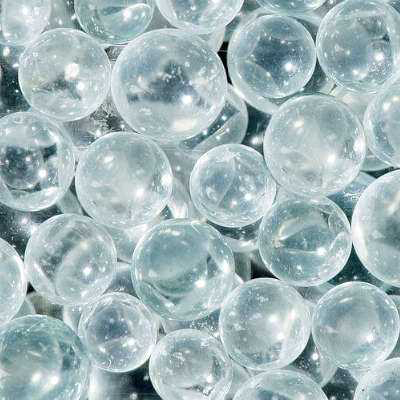}\\
 (c) & (d) \\
\multicolumn{2}{c}{\includegraphics[width=0.96\columnwidth]{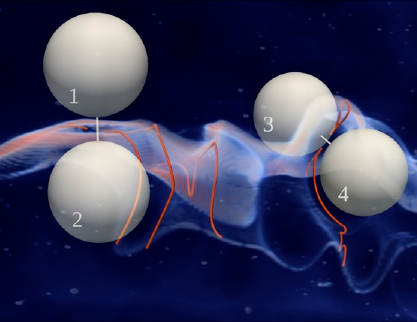}}\\
\multicolumn{2}{c}{(e)}
\end{tabular}
\end{centering}
\caption{Various granular matter as examples of \emph{discrete} porous media: (a) granular sandstone ~\cite{El-Bied:2002aa}, (b) corn kernels, (c) packed corks, (d) glass beads.  (e) Mixing of a continuously injected dye plume through a random glass bead pack~\cite{Heyman:2020aa}. Fluid is index-matched to the beads and only a few beads are shown in grey at 40\% of their true diameter.
}\label{fig:discrete}
\end{figure}

\begin{figure*}[ht]
\centering\includegraphics[width=1.95\columnwidth]{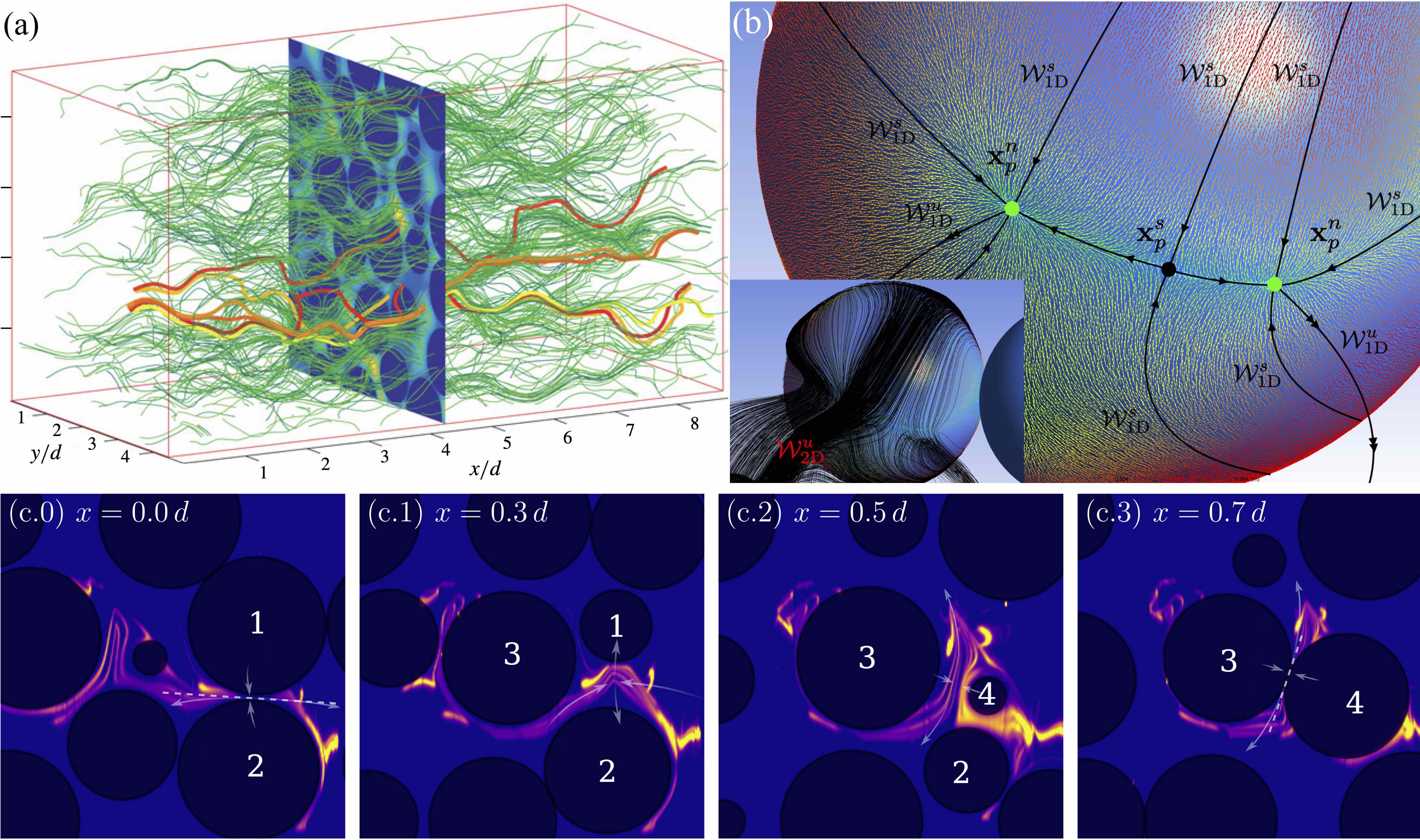}
\caption{Characteristics of chaotic mixing in discrete porous media. (a) Numerically reconstructed trajectories of tracer particles, taken from PIV experiments within a glass bead pack (adapted from \cite{Souzy:2020aa}). (b) Numerically computed skin friction field $\mathbf{u}(\mathbf{x})$ over the surface of a sphere for steady 3-D Stokes flow within a bead pack (other spheres not shown) with node $\mathbf{x}_p^n$ (green) and saddle $\mathbf{x}_p^s$ (black) points, and 1D stable $\mathcal{W}_{1\text{D}}^s$ and unstable $\mathcal{W}_{1\text{D}}^u$ manifolds (black lines). Inset: the same sphere with streamlines shown close to the surface, indicating separation of streamlines in the vicinity of the 2D unstable manifold $\mathcal{W}_{2\text{D}}^U$. Image courtesy Regis Turuban, Scuola Internazionale Superiore di Studi Avanzati, Italy. (c) Sequences of experimental dye trace images for steady flow in a random bead pack at different distances $x$ downstream from the injection point in terms of the bead diameter $d$. These images show that bead contacts systematically trigger stretching and folding of fluid elements leading to the formation of sharp cusps in the dye filament. Numbers label fixed spheres and arrows depict directions of fluid stretching (adapted from \cite{Heyman:2020aa}).}
\label{fig:disc_comp}
\end{figure*}

\begin{figure*}[ht]
\centering\includegraphics[width=1.95\columnwidth]{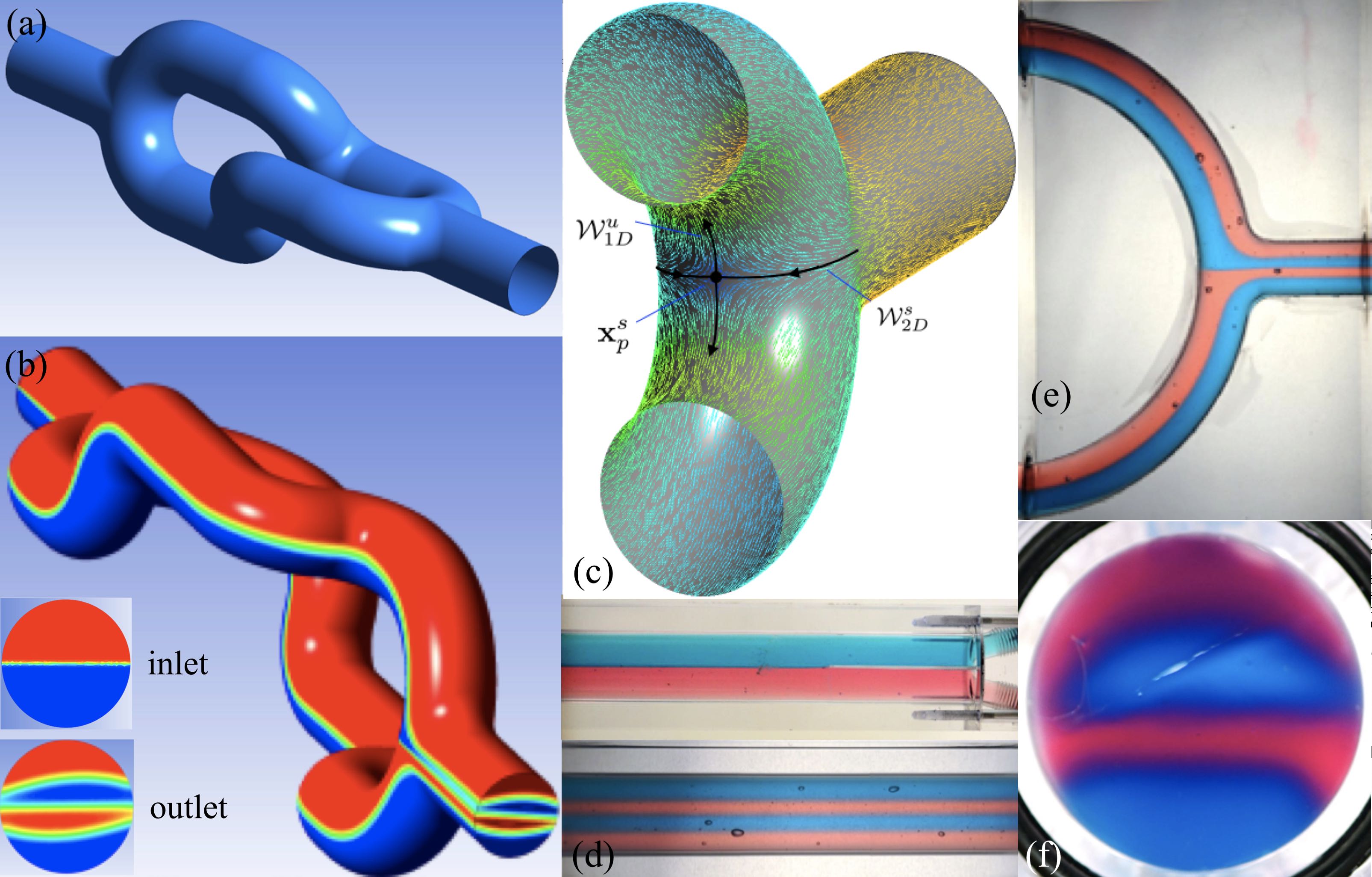}
\caption{Characteristics of chaotic mixing in continuous porous media. (a) An archetypal element of an open (continuous) porous network involving a connected pore branch and merger. (b) Numerical simulation of fluid mixing of a diffusive scalar in (a), illustrating the formation of striated material distributions due to fluid stretching and folding which arises at (c) the saddle-type stagnation point ($\mathbf{x}_p^s$) in the skin friction field. (d) Experimental images of dyed fluids at the inlet (top) and outlet (bottom) of the macroscopic analogue of the  pore branch and merger shown in (a). (e) Detail of dyed fluid distributions near the macroscopic pore merger and (f) cross-section of the dye distribution exiting the pore merger, which agrees well with the scalar distribution shown in (b) (adapted from \cite{Lester:2019aa}).}
\label{fig:cts_comp}
\end{figure*}

Porous media may be broadly classified into two distinct classes based on continuity of the pore-scale solid phase. The first group, \emph{continuous} porous media, has a solids phase which is continuous, such as porous networks. Examples of continuous porous media are shown in Fig.~\ref{fig:continuous}, and include (a) biological materials, (b) tissue scaffolds, (c) ceramic and metallic foams, (d) vascular networks, (e) microfluidic and static mixers and catalyst supports. Chaotic mixing arises in continuous porous media through the continual splitting and recombination of fluid elements at pore branches and mergers, leading to efficient mixing of fluids such as that shown in Fig.~\ref{fig:continuous}e.
The second group, classed as \emph{discrete} porous media, involve a solids phase which consists of a jammed array of discrete particles that have point-wise contacts. Such granular matter includes gravels and sands, packed and jammed media shown in Fig.~\ref{fig:discrete}. Chaotic mixing in discrete porous media arises due to continual distortion of fluid elements as they flow through highly tortuous paths between grains, leading to rapid deformation and mixing of a dye plume, such as that shown in Fig.~\ref{fig:discrete}(e).

This classification into discrete and continuous covers most types of porous media except for fractured media which have constrained mixing dynamics due to the pseudo-2D nature of the fractures, and heterogeneous systems such as granular assemblies of porous particles, which may be considered as multi-scale combinations of continuous and discrete porous media. The fundamental differences in pore-scale architecture between continuous and discrete porous media is an important distinction as the fluid dynamical features at the fluid/solid interface drives chaotic mixing in both continuous~\citep{Lester:2013aa,Lester:2016aa} and discrete~\citep{Turuban:2018aa, Turuban:2019aa} porous media. The characteristics of chaotic mixing in these distinct classes are illustrated in Fig.s~\ref{fig:disc_comp} and \ref{fig:cts_comp}, which show the evolution of particle trajectories and dye plumes in these media as well as the invariant structures (skin friction field, hyperbolic manifolds, critical lines and points) that govern chaotic mixing, as shall be explained in Section~\ref{sec:background}. 

Recent experiments have directly~\cite{Kree:2017aa, Heyman:2020aa} and indirectly~\cite{Souzy:2020aa,Heyman:2021aa} observed chaotic mixing in discrete porous media. Fig.~\ref{fig:disc_comp}(c) shows that chaotic mixing in granular media are generated at contact points between grains, leading to repeated stretching and folding motions of fluid elements that rapidly generate highly elongated and ramified material distributions within the pore space. Similarly, pore-scale chaotic mixing has been observed in the macroscopic analogue of an open porous network~\cite{Lester:2019aa}, and is generated at saddle-type stagnation points on the pore boundary (Fig.~\ref{fig:cts_comp}(e-f)). Both sets of observations are important as chaotic mixing generates complex, highly striated material distributions that can quickly obscure their mechanistic origins. In both cases, saddle or contact points appear to generate fluid stretching and folding motions in the fluid bulk, the hallmarks of chaotic mixing. Despite these detailed observations, there exist important outstanding questions regarding the nature and origins of chaotic mixing in continuous and discrete porous media:

\begin{enumerate}[label=(\roman*)]
\setlength\itemsep{-0.5em}
\item An outstanding challenge is to obtain a unified framework for the description of chaotic mixing across both discrete and continuous porous media.
\item The non-smooth geometry local to contact points in discrete porous media invalidates the topological theory~\cite{Lester:2013aa} that describes chaotic mixing in continuous porous media.
\item It is unknown how exponential fluid stretching (a characteristic of chaotic mixing) arises at contact points in discrete porous media~\cite{Turuban:2019aa}.
\item Although the mechanisms of fluid stretching in continuous porous media are well-understood, an understanding of how folding of fluid elements~\cite{Thiffeault:2004aa} arises in these flows is incomplete.
\item Experimental and numerical studies show evidence of \emph{discontinuous mixing} (involving cutting and shuffling of fluid elements) in continuous porous media that is not understood.
\item Although accurate models for the prediction of the rate of exponential stretching have been proposed for continuous porous media~\cite{Lester:2013aa}, no such analogue exists for discrete porous media.
\end{enumerate}

These questions (i)-(vi) highlight that current understanding of chaotic mixing in both continuous and discrete porous media is incomplete. In this study we address these outstanding questions and develop a unified description of chaotic mixing in discrete and continuous porous media. We highlight the differences and similarities between the mixing mechanisms in both classes of porous media, and connect these theories to the observations and dynamical structures shown in Fig.s~\ref{fig:disc_comp} and \ref{fig:cts_comp}. This unified description of chaotic mixing in porous media provides deep insights into the generation of chaos, and facilitates prediction and optimization of mixing and transport across a wide range of porous materials.

To simplify exposition we limit scope to idealised pore networks and granular media and ignore factors such as surface roughness, distributions of particle pore shapes and sizes. In Section~\ref{sec:conclusions} we discuss the implications of these factors upon the mechanisms of chaotic mixing. The remainder of this paper is organised as follows. To provide a foundation for later results, a brief review of fluid stretching in continuous porous media is given in Section~\ref{sec:background}. Topological equivalence of both porous media classes is established and a unified description of chaotic mixing is developed in Section~\ref{sec:topology}, addressing question (i) above. The mechanisms of fluid stretching in both porous media classes is considered in Section~\ref{sec:stretching}, addressing questions (ii) and (iii). The mechanisms of fluid folding in both porous media classes is considered in Section~\ref{sec:folding}, addressing question (iv). The origins and implications of discontinuous fluid mixing are investigated in Section~\ref{sec:disco}, addressing question (v). In Section~\ref{sec:lyapunov} predictive models for fluid stretching in both porous media classes are developed, addressing question (vi), and conclusions are given in Section~\ref{sec:conclusions}.

\section{Background}
\label{sec:background}

\subsection{Fluid Stretching in Continuous Porous Media}

To facilitate address of questions (i)-(vi) above, we first briefly review the theory~\cite{Lester:2013aa,Lester:2016aa} of fluid stretching in continuous porous media. To distinguish between the dynamical systems notion of ergodic \emph{mixing} of non-diffusive fluid particles and physical mixing of a diffusive solute, throughout we use ``fluid mixing'' or ``chaotic mixing'' to describe the former process and ``diffusive mixing'' for the latter.

\begin{figure}[t]
\begin{centering}
\begin{tabular}{c c}
\includegraphics[width=0.38\columnwidth]{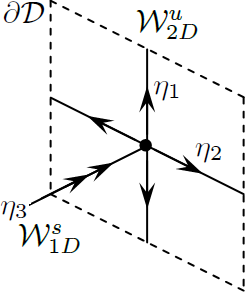}&
\includegraphics[width=0.38\columnwidth]{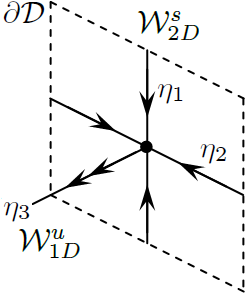}\\
I & II \\
\includegraphics[width=0.48\columnwidth]{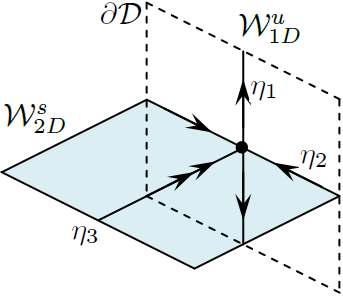}&
\includegraphics[width=0.48\columnwidth]{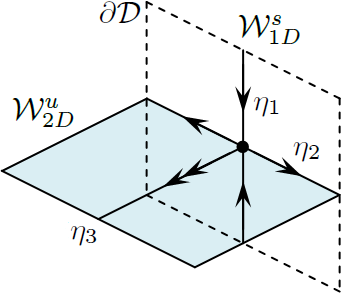}\\
 III & IV
\end{tabular}
\end{centering}
\caption{Schematic of the structure of the skin friction field $\mathbf{u}$ surrounding type I-IV critical points (black dots) on a portion (bounded by the dotted lines) of the fluid boundary $\partial\Omega$ and the associated stable $\mathcal{W}^s$ and unstable $\mathcal{W}^u$ manifolds. The interior 2D manifolds for type III, IV critical points are shown as light blue surfaces. Arrows indicate the eigenvectors of the skin friction gradient tensor, and the double arrows on the streamlines reflect the sum $\eta_1+\eta_2+2\eta_3=0$. Adapted from \cite{Lester:2016aa}.}\label{fig:dist}
\end{figure}

We denote $\mathbf{v}(\mathbf{x})$ as the steady divergence-free 3D fluid velocity field in the pore space $\Omega$ of the continuous porous medium, which satisfies the no-slip condition $\mathbf{v}=\mathbf{0}$ on the exterior fluid/solid boundary $\partial\Omega$. We also define $\mathbf{u}(\mathbf{x})\equiv \partial\mathbf{v}/\partial x^\prime_3$ as the skin friction field on $\partial\Omega$, where $x^\prime_3$ is the spatial coordinate normal to $\partial\Omega$. Mackay~\cite{Mackay:1994aa} shows that critical points $\mathbf{x}_p$ of the skin friction field (where $\mathbf{u}(\mathbf{x}_p)=\mathbf{0}$) play a central role in the generation of chaotic mixing. These critical points generate local exponential fluid stretching if they are \emph{non-degenerate}, i.e., the eigenvalues $\eta_i$, $i=1:3$ of the skin friction gradient tensor $\mathbf{A}\equiv\partial\mathbf{u}/\partial\mathbf{x}$ at $\mathbf{x}=\mathbf{x}_p$ are all non-zero. Without loss of generality we denote the eigenvalues tangent to $\partial\Omega$ as $\eta_1$, $\eta_2$  (with $\eta_1\leqslant\eta_2$), and $\eta_3$ is the interior eigenvalue whose eigenvector is normal to $\partial\Omega$. The divergence-free condition means that the eigenvalues $\eta_i$ satisfy 
\begin{equation}
\eta_1+\eta_2+2\eta_3=0.\label{eqn:skinfr_evals}
\end{equation}
As such, non-degenerate critical points $\mathbf{x}_p$ consist of one stable and two unstable eigenvalues or vice-versa, and form either reattachment ($\eta_3<0$) or separation ($\eta_3>0$) points (Fig.~\ref{fig:dist}).
The four different critical point types depicted in Fig.~\ref{fig:dist} are then:
\begin{enumerate}[label=(\roman*)]
\setlength\itemsep{-0.5em}
\item[I:] separation point $\eta_3>0$, attractor for $\mathbf{u}(\mathbf{x})$, $\eta_1<0$, $\eta_2<0$,
\item[II:] reattachment point $\eta_3<0$, repeller for $\mathbf{u}(\mathbf{x})$, $\eta_1>0$, $\eta_2>0$,
\item[III:] separation point $\eta_3>0$, saddle for $\mathbf{u}(\mathbf{x})$, $\eta_1<0$, $\eta_2>0$,
\item[IV:] reattachment point $\eta_3<0$, saddle for $\mathbf{u}(\mathbf{x})$, $\eta_1<0$, $\eta_2>0$.
\end{enumerate}

Type I and II critical points are node points, whereas type III and IV critical points form saddle points. If, subject to technical conditions, the critical points are \emph{strongly hyperbolic}~\cite{Surana:2006aa}, the eigenvector directions indicated in Fig.~\ref{fig:dist} impart local exponential stretching of the fluid which persists away from the critical points as a continuation of these eigenvectors in the form of 1D or 2D stable $\mathcal{W}^s$ and unstable $\mathcal{W}^u$ hyperbolic manifolds. Hence the stable and unstable manifolds from type III and IV saddle points represent material surfaces that are respectively exponentially contracting or expanding in the fluid bulk. Unless symmetry conditions are imposed~\cite{Haller:1998ab}, these invariant 2D hyperbolic manifolds intersect transversely in the fluid domain, forming a \emph{heteroclinic tangle}, the hallmark of chaotic dynamics in Hamiltonian systems~\cite{Ott:2002aa}. In practice, a single transverse intersection of stable $\mathcal{W}^s_{\text2D}$ and unstable $\mathcal{W}^u_{\text2D}$ 2D manifolds implies a chaotic tangle of infinitely many~\cite{Ottino:1989aa}, leading to strong fluid stretching and folding and chaotic mixing. 

MacKay~\cite{Mackay:1994aa} shows that interior 2D unstable manifolds form a \emph{skeleton of the flow} that is comprised of these surfaces of locally minimal transverse flux for diffusive solutes and so organise both fluid and solute transport and mixing. Conversely, if the interior hyperbolic manifolds are 1D, their impact on fluid transport is minimal. Thus, only saddle points give rise to interior 2D hyperbolic manifolds and chaotic mixing.

\begin{figure}[t]
\begin{centering}
\begin{tabular}{c}
\includegraphics[width=0.95\columnwidth]{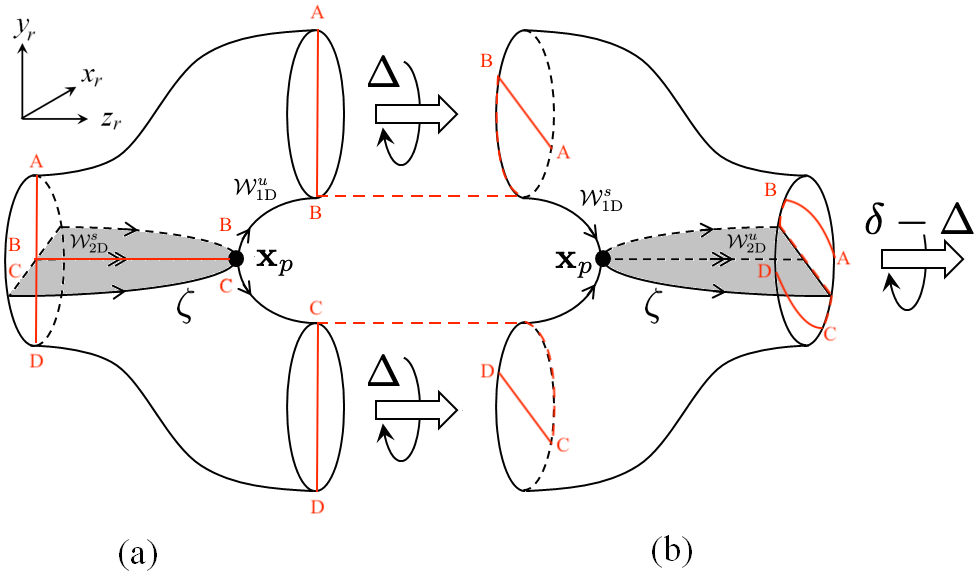}
\end{tabular}
\end{centering}
\caption{Schematic of a pore branch (a) and merger (b) in continuous porous media. Non-degenerate critical points $\mathbf{x}_p$ (black dots) generate 2D hyperbolic stable $\mathcal{W}^s_{2\text{D}}$ and unstable $\mathcal{W}^u_{2\text{D}}$ manifolds (grey) which are surfaces of locally minimum transverse flux. The angles $\Delta$, $\delta$ characterise the relative orientation of pore branch and merger elements in the pore network. The red lines pertain to Section~\ref{sec:disco}, and depict evolution of a continuously injected material line (red). Segments AB and CD of this material line are separated by the critical line $\zeta$ in the pore branch and are advected through different branches of these pores. Dotted red lines indicate connected material elements that are not resolved by the spatial maps $\mathcal{M}$, $\mathcal{M}^{-1}$ defined in equation (\ref{eqn:Mmap}).}\label{fig:sheet_cut}
\end{figure}

The prevalence of saddle points in $\partial\Omega$ can be related to the topological complexity of the continuous porous medium, given by the number density $\rho(\chi)$ of the Euler characteristic $\chi(\Omega)$, which is given~\cite{Vogel:2002aa} in terms of the numbers of pores $N$, redundant connections $C$ and completely enclosed cavities $H$ in the medium as
\begin{equation}
\chi(\Omega)=N-C+H.\label{eqn:chi}
\end{equation}
Typically $\rho(\chi)$ is strongly negative (-200 -- -500 mm$^{-3}$) \cite{Scholz:2012aa,Vogel:2002aa}, reflecting the inherent topological complexity of all porous materials. For closed bounded manifolds $\Omega$, the Euler characteristic of $\Omega$ is related~\cite{Armstrong:2019aa} to that of its boundary $\delta\Omega$ as 
\begin{equation}
\chi(\delta\Omega)=2\chi(\Omega),\label{eqn:genusbdry}
\end{equation}
hence the Euler characteristic of the pore boundary is also strongly negative. The Poincar\'e-Hopf theorem connects the Euler characteristic $\chi(\delta\Omega)$ of the pore boundary to the sum of indices $\gamma_p$ of stagnation points $\mathbf{x}_p$ in $\delta\Omega$ as 
\begin{equation}
\chi(\delta\Omega)=2(1-g)=\sum\gamma_p(\mathbf{x}_p)=n_n-n_s,\label{eqn:P_Hopf}
\end{equation}
where $g$ is the topological genus of the pore boundary, $\gamma_p=+1$ for node points, and $\gamma_p=-1$ for saddle points, and so $n_n$ and $n_s$ respectively denote the number of node and saddle points. As such, $|\rho(\chi)|$ provides a lower bound for the number density of saddle points which, as shown in Fig.~\ref{fig:sheet_cut}, naturally arise at pore branches and mergers. Although the specific geometry of pore branches and mergers may vary significantly, the basic topology shown in Fig.~\ref{fig:sheet_cut} that drives chaotic mixing is universal to all continuous porous media.

As indicated by the angle $\theta\equiv\Delta-\delta$ in Fig.~\ref{fig:sheet_cut}, the 2D stable and unstable manifolds emanating from the saddle points may either intersect smoothly ($\theta=0$) or transversely ($\theta\neq0$), leading respectively to either zero net stretching or chaotic mixing. Indeed, orthogonally oriented manifolds ($\Delta=\pi/2$, $\delta=0$) correspond to a 3D fluid mechanical analogue of the baker's map~\cite{Carriere:2007aa,Lester:2013aa}.

\subsection{Advective Mapping in Continuous Porous Media}\label{subsec:advmap_cts}

\begin{figure}[t]
\begin{centering}
\begin{tabular}{ccc}
\hline\\
\multicolumn{3}{l}{(i): ordered network (3D baker's map), $\lambda_\infty=\ln 2$}\\
\includegraphics[width=0.31\columnwidth]{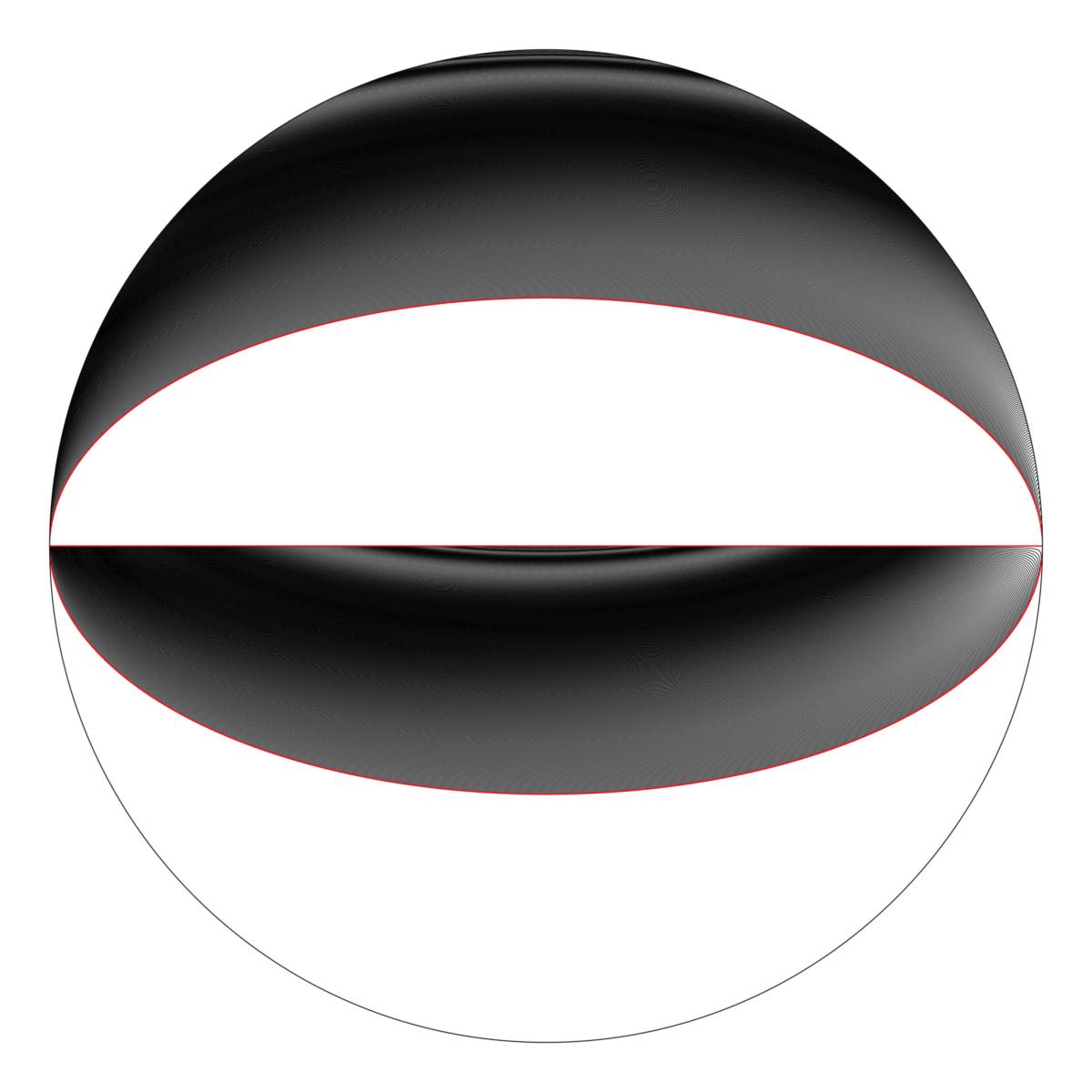}&
\includegraphics[width=0.31\columnwidth]{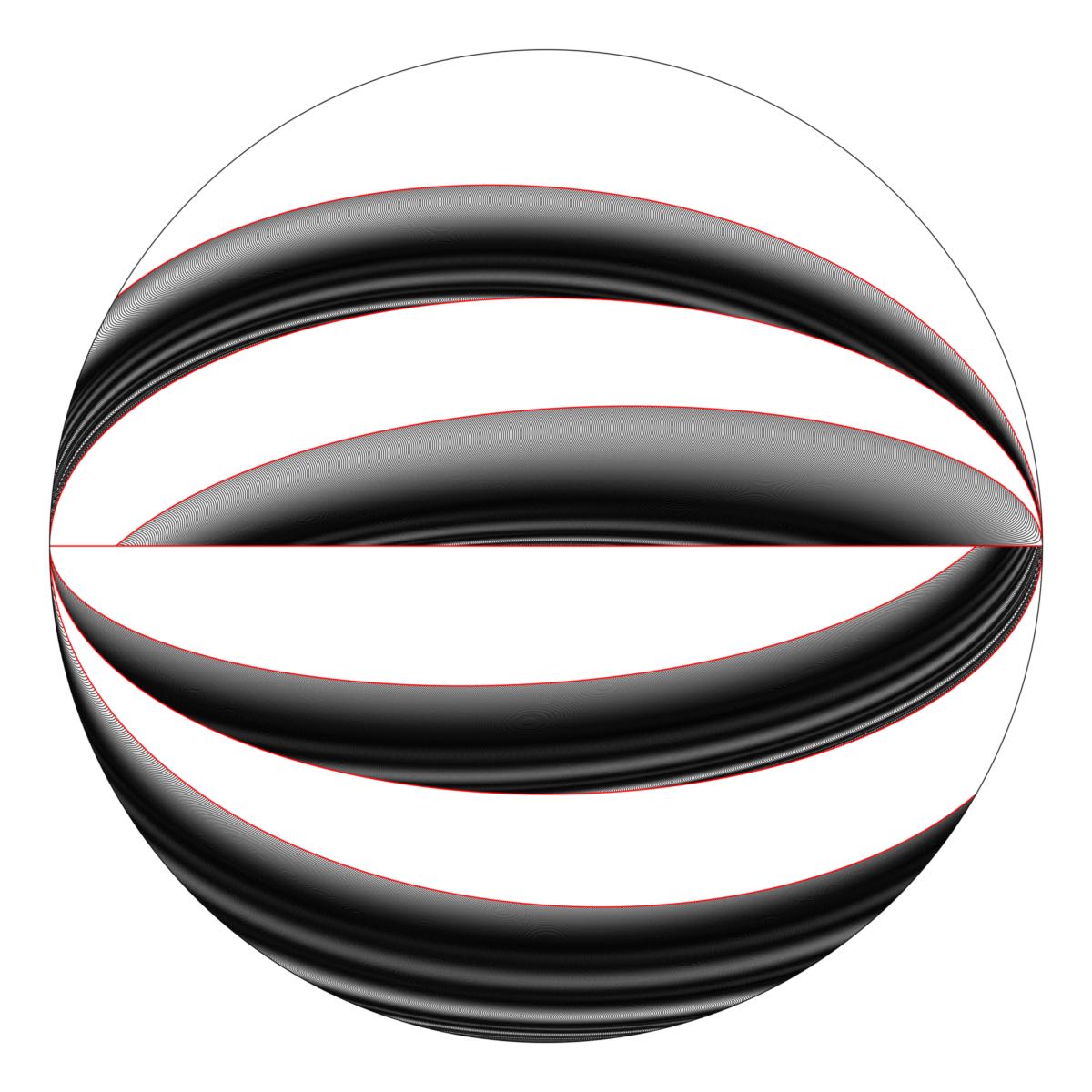}&
\includegraphics[width=0.31\columnwidth]{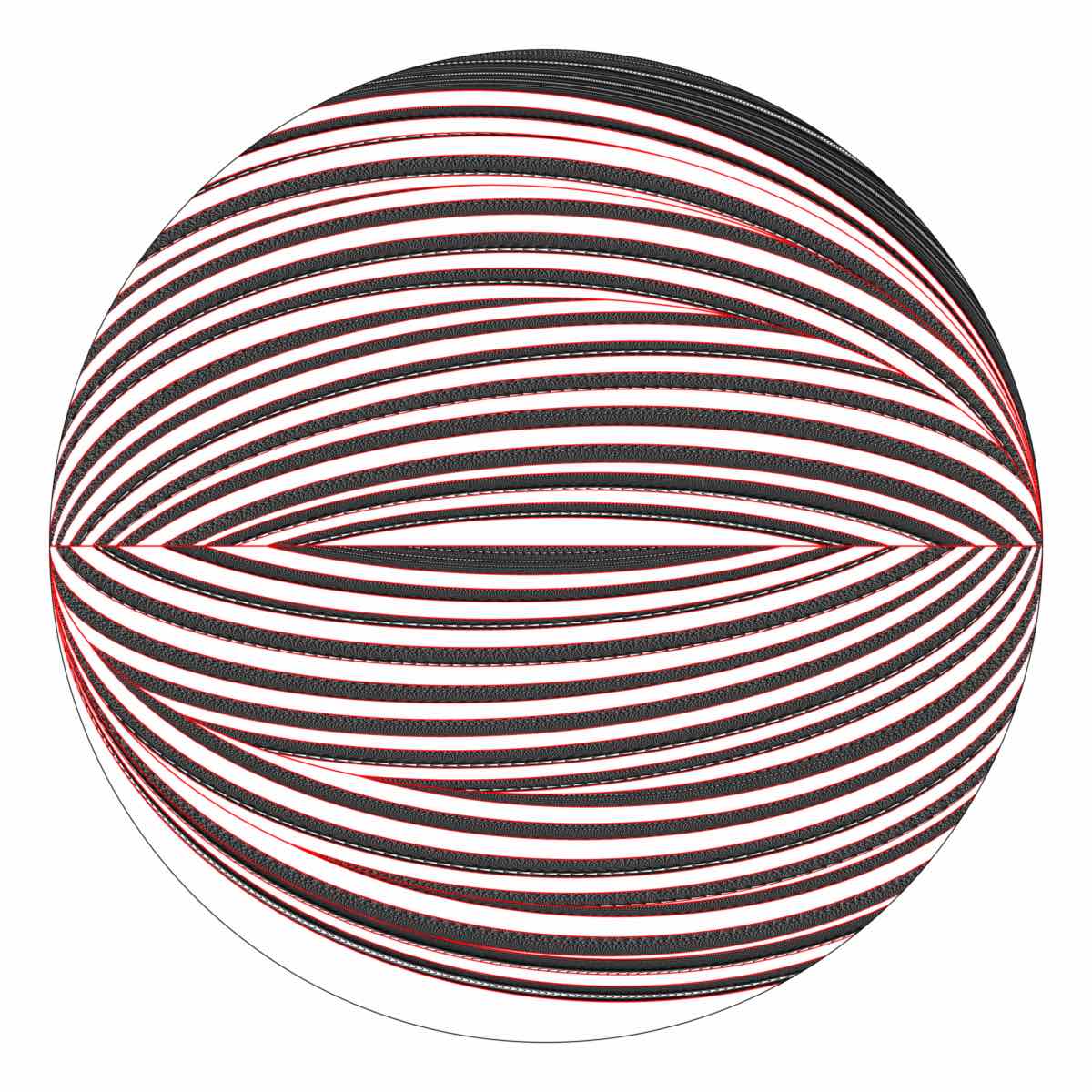}\\
n=1 & n=2 & n=5\\
\hline\\
\multicolumn{3}{l}{(ii): ordered network, chaotic mixing, $\lambda_\infty\approx 0.403$}\\
\includegraphics[width=0.31\columnwidth]{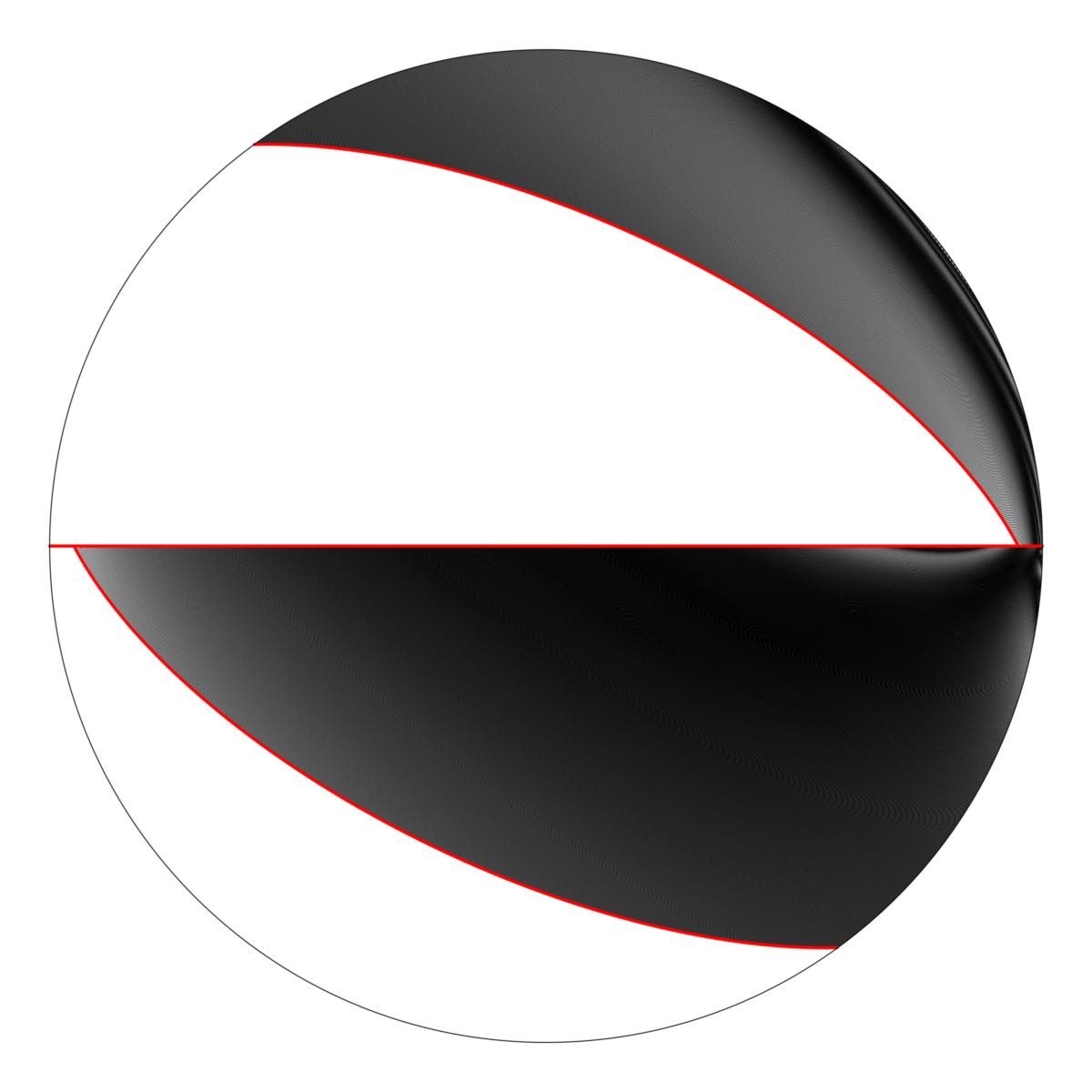}&
\includegraphics[width=0.31\columnwidth]{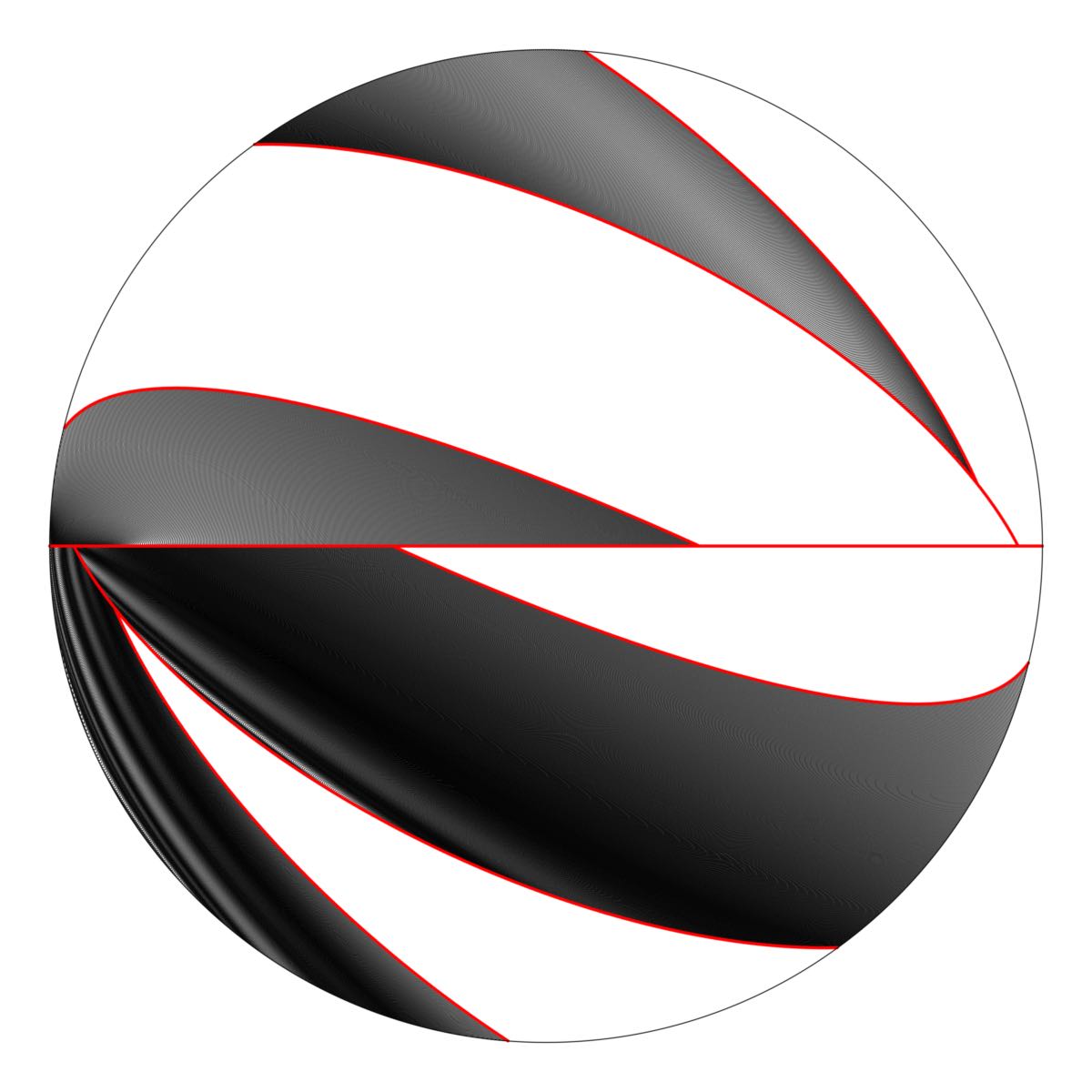}&
\includegraphics[width=0.31\columnwidth]{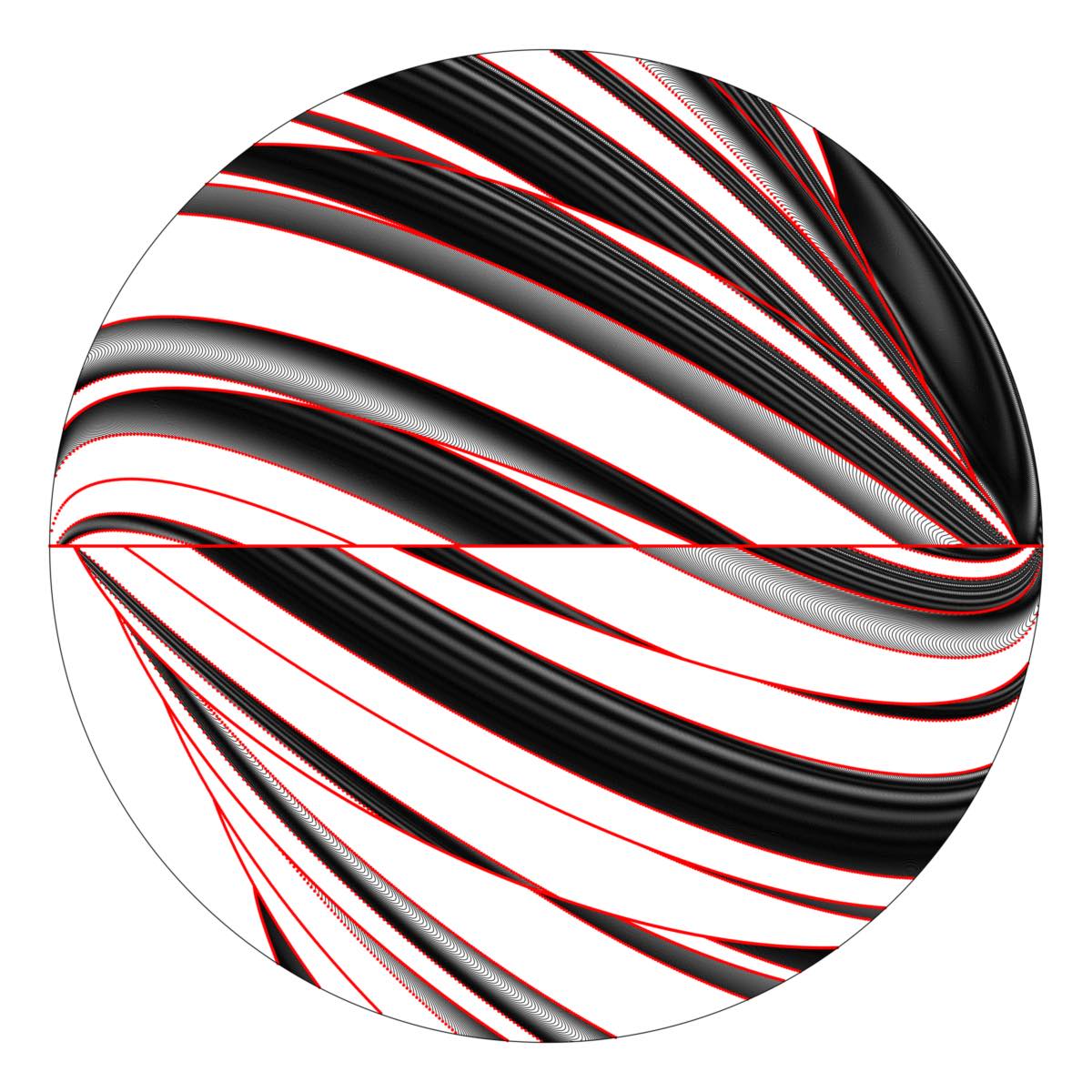}\\
n=1 & n=2 & n=5\\
\hline\\
\multicolumn{3}{l}{(iii): ordered network, non-chaotic, $\lambda_\infty=0$}\\
\includegraphics[width=0.31\columnwidth]{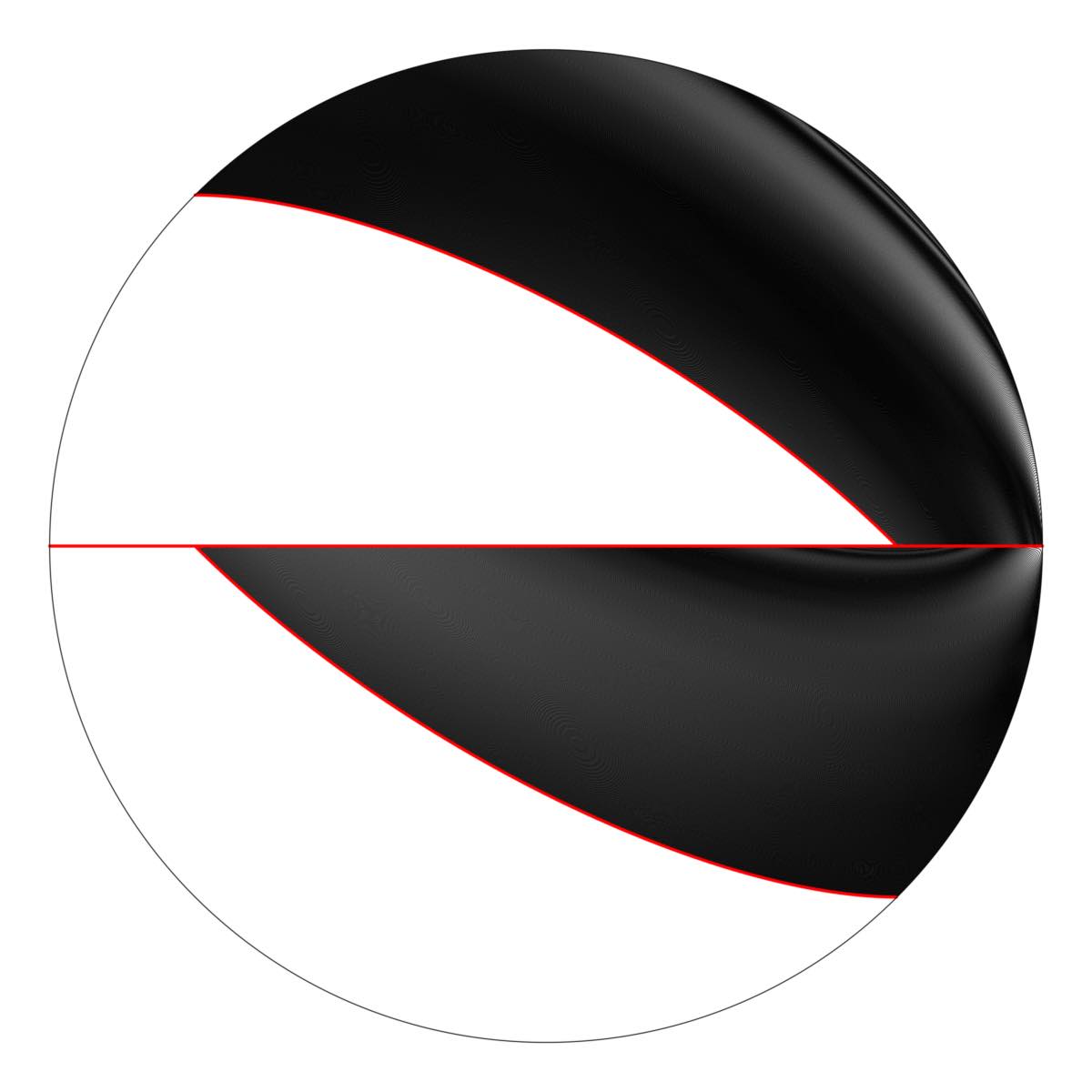}&
\includegraphics[width=0.31\columnwidth]{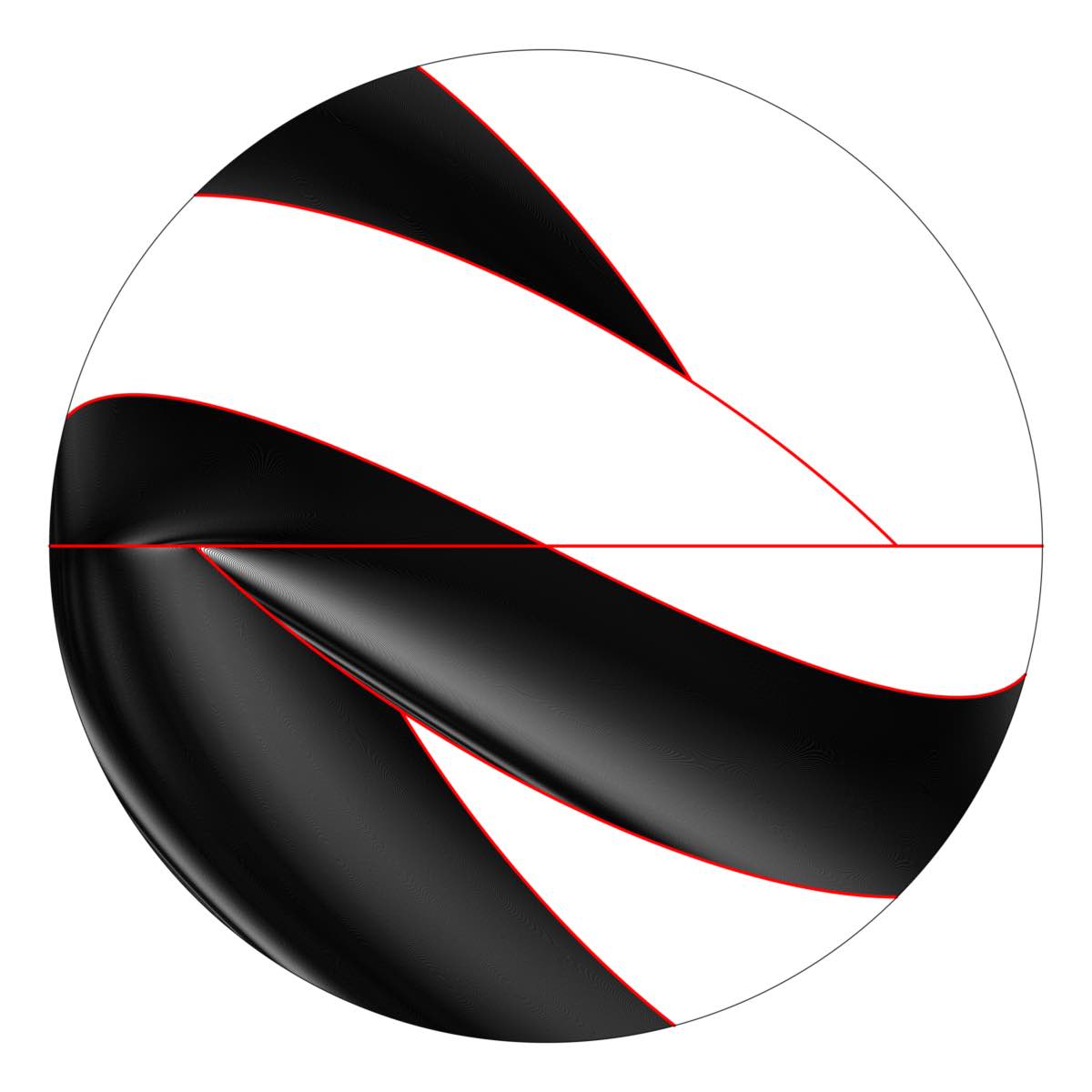}&
\includegraphics[width=0.31\columnwidth]{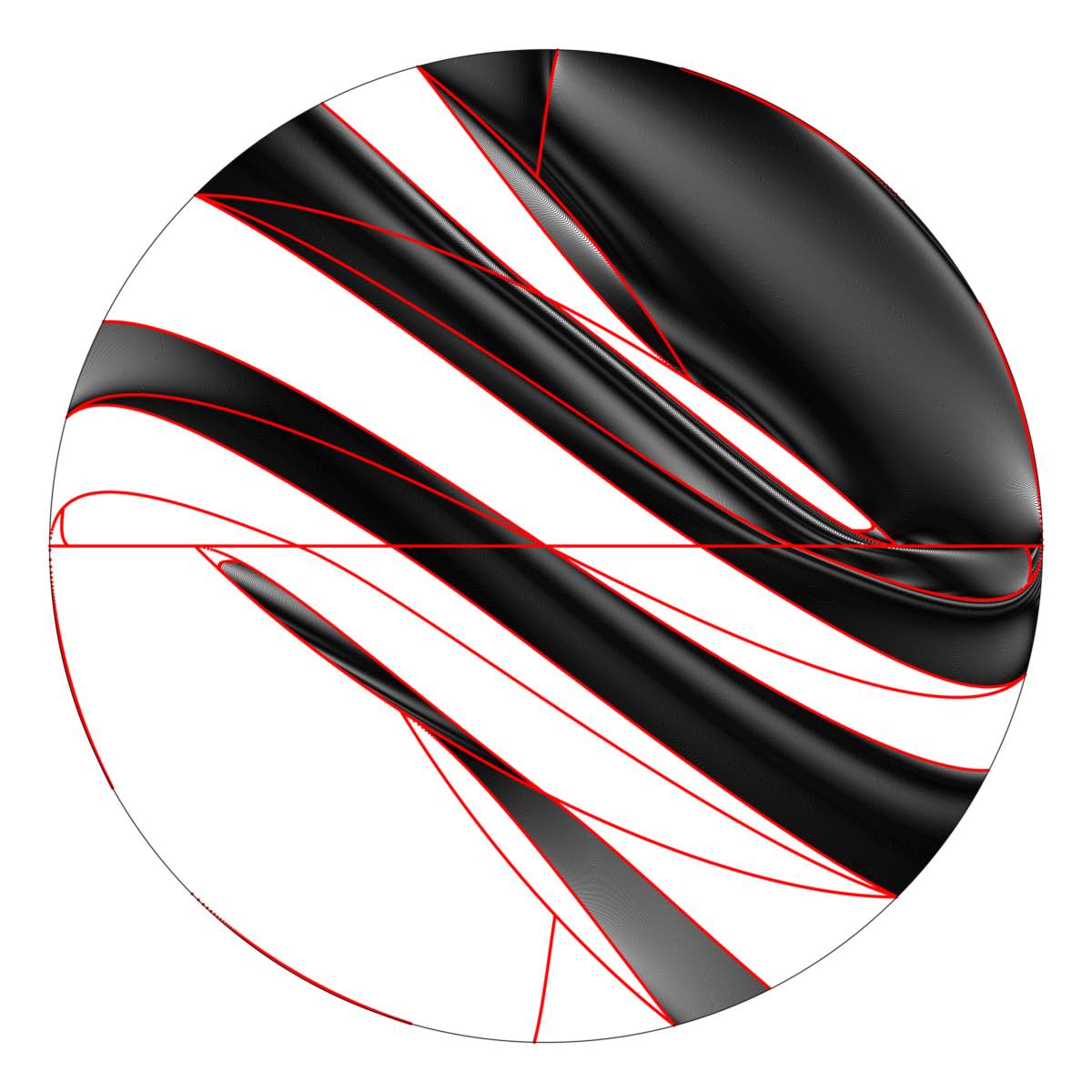}\\
n=1 & n=2 & n=5\\
\hline\\
\multicolumn{3}{l}{(iv): random network, chaotic mixing, $\bar\lambda_\infty\approx 0.118$}\\
\includegraphics[width=0.31\columnwidth]{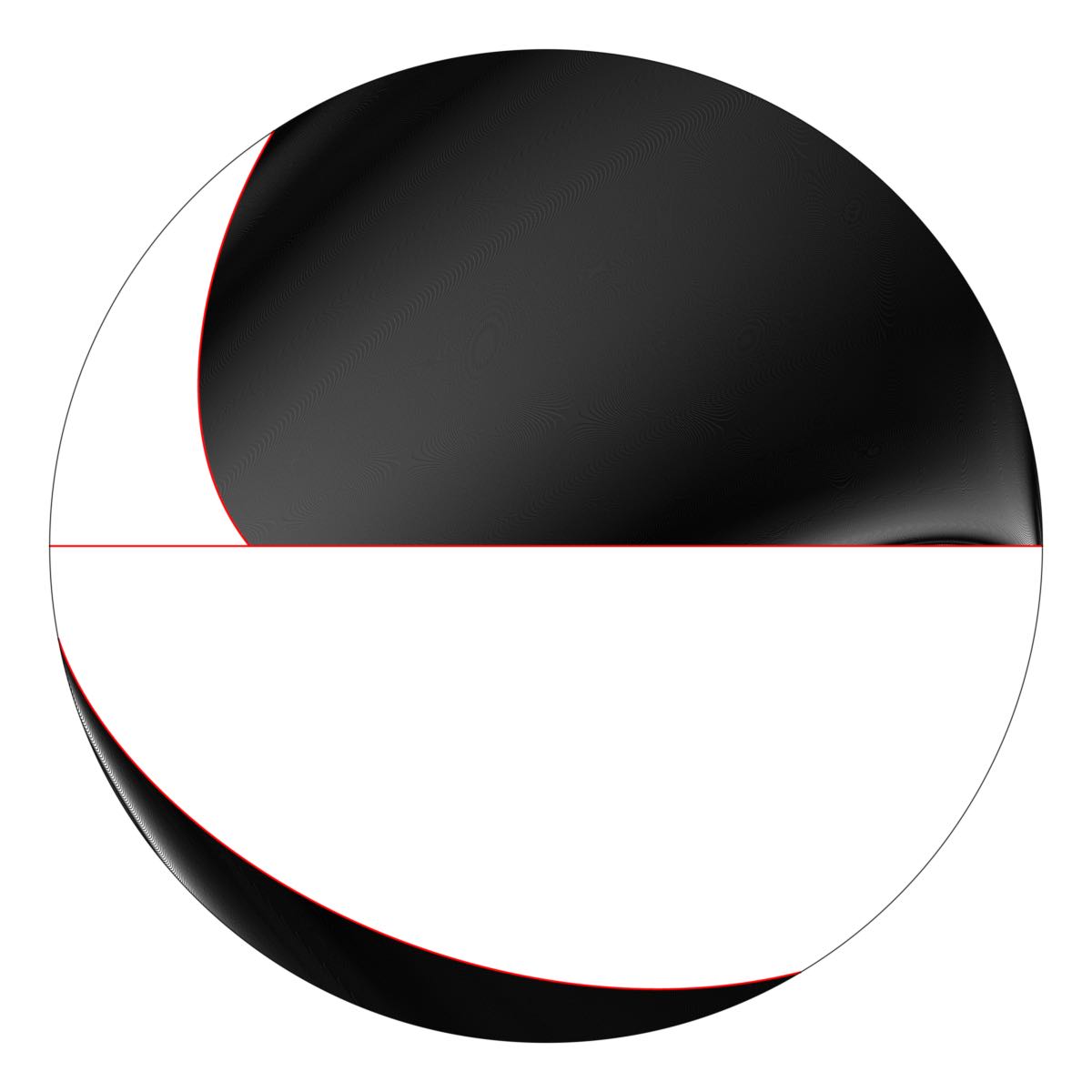}&
\includegraphics[width=0.31\columnwidth]{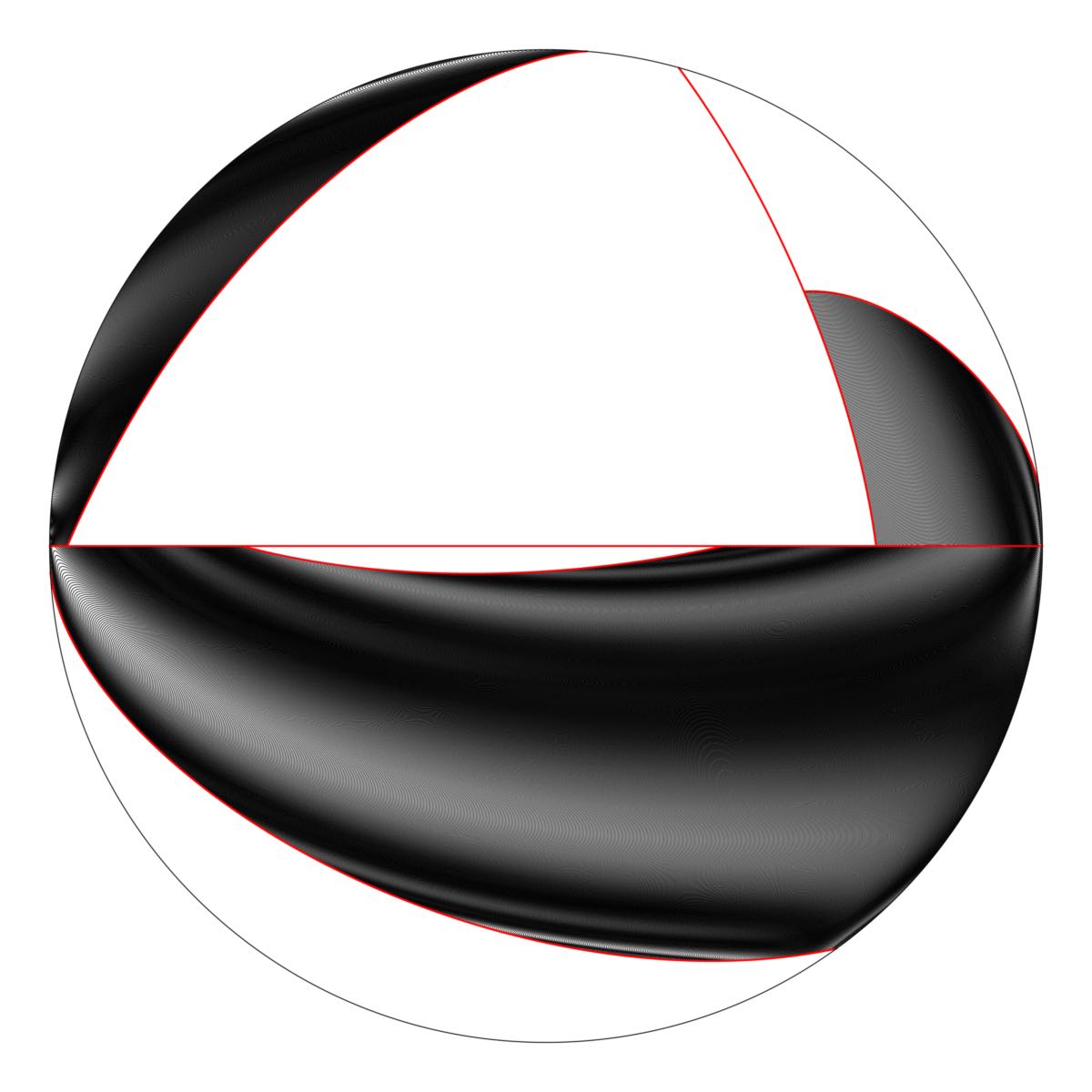}&
\includegraphics[width=0.31\columnwidth]{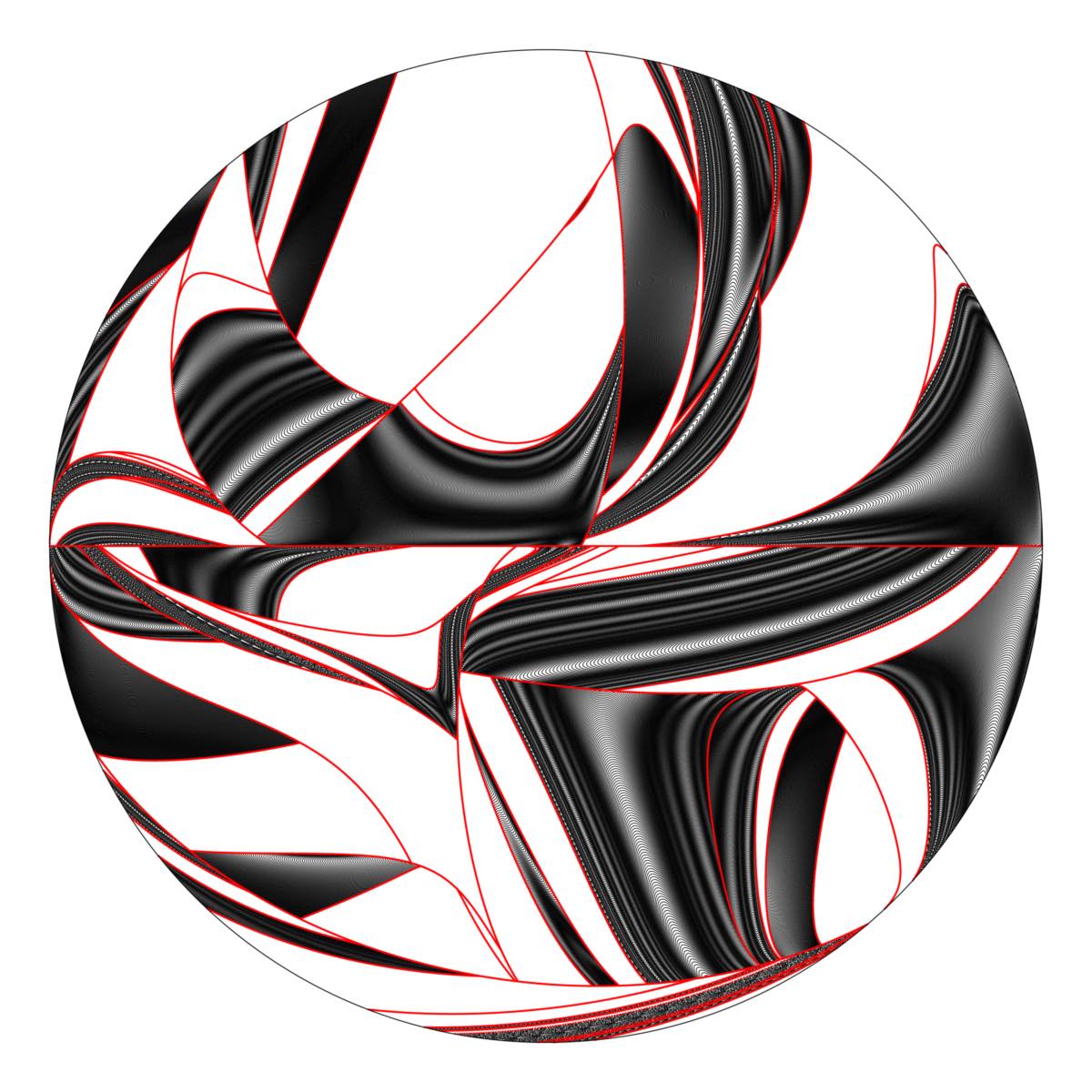}\\
n=1 & n=2 & n=10
\end{tabular}
\end{centering}
\caption{Mixing of fluid elements (black/white) and evolution of web of discontinuities (red, discussed in Section VII) via the map $L(\delta,\Delta)$ with pore number $n$ for ordered (i-iii) and random (iv) pore networks.}\label{fig:web}
\end{figure}

This basic mechanism of fluid stretching has been used to develop a simple yet remarkably accurate model of advection in pore networks. Numerical simulations~\cite{Lester:2014ab,Lester:2016ab} of Stokes flow through the pore branch shown in Fig.~\ref{fig:sheet_cut}(a) have shown that particle advection from a pore branch inlet to either outlet is well-approximated by the dimensionless spatial map $\mathcal{M}=\mathcal{M}_n\circ\mathcal{M}_a$
\begin{align}
&\mathcal{M}:(x_r,y_r)\mapsto
\begin{cases}
(x_r,2y_r-\sqrt{1-x_r^2}),\,\,\,\,&\text{if}\,\,y_r>0,\\
(x_r,2y_r+\sqrt{1-x_r^2}),\,\,\,\,&\text{if}\,\,y_r\leqslant 0,
\end{cases}\label{eqn:Mmap}
\end{align}
which is comprised of affine $\mathcal{M}_a$ and non-affine $\mathcal{M}_n$ parts
\begin{align}
& \mathcal{M}_a:(x_r,y_r)\mapsto(x_r,2y_r),
\label{eqn:M_aff}\\
&\mathcal{M}_n:(x_r,y_r)\mapsto
\begin{cases}
(x_r,y_r-\sqrt{1-x_r^2}),\,\,&\text{if}\,\,y_r>0,\\
(x_r,y_r+\sqrt{1-x_r^2}),\,\,&\text{if}\,\,y_r\leqslant 0,
\end{cases}
\label{eqn:M_naff}
\end{align}
where $(x_r,y_r)$ are the local transverse coordinates within a pore branch inlet or outlet, such that $x_r^2+y_r^2=0,1$ respectively correspond to the pore centre and boundary. The inverse spatial map $\mathcal{M}^{-1}=\mathcal{M}_a^{-1}\circ\mathcal{M}_n^{-1}$ accurately approximates advection of fluid particles over a pore merger, and the temporal $\mathcal{T}^*$ map also accurately approximates the advection time $t$ of fluid particles over a pore branch as
\begin{align}
&\mathcal{T}^*=\mathcal{T}\circ\mathcal{M},\quad\mathcal{T}:t\mapsto t+\frac{1}{1-x_r^2-y_r^2}\label{eqn:Tstar},
\end{align}
whereas the inverse temporal map $\mathcal{T}\circ\mathcal{M}^{-1}$ approximates advection of fluid particles over a pore merger. These maps may be extended to pore branches and mergers with respective orientations $\varphi_b$, $\varphi_m$ in the $x_r-y_r$ plane via the reoriented spatial maps
\begin{align}
&\mathcal{S}_b(\varphi_b)=R(\varphi_b)\circ\mathcal{M}\circ R^{-1}(\varphi_b),\\
&\mathcal{S}_m(\varphi_m)=R(\varphi_m)\circ\mathcal{M}^{-1}\circ R^{-1}(\varphi_m),\label{eqn:spatial_maps}
\end{align}
where $R$ is the rotation operator about the pore centre. The reorientation angle $\Delta$ between the pore branch and merger in Fig.~\ref{fig:sheet_cut} is $\Delta\equiv\varphi_m-\varphi_b$. Rapid advection of fluid particles in ordered and random model porous networks is performed via the map $\mathcal{S}$ over a coupled pore branch and merger (termed a couplet)
\begin{equation}
\begin{split}
\mathcal{S}=&\mathcal{S}_b(\varphi_b)\circ\mathcal{S}_m(\varphi_m),\\
=&R(\varphi_b)\circ \mathcal{D}(\Delta)\circ R^{-1}(\varphi_b),\label{eqn:S}
\end{split}
\end{equation}
where $\mathcal{D}(\Delta)=\mathcal{M}\circ R(\Delta)\circ\mathcal{M}^{-1}\circ R^{-1}(\Delta)$. For a series of $n$ concatenated couplets, the composite map $\Lambda_n$ is then 
\begin{equation}
\begin{split}
\Lambda_n
=&R(\varphi_{b,n})\circ\left(\prod_{j=1}^n L(\delta_j,\Delta_j)\right)\circ R^{-1}(\varphi_{b,n}),\label{eqn:Lambda_map}
\end{split}
\end{equation}
where $L(\delta,\Delta)\equiv R(\delta)\circ\mathcal{D}(\Delta)$ and $\delta$ is the angle between couplets: $\delta_j=\varphi_{b,j+1}-\varphi_{b,j}$ for $j=1:n-1$ and $\delta_n=\varphi_{b,1}-\varphi_{b,n}$. Previous studies~\cite{Lester:2014aa,Lester:2014ab} have considered both random pore networks with a uniform distribution of the reorientation angles $\delta_j\sim U[0,2\pi]$, $\Delta_j\sim U[0,2\pi]$ and ordered pore networks with deterministic sequences of angles $\delta_j$, $\Delta_j$. Consider the following ordered and random pore network types that exhibit chaotic and non-chaotic behaviour:
\begin{enumerate}[label=(\roman*)]
\setlength\itemsep{-1em}
\item ordered network corresponding to 3D baker's map: $\Delta=\pi/2$, $\delta=0$, $\lambda_\infty=\ln 2\approx 0.693$\\
\item ordered network with chaotic mixing: $\Delta=\pi/2$, $\delta=\pi/6$, $\lambda_\infty=\ln[(5\sqrt{3}+\sqrt{11})/8]\approx 0.403$\\
\item ordered network with non-chaotic mixing: $\Delta=\pi/4$, $\delta=\pi/4$, $\lambda_\infty=0$\\
\item random network with chaotic mixing: $\quad\quad\quad\quad\Delta\sim U[0,\pi]$, $\delta\sim U[0,\pi]$, $\bar\lambda_\infty\approx 0.118$
\end{enumerate}
As expected, the ordered network (i) shown in Fig.~\ref{fig:web} corresponding to the Baker's map  exhibits the fastest mixing, whereas the random network (iv) is significant slower.

\subsection{Lyapunov Models in Continuous Porous Media}
\label{subsec:lyapunov_cts}

These advective maps can also generate accurate predictions of the fluid stretching in continuous porous media, as quantified by the dimensionless (infinite-time) Lyapunov exponent
\begin{equation}
\lambda_\infty\equiv\hat{\lambda}_\infty\frac{\ell}{\langle v_1\rangle},
\end{equation}
where $\ell$ the characteristic length scale of the medium (such as the length of a pore branch/merger) and $\langle v_1\rangle$ is the mean longitudinal velocity. Linearisation of $\mathcal{M}$ in (\ref{eqn:Lambda_map}) yields~\cite{Lester:2013aa} an expression for $\lambda_\infty$ in ordered porous media with fixed $\delta$, $\Delta$
\begin{equation}
\lambda_\infty(\delta,\Delta)=\ln|\xi+\sqrt{\xi^2-1}|,\,\,\xi=\frac{9}{8}\cos\delta-\frac{1}{8}\cos(2\Delta+\delta),\label{eqn:Lyapunov_cts_ordered}
\end{equation}
where $\lambda_\infty=0$ for $|\xi|\leqslant 1$, and maximum deformation ($\lambda_\infty=\ln 2$) occurs for $\Delta=(n+1/2)\pi$, $\delta=n\pi$, $n=0,1,\dots$. Hence ordered 3D porous networks represent extreme cases with respect to fluid stretching and deformation. As shown in Fig.~\ref{fig:web}, while a large class ($|\xi|\leqslant 1$) of ordered media do not exhibit chaotic advection ($\lambda_\infty=0$), some ordered networks exhibit maximal stretching ($\lambda_\infty=\ln 2$ ) associated with the baker's map. Chaotic mixing also occurs in random porous networks with uniformly distributed reorientation angles ($\Delta, \delta\sim U[0,2\pi]$), where the Lyapunov exponent is given by the ensemble average
\begin{equation}
\bar{\lambda}_\infty\equiv\frac{1}{4\pi^2}\int_0^{2\pi}\int_0^{2\pi}\lambda_\infty(\delta,\Delta)\,d\delta\, d\Delta\approx 0.1178,\label{eqn:Lyapunov_cts_random}
\end{equation}
which agrees very well with direct numerical simulations~\cite{Lester:2013aa}. For random pore networks, persistent fluid stretching arises because material lines undergoing stretching rotate toward the stretching direction which amplifies stretching, whereas material lines undergoing contraction rotate away from the contracting direction, retarding contraction. The Lyapunov exponent $\bar\lambda_\infty\approx0.1178$ is thus a manifestation of the asymmetry between these stretching and contraction processes.

While the Lyapunov exponent characterises fluid stretching that is integral to continuous mixing, it does not account for contributions to discontinuous mixing, which arise from cutting and shuffling (CS) of fluid elements rather than stretching and folding (SF). Indeed, while case (iii) in Fig.~\ref{fig:web} is non-chaotic and has zero Lyapunov exponent, significant mixing of fluid elements can still be observed due to the cutting and shuffling of fluid elements. Similarly, cases (i-ii) and (iv) exhibit a combination of continuous (SF) and discontinuous (CS) mixing. The results presented in Fig.~\ref{fig:web} will be used in Section VII to explore the interplay of SF and CS in continuous porous media.


\section{Topology of Discrete and Continuous Porous Media}
\label{sec:topology}

\begin{figure*}[ht]
\begin{centering}
\begin{tabular}{c c c}
\includegraphics[width=0.5\columnwidth]{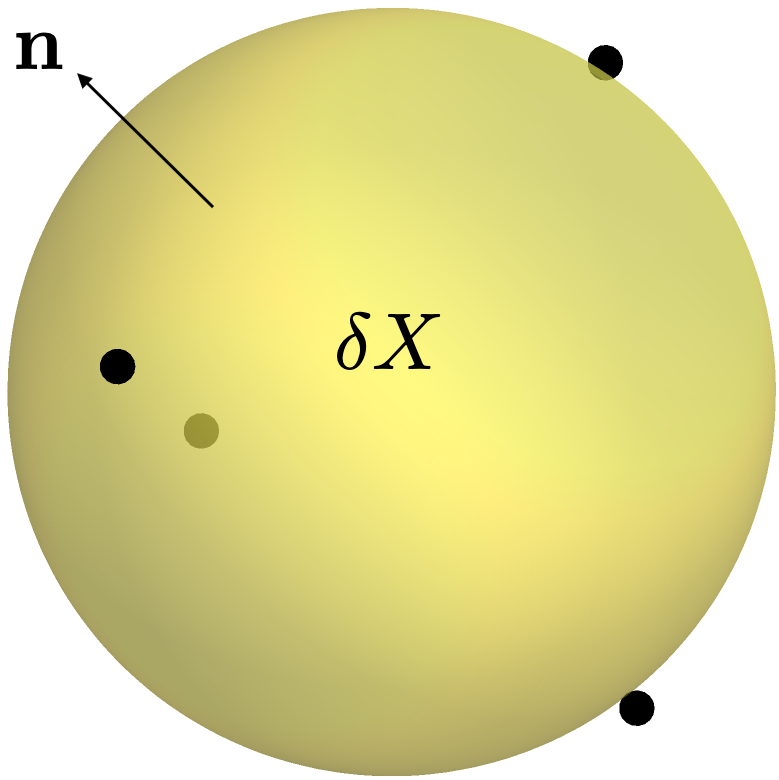}&
\hspace{10mm}
\includegraphics[width=0.5\columnwidth]{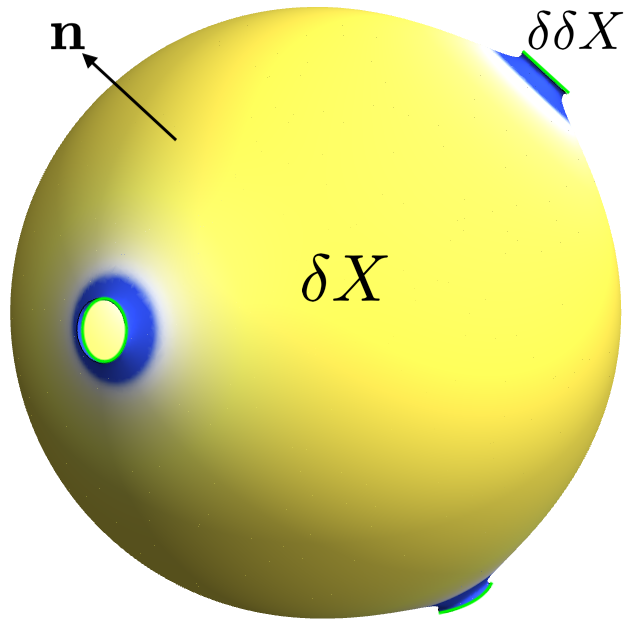}&
\hspace{10mm}
\includegraphics[width=0.62\columnwidth]{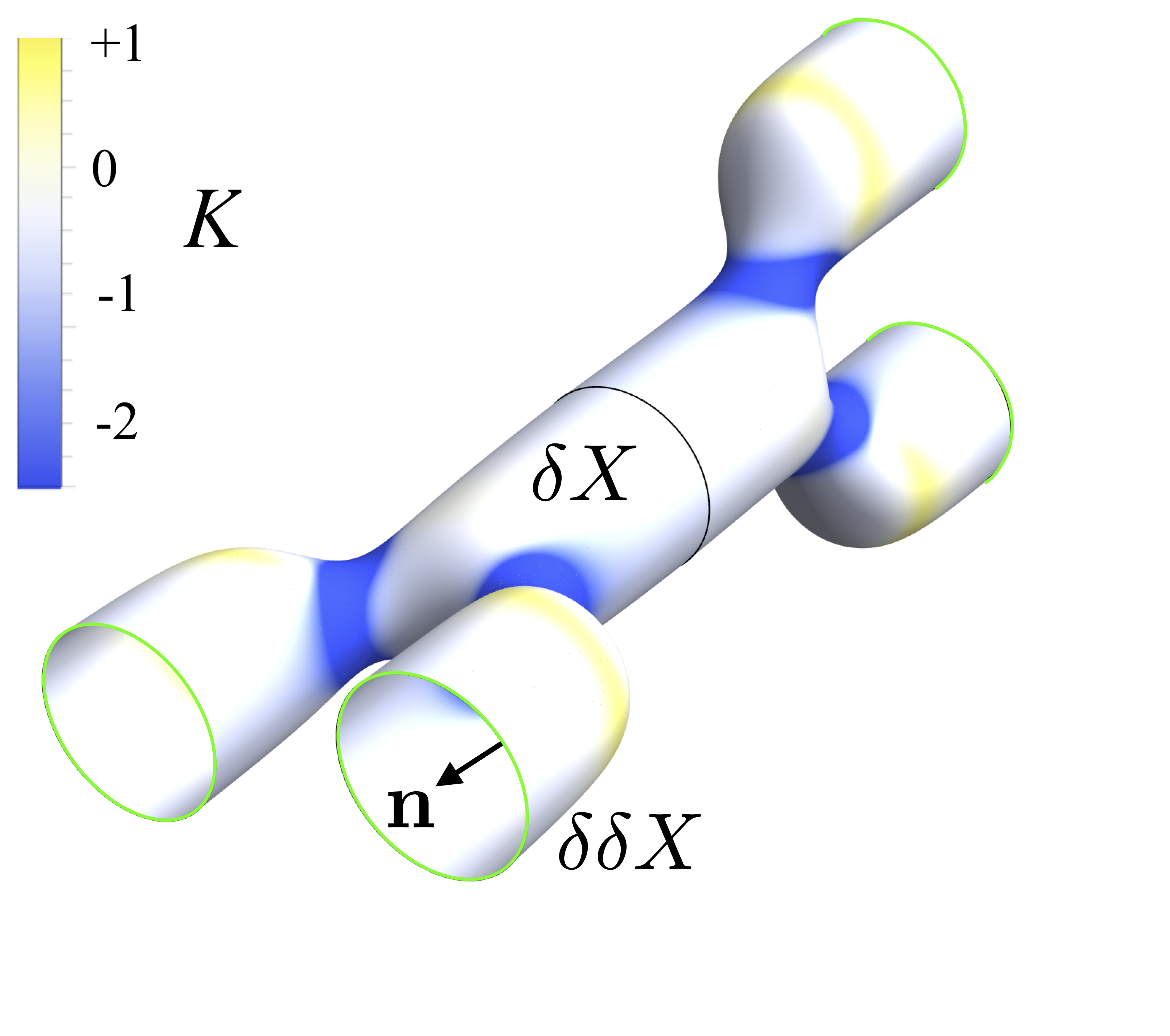}\\
(a)  & \hspace{10mm} (b) & \hspace{10mm} (c)
\end{tabular}
\end{centering}
\caption{Topological equivalence of the 2D pore boundary $\delta\Omega$ separating the fluid $\Omega$ and solid domains of the fundamental elements of (a) discrete and (c) continuous porous media. The normal vector $\mathbf{n}$ indicates the normal vector pointing into the fluid domain $\Omega$ from $\delta\Omega$, and $\delta\delta\Omega$ is 1D boundary of the pore boundary $\delta\Omega$ (green lines). $\delta\Omega$ is coloured according to its local Gaussian curvature $K$. (a) Pore boundary $\delta\Omega$ of single spherical grain (semi-transparent) with four contact points (black) and uniform positive curvature ($K=+1$) in discrete porous media. (b) Pore boundary of the same grain as (a) but with the cusp-shaped contact points smoothed to form finite connections between contacting grains, forming boundaries $\delta\delta\Omega$. (c) Pore boundary $\delta\Omega$ for a connected pore branch and merger associated with continuous porous media.}\label{fig:topology}
\end{figure*}

\subsection{Topological Complexity of Discrete and Continuous Porous Media}

In this Section we establish the topological equivalence of continuous and discrete porous media, which forms the basis for a unified description of chaotic mixing. The fundamental elements of continuous and discrete media are respectively the solid grain and the connected pore branch and merger shown in Fig.~\ref{fig:topology}a and c. Connections of pore bifurcations (Fig.~\ref{fig:topology}c) form an extensive 3D pore network similar to those shown in Fig.~\ref{fig:continuous}, whereas assemblies of grains (Fig.~\ref{fig:topology}a) form granular media similar to that shown in Fig.~\ref{fig:discrete}.

Although pore networks can differ with respect to pore topology and pore size and shape distributions, they all share the basic feature of many branching and merging pores that can be represented by the basic pore bifurcation shown in Fig.~\ref{fig:topology}c. The connection of many such bifurcations renders porous networks \emph{topologically complex} in that the topological genus $g$ of the pore-boundary $\delta\Omega$ is large, where $g$ characterises the number of ``holes’’ or ``handles’’ of the pore space $\Omega$ enclosed by the pore boundary $\delta\Omega$. As per  (\ref{eqn:chi}) and (\ref{eqn:genusbdry}), the associated Euler characteristic $\chi(\delta\Omega)=2(1-g)$ of the pore boundary is strongly negative~\cite{Armstrong:2019aa}, guaranteeing existence of saddle points on the pore boundary $\delta\Omega$ that generate chaotic mixing in the pore space $\Omega$~\cite{Lester:2013aa}.

For porous networks, the topological genus $g$ and Euler characteristic $\chi(\delta\Omega)$ of the pore network boundary is given by the number $n_b$ of pore bifurcations as
\begin{equation}
g=1-\frac{\chi(\delta\Omega)}{2}=1+\frac{n_b}{2}.\label{eqn:genus_cts}
\end{equation}
Hence, a straight pore ($n_b=0$) with periodic boundary conditions (to avoid trivial complications associated with macroscopic boundaries of the porous medium) is topologically equivalent to a torus and so has genus $g=1$, while a periodically connected pore branch and merger ($n_b=2$) forms an additional handle and so has genus $g=2$. Thus the pair of bifurcations shown in Fig.~\ref{fig:topology}c also has genus $g=2$ and Euler characteristic $\chi(\delta\Omega)=-2$. The addition of more pairs of pore branches and mergers in any topological configuration then increases the genus as per (\ref{eqn:genus_cts}). Hence all continuous porous media are topologically complex in that they possess a large topological genus $g$ number density, which is critical to chaotic mixing.

Similarly, close-packing of many grains such as those shown in Fig.~\ref{fig:topology}a generates extensive 3D granular assemblies that are topologically complex, which has also been shown~\cite{Turuban:2019aa} to be critical in generating chaotic mixing. Although various granular matter may differ with respect to the distributions of particle sizes, shapes and number of contacts, they all share the same basic topology. Under periodic boundary conditions, the topological genus $g$ and Euler characteristic $\chi(\delta\Omega)$ of the grain boundary $\delta\Omega$ of granular assemblies is given in terms of the total number of grain contacts $n_c$ as~\cite{Turuban:2019aa}
\begin{equation}
g=1-\frac{\chi(\delta\Omega)}{2}=n_c.\label{eqn:genus_discrete}
\end{equation}
As such, subject to periodic boundary conditions, the grain with four contact points shown in Fig.~\ref{fig:topology}c has $n_c=2$ (as a contact point is shared between two spheres) and  genus $g=2$ and Euler characteristic $\chi(\delta\Omega)=-2$. From (\ref{eqn:genus_discrete}), granular assemblies are also topologically complex in that they posses large topological genus $g$ per unit volume.

\subsection{Topological Equivalence of Discrete and Continuous Porous Media}

Topological equivalence between these seemingly disparate classes of porous media may be established by smoothing the cusp-shaped contacts between grains as shown in Fig.~\ref{fig:topology}b. This smoothed grain has the same topology ($g=2$, $\chi(\delta\Omega)=-2$) as the discrete grain shown in Fig.~\ref{fig:topology}a, and is topologically equivalent (i.e. it can be morphed solely by stretching and deforming) to the coupled pore branch and merger shown in Fig.~\ref{fig:topology}c, but with the distinct difference that the fluid domain (indicated by the outward normal $\mathbf{n}$) is external to the pore boundary $\delta\Omega$ shown in Fig.~\ref{fig:topology}a,b), whereas the fluid domain is internal to the pore branch and merger shown in Fig.~\ref{fig:topology}c. Hence, discrete porous media with smoothed contacts may be considered as the phase inverse of continuous porous media.

Despite this inverse relationship, the boundaries $\delta\Omega$ in continuous and smoothed discrete porous media are topologically equivalent, as reflected by their topological genus $g$. Thus, the Poincar\'e-Hopf theorem (\ref{eqn:P_Hopf}) also applies to the skin friction field $\mathbf{u}$ on the boundary $\delta\Omega$ of smoothed discrete porous media, and so ensures the existence of saddle points and 2D hyperbolic manifolds regardless of the orientation of $\mathbf{n}$. As the pore boundary $\delta\Omega$ of smoothed discrete and continuous porous media are topologically equivalent, the same basic mechanism drives chaotic mixing in these different porous media classes. The question as to what extent this connection persists for discrete porous media with non-smooth contacts shall be addressed in Sections~\ref{sec:stretching} and \ref{sec:folding}.

Although discrete and continuous porous media have equivalent topology, their geometry is markedly different, as reflected by the distribution of local Gaussian curvature $K=\kappa_1\kappa_2$ (where $\kappa_1=1/r_2$, $\kappa_2=1/r_2$ are the principal (maximum and minimum) local curvatures with corresponding radii $r_1$, $r_2$) over the surfaces shown in Fig.~\ref{fig:topology}. The Gaussian curvature $K$ of a plane and a cylinder is zero (as at least one of the principal curvatures is zero), but is positive for a sphere and negative for a hyperbolic surface such as a saddle. However, as these fundamental elements share the same topology, their total curvature is equivalent, as reflected by the Gauss-Bonnet theorem
\begin{equation}
\int_{\delta\Omega}KdA+\int_{\delta\delta\Omega}k_g ds=2\pi\,\chi(\delta\Omega)=-4\pi,\label{eqn:GB}
\end{equation}
where $k_g$ is the geodesic curvature of the 1D boundary $\delta\delta\Omega$. For the pore branch/merger (Fig.~\ref{fig:topology}c) and the smoothed grain (Fig.~\ref{fig:topology}b), the net contribution of $\delta\delta\Omega$ to (\ref{eqn:GB}) is zero due to periodicity, but for the discrete grain (Fig.~\ref{fig:topology}a), the boundary $\delta\Omega$ terminates at the contact points $\mathbf{x}_c$, hence there is a contribution of angle $-2\pi$ to (\ref{eqn:GB}) at each contact point~\cite{Grinfeld:2013aa} due to the cusp-shaped angle $\pi$ formed with each contacting sphere. As the spherical discrete grain in Fig.~\ref{fig:topology}a has uniform Gaussian curvature $K=1/r^2$, the first integral in (\ref{eqn:GB}) is $4\pi$, whereas the second integral is $-8\pi$. Thus, whilst the discrete grains have positive curvature, grain contacts act as infinitesimal regions of infinite negative curvature, giving total negative curvature of $-4\pi$. The smoothed grain shown in Fig.~\ref{fig:topology}c shows the finite contact regions have finite negative curvature, which yields the total negative curvature of $-4\pi$. Conversely, the branch/merger shown in Fig.~\ref{fig:topology}a has predominantly negative or zero curvature.

Although the total curvature of these elements is preserved as per (\ref{eqn:GB}), the distribution of Gaussian curvature $K$ impacts the number and type of critical points as saddle (node) points tend to arise in negative (positive) curvature regions. Hence, saddle points arise in continuous porous media near pore branch/merger junctions, whereas smoothed grains admit node points on the grain surface and saddle points near contact points. 

Thus, the simplest arrangement of critical points for the branch/merger shown in Fig.~\ref{fig:topology}a involves two saddle points $\mathbf{x}_p^s$ near the pore junctions, satisfying (\ref{eqn:P_Hopf}) as $\sum\gamma_p(\mathbf{x}_p^s)=-n_s=\chi(\delta\Omega)$. Conversely, the simplest arrangement of critical points for the grain with smoothed connections shown in Fig.~\ref{fig:topology}c involves two node points $\mathbf{x}_p^n$ on the grain surface and two saddle points $\mathbf{x}_p^s$ per contact ($n_s=4$ once periodicity of contacts is accounted for), satisfying (\ref{eqn:P_Hopf}) as $\sum\gamma_p(\mathbf{x}_p^s)=n_c-n_s=\chi(\delta\Omega)$. Therefore, despite topological equivalence, the different geometry of smoothed discrete porous media tends to generate different types and numbers of critical points than continuous porous media. 

Hence, when the contacts in discrete porous media are smoothed, chaotic mixing arises via the same fundamental mechanism as continuous porous media, but it is unknown whether this extends to discrete porous media in general. \citet{Turuban:2018aa, Turuban:2019aa} and \citet{Heyman:2020aa} have respectively identified that fluid stretching and folding is generated at the contact points of discrete porous media. In Sections~\ref{sec:stretching} and \ref{sec:folding} respectively, the detailed mechanisms that govern fluid stretching and folding are uncovered. 

\section{Fluid Stretching in Continuous and Discrete Porous Media}
\label{sec:stretching}

\begin{figure*}[ht]
\begin{centering}
\begin{tabular}{c c}
\includegraphics[width=0.74\columnwidth]{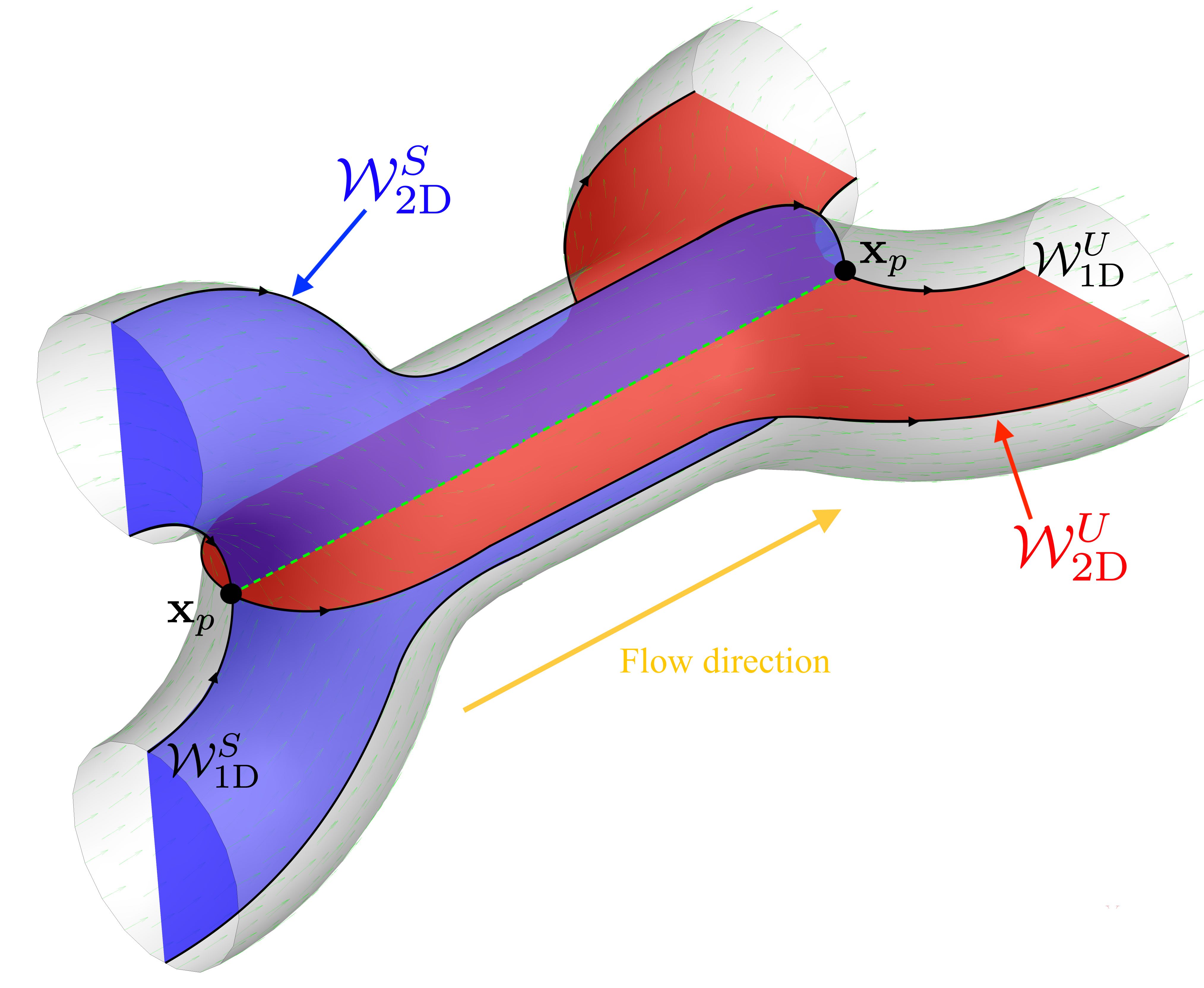}&
\includegraphics[width=1.24\columnwidth]{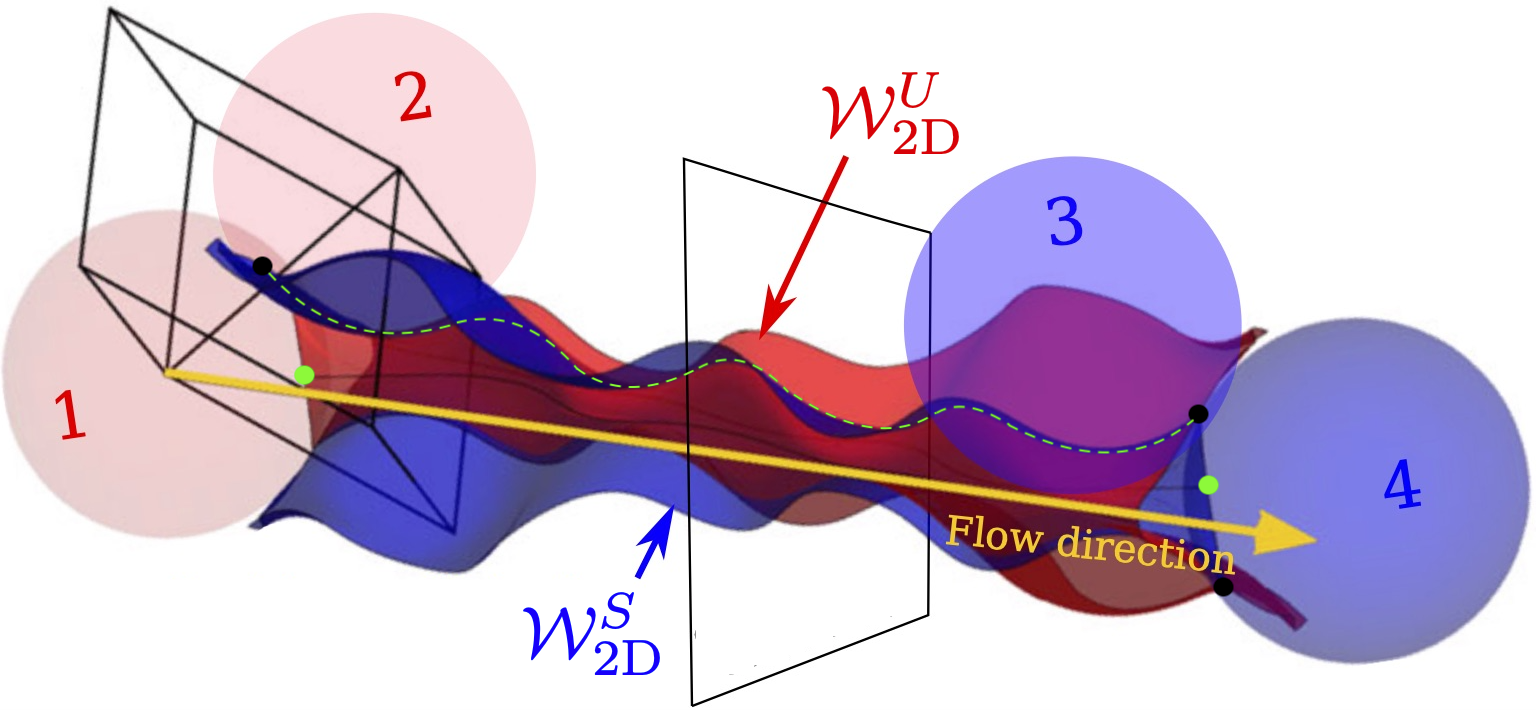}\\
(a) & (b)
\end{tabular}
\end{centering}
\caption{Hyperbolic manifolds, critical points and lines in (a) continuous and (b) discrete porous media. Intersection (dotted green line) of stable (blue surface) and unstable (red surface) 2D hyperbolic manifolds in (a) pore branch and merger (grey volume) in an open porous network and (b) numerical simulation of Stokes flow through a body-centred cubic (bcc) lattice of monodisperse spheres (modified from \cite{Turuban:2019aa}). In the manifold intersection connects the saddle points (black dots) of the branch/merger in (a), and the contact points (black dots) between spheres in (b). In (b), the 2D manifolds emerge from the skin friction field of the two spheres labelled 1 and 4, and the green points indicate node points. The open plane indicates the orientation of transverse cross-sections in Fig.~\ref{fig:3Dmanifold}, and the black cell is the BCC unit cell.}
\label{fig:pore_manifolds}
\end{figure*}

\subsection{Comparison of Fluid Stretching in Discrete and Continuous Porous Media}

A schematic of fluid deformation in continuous porous media is shown in Fig.~\ref{fig:pore_manifolds}a. This figure shows a connected pore branch/merger with offset angle $\Delta=\pi/2$ and the associated 1D and 2D stable and unstable manifolds which emanate from the saddle points $\mathbf{x}_p$ situated at the pore branch and merger junctions. The intersection of the 2D stable and unstable manifolds forms a 1D \emph{critical line} that connects the saddle points. If many of these pore branch/mergers are connected in series in a manner that maintains transverse connections, then multiple 2D stable and unstable manifolds emanate respectively from the down- and up-stream saddle points into the element shown in Fig.~\ref{fig:pore_manifolds}a, producing many more manifold intersections, a heteroclinic tangle and chaotic mixing. Conversely, if the stable and unstable 2D manifolds connect smoothly and tangentially, they cancel each other out and don't produce persistent exponential stretching.

As shown in Fig.~\ref{fig:pore_manifolds}b, a similar mechanism arises in Stokes flow over a body-centred cubic (bcc) lattice of smooth spheres~\cite{Turuban:2019aa}. The same basic invariant features arise in this flow with contact points playing a similar role to saddle points, and stable and unstable 2D manifolds emanate from the sphere skin friction fields and intersect along 1D critical lines that connect contact points between spheres. Multiple manifold intersections also occur from 2D hyperbolic manifolds emanating from other spheres up- and down-stream of those shown in Fig.~\ref{fig:pore_manifolds}b, leading to a heteroclinic tangle and chaos.

Fig.~\ref{fig:pore_manifolds}b also shows isolated saddle points which lie along the 1D intersection of the 2D manifolds with the sphere surfaces. An important difference to continuous porous media is that fluid deformation at contact points is limited by the local cusp-shaped geometry, potentially rending these points \emph{degenerate} (meaning at least one of eigenvalue $\eta_i$ of $\mathbf{A}$ is zero) with non-exponential stretching. Conversely, the smoothed grains shown in Fig.~\ref{fig:topology}c) admit non-degenerate saddle points with exponential fluid stretching. Hence it is unclear how chaotic mixing is generated in discrete porous media.

\subsection{Fluid Stretching at Contact Points}

\begin{figure}[ht]
\centering\includegraphics[width=\linewidth]{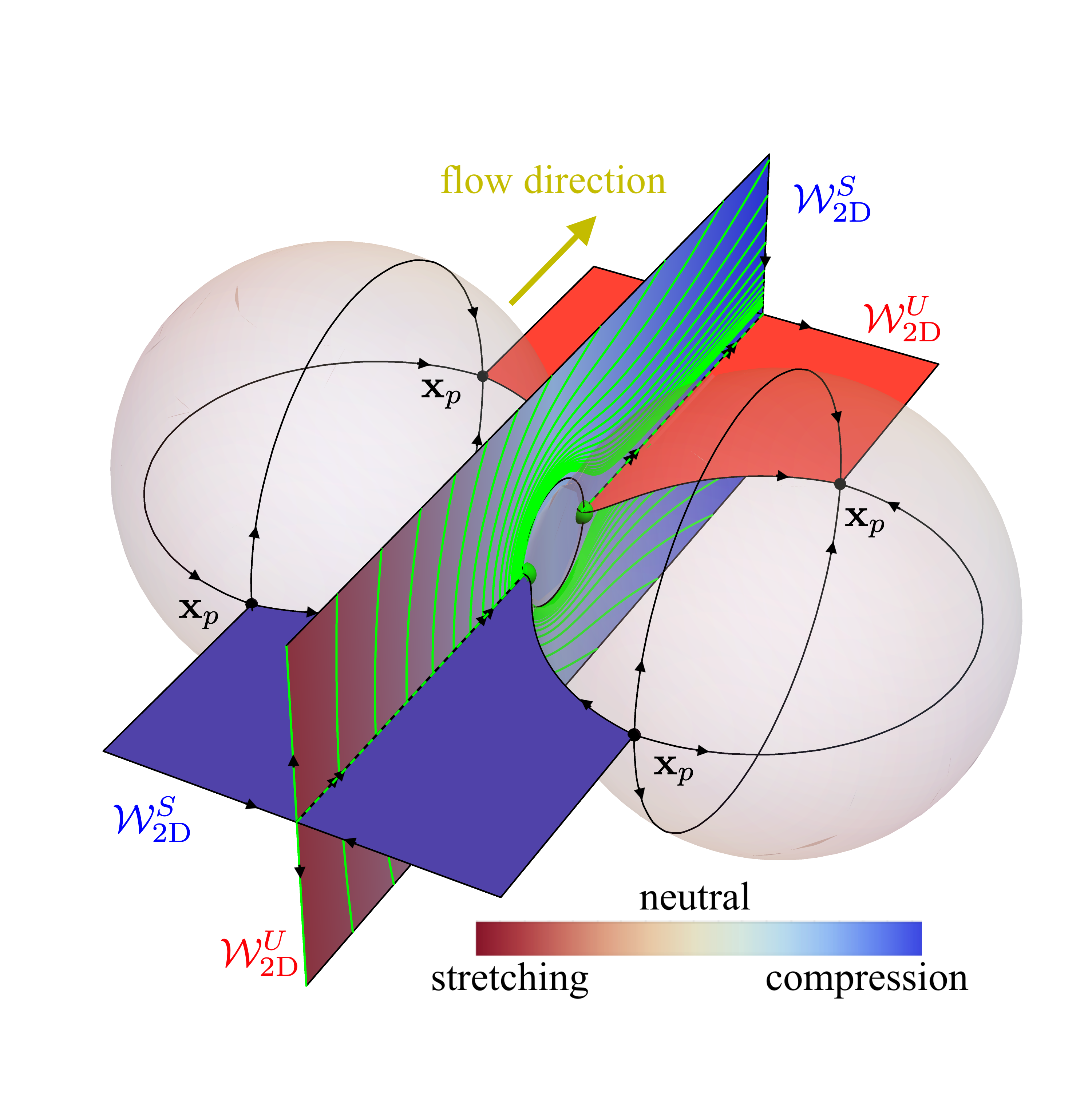}
\caption{Schematic of evolving stable $\mathcal{W}^S_{2\text{D}}$ (blue surfaces) and unstable $\mathcal{W}^U_{2\text{D}}$ (red surfaces) manifolds as they are advected over a finite-sized connection between two spheres (grey). Black arrows indicate fluid stretching directions in the bulk fluid and skin friction field. Black and green points respectively indicate saddle and node points. These manifolds become degenerate in the neighbourhood of the connection (indicated by transition to grey colour) and exchange stability as they pass over, generating non-affine folding of material lines (green).}
\label{fig:meth:contactmanifold}
\end{figure}

\begin{figure}[ht]
\begin{centering}
\begin{tabular}{c c}
\includegraphics[width=0.48\columnwidth]{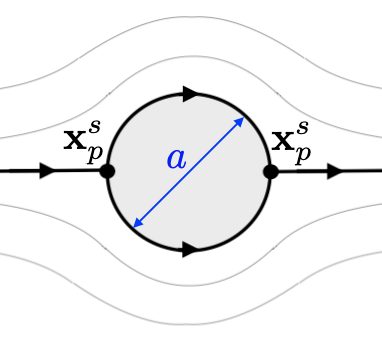}&
\includegraphics[width=0.48\columnwidth]{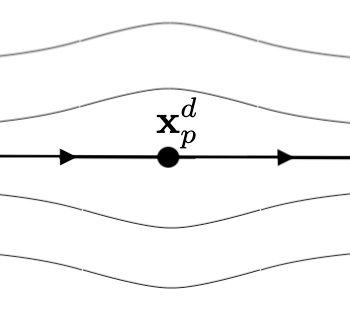}\\
(a) & (b)
\end{tabular}
\end{centering}
\caption{Streamlines (thin) and critical lines (thick) for 2D velocity field $\mathbf{v}_{\text{2D}}(\mathbf{x})$ in the symmetry plane between two spheres connected by (a) a smoothed contact of diameter $a$ and (b) and infinitesimal contact point. Critical points are denoted as either a saddle point $\mathbf{x}_p^s$ or a degenerate topological saddle $\mathbf{x}_p^d$. Due to the no-slip condition, in both cases the skin friction field $\mathbf{u}(\mathbf{x})\equiv\partial\mathbf{v}/\partial x_3^\prime$ on the grain boundaries local to this contact shares the same topology as the velocity field $\mathbf{v}_{\text{2D}}(\mathbf{x})$.}\label{fig:symm}
\end{figure}

Fig.~\ref{fig:meth:contactmanifold} depicts the stable and unstable manifolds associated with a smoothed contact between two spheres (analogous to Fig.~\ref{fig:topology}c) whose centres are oriented normal to the far-field flow direction. Due to the symmetries of this flow, the velocity field $\mathbf{v}(\mathbf{x})$ in the symmetry plane between these two spheres has no transverse component. As shown in Fig.~\ref{fig:symm}a, this symmetry plane contains an inclusion from the smoothed contact that is a diameter $a$ disc with saddle points $\mathbf{x}_p$ on the upstream and downstream sides that are connected to critical lines. Fig.~\ref{fig:symm}b shows that as this finite contact shrinks to an infinitesimal point-like contact ($a\rightarrow 0$), these two critical points coalesce into a single critical point. As shown in Fig.~\ref{fig:symm}b, this critical point appears to be degenerate as the associated steady 2D flow in the symmetry plane between the spheres cannot generate exponential stretching.

\subsection{Degeneracy of Contact Points}

\citet{Turuban:2019aa} proposed that this critical point is a \emph{topological saddle}~\cite{Brons:1999aa,Bakker:1991aa}, which is degenerate and so does not generate exponential fluid stretching. As a topological saddle has Poincar\'{e} index $\gamma=-2$, then from (\ref{eqn:P_Hopf}) the Euler characteristic in discrete porous media is related to the number $n_d$ of degenerate topological saddles, nodes and regular saddles as
\begin{equation}
\chi(\Omega)=2(1-g)=n_n-n_s-2n_d.
\end{equation}
As the topological genus per grain $g$ is equal to the number of contacts $n_c$, then under the assumption that all contact points are topological saddles $n_c=n_d$, the number of non-degenerate nodes $n_n$ and saddle $n_s$ points on each grain is independent of the Euler characteristic $\chi(\Omega)$; $n_s=n_n-2$. Hence it was proposed in \cite{Turuban:2019aa} that the simplest possible skin friction field in discrete porous media involves two isolated node points ($n_n=2$, $\gamma=+1$) and no saddle points ($n_s=0$) and therefore no hyperbolic 2D manifolds in the fluid bulk and no chaotic mixing. Hence it was concluded that chaotic mixing in discrete porous media is not ubiquitous as it requires bifurcation of the skin friction field beyond its simplest possible state.

However, this conclusion is not supported by the saddle-like flow structures observed in Fig.~\ref{fig:3Dmanifold} which arise at the intersection of the non-degenerate stable $\mathcal{W}^s_{\text{2D}}$ and unstable $\mathcal{W}^u_{\text{2D}}$ 2D manifolds that emanate from contact points. These flow structures clearly indicate the hyperbolic nature of the manifolds as they pass through contact points and intersect with each other (also shown in Fig.~\ref{fig:pore_manifolds}b), whereas degenerate points are not associated with exponential (hyperbolic) stretching~\cite{Turuban:2019aa}. 

However, deeper inspection reveals that contact points are only degenerate with respect to the 2D flow $\mathbf{v}_{\text{2D}}(\mathbf{x})$ in the symmetry plane shown in Fig.~\ref{fig:meth:contactmanifold}, but still exhibit exponential stretching in the direction transverse to the symmetry plane between contacting grains. In Section~\ref{sec:background} it was shown that isolated node points (type III and type IV critical points in Fig.~\ref{fig:dist}) cannot generate 2D hyperbolic manifolds in the fluid interior. However, the interaction between a node and contact point can produce an interior 2D hyperbolic manifold, as shall first be shown for finite-sized contacts. 

\subsection{Hyperbolic Manifolds at Finite-Sized Contacts}

Fig.~\ref{fig:meth:contactmanifold} shows that a saddle point (green) and a pair of node points (black) arises on the upstream and downstream sides of a smoothed connection and each sphere. The 1D manifold along the sphere surfaces connecting the upstream nodes and saddles is unstable with respect to these node points and stable with respect to the saddle points. The combination of this stable 1D manifold with the stable 1D manifold in the fluid interior that connects to the upstream saddle point generates the 2D stable manifold $\mathcal{W}_{\text{2D}}^S$ that is shown in Fig.~\ref{fig:meth:contactmanifold} without material (green) lines. Similarly, a downstream 2D unstable manifold $\mathcal{W}_{\text{2D}}^U$ (Fig.~\ref{fig:meth:contactmanifold} without material (green) lines) is generated that connects the downstream saddle and node points. Both of these 2D manifolds are oriented parallel to the line connecting the sphere centres and so are termed \emph{parallel} 2D manifolds. Conversely, the manifolds transverse to this connecting line (shown in  Fig.~\ref{fig:meth:contactmanifold} with material (green) lines) are termed \emph{transverse} 2D manifolds.

\begin{figure*}[!htb]
\centering\includegraphics[width=0.9\linewidth]{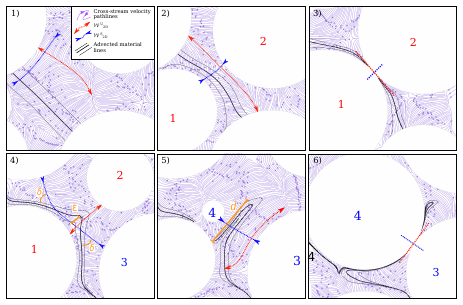}
\caption{Series of cross-sections with increasing downstream distance of Stokes flow through a bcc lattice of spheres between the two bead pairs ((1-2) and (3-4)) shown in Fig.~\ref{fig:pore_manifolds}b. Purple lines show vector field lines of the 2D velocity field transverse to each cross-section, thick and thin black lines indicate material lines advected by the flow, and red and blue lines respectively represent 2D unstable and stable manifolds of the flow. Intersection of the 2D stable and unstable manifolds forms a 1D curve which connects the contact points of the (1-2) and (3-4) bead pairs. Modified from \cite{Turuban:2019aa}.}
\label{fig:3Dmanifold}
\end{figure*}

The local cusp-shaped geometry near contact points constrains any (un)stable 2D manifolds that are generated (up)downstream to pass over this connection as transverse 2D manifolds, as per Fig.~\ref{fig:pore_manifolds}b. The analogous situation for a porous network is shown in Fig.~\ref{fig:pore_manifolds}a, where the 2D (un)stable manifold is generated by a (up)downstream saddle point and connects with an (down)upstream saddle point. This constrained geometry forces the transverse 2D stable and unstable manifolds shown in Fig.~\ref{fig:meth:contactmanifold} with material (green) lines to connect tangentially over the sphere connection. The 1D intersection (green dotted line in Fig.~\ref{fig:meth:contactmanifold}) of the transverse and parallel 2D manifolds are responsible for the saddle-type flow structures shown in Fig.~\ref{fig:3Dmanifold}.

\subsection{Hyperbolic Manifolds at Contact Points}

If the connection between the spheres in Fig.~\ref{fig:meth:contactmanifold} shrinks to an infinitesimal contact point, the up- and down-stream saddle points coalesce into a single saddle point which is degenerate with respect to the transverse manifolds, but non-degenerate with respect to the parallel manifolds. The interior (un)stable 1D manifolds that connect from (down)upstream to the contact point and the exterior (un)stable 1D manifolds that connect the (down)upstream node points to the contact point both persist as this contact shrinks to a point. Hence the parallel 2D stable and unstable hyperbolic manifolds persist for a contact point, and still impart exponential stretching to the fluid bulk.

Hence node points and saddle-like contact points between grains generate 2D hyperbolic manifolds and chaotic mixing in in discrete porous media. This basic mechanism is very similar to that of continuous porous media, where connected contact and node points play the role of saddle points in continuous porous media. As there is no need for isolated saddle points (away from contacts), chaotic mixing is inherent to discrete porous media.

\section{Fluid Folding in Discrete and Continuous Porous Media}
\label{sec:folding}

\subsection{Fluid Folding in Discrete Porous Media}

In this section we uncover the mechanisms that drive fluid folding in continuous and discrete porous media, Although \citet{Heyman:2020aa} show that contact points in discrete porous media generate folding of fluid elements, the detailed mechanism is unclear. Fig.~\ref{fig:meth:contactmanifold}, shows that when a 2D unstable transverse manifold impacts a contact point, this manifold becomes locally degenerate at the contact point, and transitions to a stable 2D transverse manifold that emanates from the contact. Conversely, a 2D stable parallel manifold terminates at the contact point and upstream node points, and a 2D unstable parallel manifold emanates downstream from the contact point and downstream node points.

This exchange in stability can generate strong folding of fluid elements in the symmetry plane between the spheres, as indicated by the green material lines in Fig.\ref{fig:meth:contactmanifold}. Although the schematic in Fig.\ref{fig:meth:contactmanifold} is fore-aft symmetric, this can be broken by additional grains in the assembly or contacts that are skew to the mean flow direction, meaning that folding of material elements also manifests in the plane transverse to the mean flow direction. As shown in the sequence of panels in \ref{fig:3Dmanifold}, subsequent fluid stretching elongates these folds into tight cusps and highly filamentous structures such as those observed in Figs.~\ref{fig:disc_comp}c and \ref{fig:3Dmanifold}. 

These insights provide a link between the folding dynamics described here and in \cite{Heyman:2020aa} and recent studies~\cite{Turuban:2018aa,Turuban:2019aa} that describe the mechanisms leading to pore-scale fluid stretching in discrete porous media. This means that chaotic mixing proceeds in discrete porous media via iterated stretching and folding of fluid elements as they pass over contact points between grains. 

\subsection{Fluid Folding in Continuous Porous Media}

\begin{figure}[h]
\begin{centering}
\begin{tabular}{c c c}
\includegraphics[width=0.31\columnwidth]{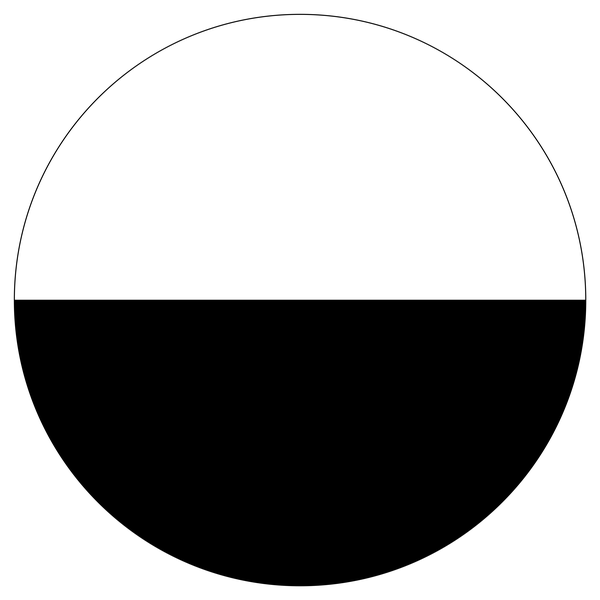}&
\includegraphics[width=0.31\columnwidth]{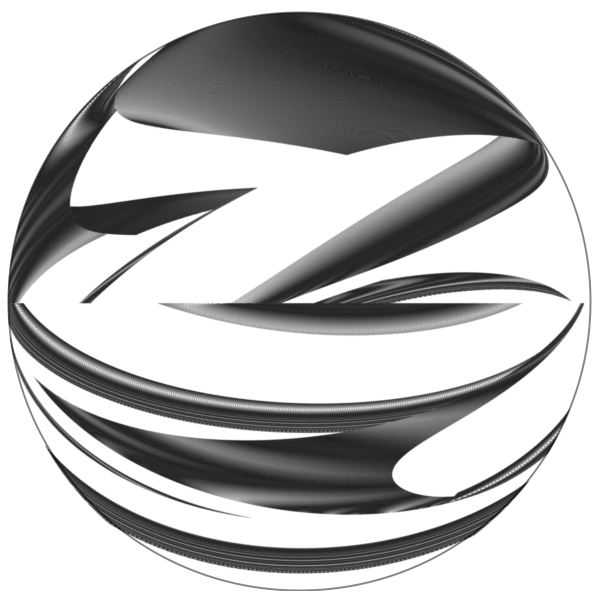}&
\includegraphics[width=0.31\columnwidth]{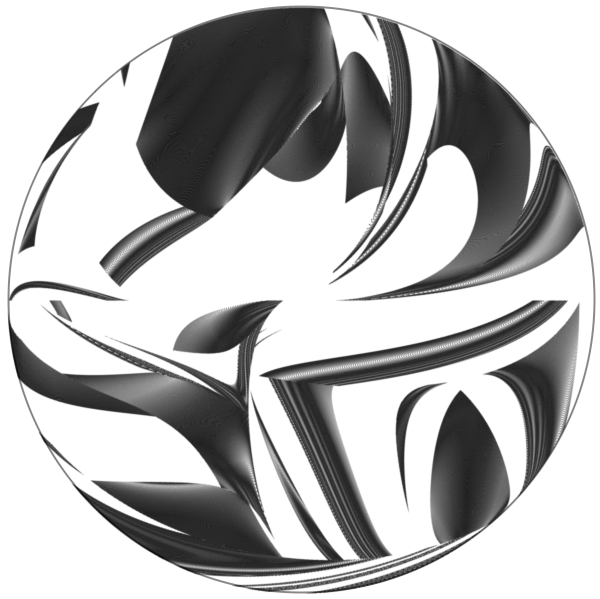}\\
n=0 & n=5 & n=10\\
\includegraphics[width=0.31\columnwidth]{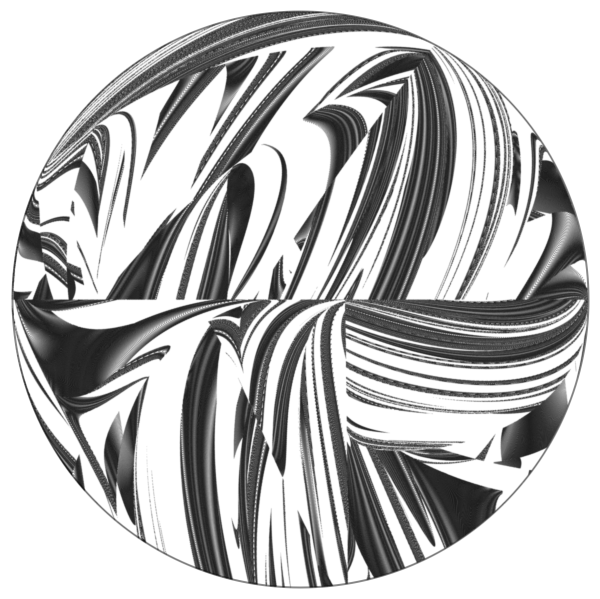}&
\includegraphics[width=0.31\columnwidth]{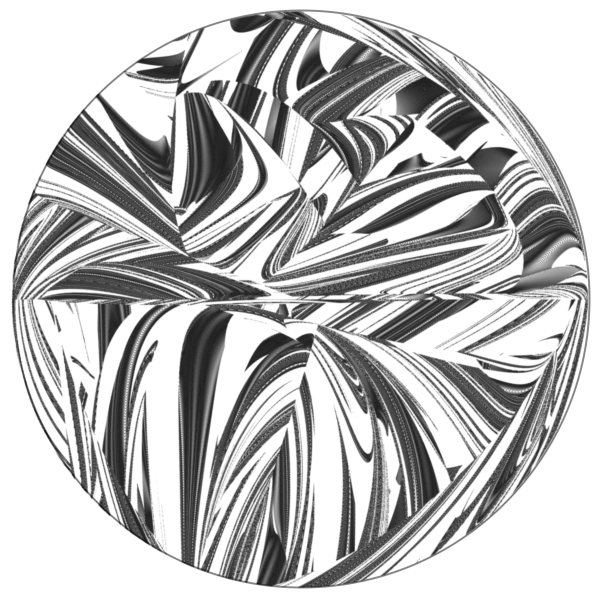}&
\includegraphics[width=0.31\columnwidth]{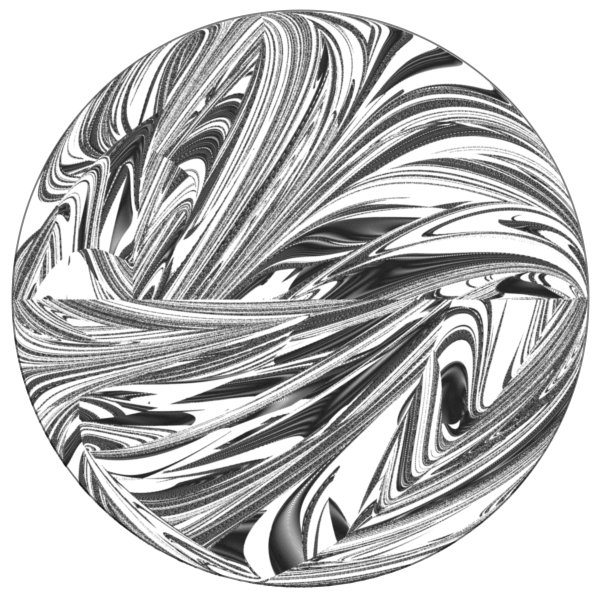}\\
n=20  & n=30 & n=40
\end{tabular}
\end{centering}
\caption{Mixing of fluid elements with pore number $n$ in a random pore network given by the spatial map $\Lambda_n$ in (\ref{eqn:Lambda_map}) with $\delta_j\sim U[0,2\pi],\Delta_j\sim U[0,2\pi]$.}\label{fig:dye_network}
\end{figure}

The maps $\mathcal{S}_b$, $\mathcal{S}_m$ (\ref{eqn:spatial_maps}) can be used to uncover folding mechanism in continuous porous media. The dye trace images in Fig.~\ref{fig:dye_network} show evidence of cutting, shuffling, stretching and folding of material lines as they are advected through the pore-space. To unravel these motions we consider evolution of a continuously injected line source through the pore branch and merger shown in Fig.~\ref{fig:sheet_cut}.
In this figure the line source with segments marked AB and CD is continuously injected along the line $x_r=0$ before bifurcating into two separate segments at the critical point $\mathbf{x}_p$. The intersection of the 2D stable manifold $\mathcal{W}_{2\text{D}}^s$ with the boundary $\partial\Omega$ represents a repelling (i.e. $\nabla\cdot\mathbf{u}>0$) \emph{critical line} $\zeta$ of the skin friction field $\mathbf{u}$. Any injected material line that crosses the mid-plane $y_r=0$ of the inlet pore corresponding to $\mathcal{W}_{2\text{D}}^s$ is stretched and folded as it is advected downstream over $\zeta$, and fluid particles on this critical line are held there indefinitely, as shown by the temporal map $\mathcal{T}^*$ (\ref{eqn:Tstar}), which has an divergent residence time for particles entering along $y_r=0$. Folding of fluid elements over this critical line effectively splits the fluid element as it is advected into two separate pores downstream. Thus the indefinite holdup of fluid particles at the critical line $\zeta$ manifests as the distinct segments AB, CD in each outlet of the pore branch, yet these are connected via a continuous material line extending from the critical line. 

Upon exiting their respective pores, these segments are reoriented at angle $\Delta$ before entering the branch merger. The solid red lines in Fig.~\ref{fig:sheet_cut} denote parts of the material line that are resolved by the spatial map $\mathcal{M}$ (\ref{eqn:Mmap}), and the dashed red lines indicate material elements that are not resolved by this map yet remain connected in 3D due to continuity. The bifurcation and folding process is then reversed in the pore merger (Fig.~\ref{fig:sheet_cut}b), where fluid elements from separate pores are merged together over the \emph{attracting} (i.e. $\nabla\cdot\mathbf{u}<0$) critical line $\zeta$ formed by intersection of the 2D unstable manifold $\mathcal{W}_{2\text{D}}^u$ and the pore boundary $\partial\Omega$. Fluid elements that pass near this attracting critical line also experience diverging residence times, leading to further folding of material elements prior to the merger of elements arriving from different inlet pores.

When projected onto the (2D) pore outlet planes, this strong folding of fluid elements over the critical lines $\zeta$ in the 3D pore space manifests as an effective ``cutting'' of material elements as shown in Fig.~\ref{fig:dye_network}. In the following subsection we consider the nature of such discontinuous mixing and its interplay with 3D fluid stretching and folding. In addition to this strong folding that manifests as discontinuous mixing, the spatial map $\mathcal{M}$ also imparts weaker folding that is illustrated by its affine $\mathcal{M}_a$ (\ref{eqn:M_aff}) and non-affine $\mathcal{M}_n$ (\ref{eqn:M_naff}) components. 

Fluid cutting is also encoded in the non-affine map $\mathcal{M}_n$, such that fluid elements with $y_r>0$ ($y_r<0$) are mapped to the upper (lower) branch in Fig.~\ref{fig:sheet_cut}a. Weak folding of material elements in these 2D outlets arises from the nonlinear component of the non-affine map $\mathcal{M}_n$ and its inverse $\mathcal{M}_n^{-1}$, which map fluid particles along the straight line $y_r=0$ to the curved lower pore boundary. Although these maps only induce weak folding, subsequent fluid stretching amplify these folds into the sharp kinks observed in Fig.~\ref{fig:dye_network}. However,  Fig.~\ref{fig:dye_network} shows that packing of highly elongated material lines into the 2D pore outlet is dominated by cutting of material elements rather than weak non-affine folding.

Hence strong folding of fluid elements in continuous porous media arises via a similar mechanism to that of discrete porous media. In continuous media fluid elements are held up at pore junctions, whereas for discrete media fluid elements are held up at contact points, both leading to strong folding. However, for continuous porous media this hold-up generates discontinuous mixing when projected onto a 2D plane transverse to the mean flow direction. Such discontinuous mixing does not occur in discrete media due to the infinitesimal contacts. In addition, the non-affine nature of $\mathcal{M}$ imparts weak folding that may be amplified into sharp kinks by subsequent fluid stretching. As weak folding plays a secondary role, in Section~\ref{sec:disco} we focus on strong fluid folding which manifests as discontinuous mixing.  

\section{Discontinuous Mixing in Continuous Porous Media}
\label{sec:disco}

\subsection{Origins of Discontinuous Mixing}

Fig.~\ref{fig:dye_network} shows the evolution of material distributions under stretching and folding (SF) and cutting and shuffling (CS) actions in continuous porous media. The origins of these CS actions are illustrated in Fig.~\ref{fig:sheet_cut}a, where the line segments AB, CD are advected to the different outlets of the pore merger element in an apparently discontinuous manner. However, as indicated by the dashed red lines in Fig.~\ref{fig:sheet_cut}b, these segments are actually connected by a material line that is stretched and folded as it is advected over saddle point $\mathbf{x}_p$. 

In many applications, this 3D distribution of fluid elements is less important than the 2D distribution in a planar cross-section corresponding to the pore outlet planes in Fig.~\ref{fig:sheet_cut} and images in Fig.~\ref{fig:dye_network}). Material lines and sheets that are continuous in 3D can manifest as discontinuous elements in this 2D frame, even within the same pore. Henceforth we consider mixing of fluid elements in these 2D planes as being ``discontinuous’’, although we understand the fluid to undergo smooth and continuous 3D deformation, unlike e.g. granular matter or plastic materials that deform via slip planes. 

Hence chaotic mixing in continuous porous media is akin to mixed cutting and shuffling (CS) and stretching and folding (SF) actions observed during chaotic mixing of discontinuously deforming systems~\cite{Smith:2016aa,Smith:2019aa,Smith:2017aa,Smith:2017ab,Sturman:2012aa,Christov:2011aa,Juarez:2010aa}, such as granular tumblers~\cite{Christov:2011aa,Smith:2017ab} and microfluidic micro-mixers based on iterated pore branches and mergers~\cite{Sudarsan:2006aa}. The motions imparted by the spatial maps $\mathcal{S}_b$, $\mathcal{S}_m$ as fluid elements flow through each pore branch and merger shown in Fig.~\ref{fig:sheet_cut}a,b can be summarised as ``cut-stretch-fold-rotate'' in the pore branch, and ``compress-fold-glue-rotate'' in the pore merger.

For the pore branch, ``cut'' refers to splitting of fluid elements along the critical line $\zeta$, ``stretch'' refers to the affine stretching (by $\mathcal{M}_a$) by a factor of 2 in the $y_r$ direction, ``fold'' refers to non-affine folding (by $\mathcal{M}_n$) of elements to conform to the outlet pore boundaries, ``rotate'' refers to reorientation (by $R(\Delta)$) of fluid elements due to orientation of the downstream pore mergers. In the pore merger, ``compress'' refers to the compression (by $\mathcal{M}_a^{-1}$) of fluid elements by a factor of 1/2 in the $y_r$ direction, ``fold'' refers to non-affine folding (by $\mathcal{M}_n^{-1}$) to conform with the required semi-circular disc, ``glue'' refers to gluing of these semi-circular discs to form the disc-shaped pore outlet, and ``rotate'' refers to refers to reorientation (by $R(\delta)^{-1}$). As fluid elements are in general not smoothly reconnected at the gluing step, these actions lead to cutting and shuffling of material elements as shown in Figs.~\ref{fig:dye_network} and \ref{fig:web}.

\subsection{Coupled Continuous and Discontinuous Mixing}

The introduction of discontinuous CS actions significantly alters the mixing process as constraints associated with fluid continuity are relaxed. One impact of is that the Poincar\'{e} index is no longer conserved, leading to the formation of new coherent Lagrangian structures such as leaky KAM surfaces around elliptic tori, and pseudo-elliptic and pseudo-hyperbolic points~\cite{Smith:2017aa,Smith:2017ab}. The natural mathematical framework to analyse CS motions is via piecewise isometries (PWI)~\cite{Sturman:2012aa}, where fluid elements are removed (cutting) and placed in different locations and orientations (shuffling). For CS-only systems, these actions generate a \emph{web of discontinuities} (Fig.~\ref{fig:webk}) which is comprised of the forward iterates of the cutting line (corresponding to the line $y_r=0$ in the map $\mathcal{M}$ associated with bifurcation of fluid elements along the critical line $\zeta$ in the pore branch element), and complete mixing occurs if this web becomes space-filling with time. Under these conditions, pure CS actions leads to weak ergodic mixing, and associated mixing measures decay at an algebraic rate in time. 

Conversely, for pure SF actions in continuous systems, fluid stretching arises from affine deformations, whereas folding is associated with non-affine deformation. Together, these stretching and folding actions can generate chaotic advection which is characterised by strong ergodic mixing and mixing measures that decay at an exponential rate. To understand how combined SF and CS actions impact the mixing dynamics of continuous porous media in both ordered and random pore networks, we examine the four network geometries (i-iv) described in Section~\ref{subsec:advmap_cts}, whose mixing dynamics are shown in Fig.~\ref{fig:web}. In all cases these SF and CS actions lead to complete mixing of fluid elements, albeit at significantly different rates.

\subsection{Measures of Continuous and Discontinuous Mixing}

Conventional tools such as Lyapunov exponent, scalar variance decay and entropy measures are not suitable to characterise discontinuous mixing as they are based on dissipative or continuous mixing. The \emph{mix-norm} measure~\cite{Mathew:2005aa} was developed to quantify multiscale mixing across diverse applications, including non-dissipative and discontinuous mixing. The mix-norm $\Phi(c)\in[0,1]$ captures the extent of mixing of a concentration field $c(\mathbf{x},t)$ across different normalized lengthscales $s\in[0,1]$ as
\begin{align}
&\Phi(c)=\sqrt{\int_0^1\phi(c,s)^2 \mu(ds)},\label{eqn:mixnorm}\\
&\phi(c,s)=\sqrt{\int_{p\in D}d^2(c,\mathbf{p},s) \mu(d\mathbf{p})},\label{eqn:phinorm}\\
&d(c,\mathbf{p},s)=\frac{1}{\text{vol}[B(\mathbf{p},s)]}\int_{x\in B(\mathbf{p},s)}c(\mathbf{x})\mu(d\mathbf{x}),\label{eqn:davg}
\end{align}
where $\mu$ denotes the Lebesgue measure, $d(c,\mathbf{p},s)$ is the average of $c(\mathbf{x})$ over the ball $B(\mathbf{p},s):=\{\mathbf{x}:|\mathbf{x}-\mathbf{p}|<s/2\}$ centred at point $\mathbf{p}$ in the mixing domain $D$, $\phi(c,s)$ is the $L^2$-norm of $d$ over $D$, and $\Phi(c)$ is the $L^2$-norm of $\phi(c,s)$ over $s\in[0,1]$. As such, the $L^2$-norm $\phi(c,s)$ characterises the variance of the scalar field $c(\mathbf{x})$ after averaging over lengthscale $s$, and the mix-norm $\Phi(c)$ accounts for mixing across all scales by quantifying the $L^2$ norm of this measure over all $s$.

Hence $\Phi=1$ for scalar distributions that are completely segregated at the largest length scale of the system and $\Phi(c)=0$ for mixtures that are well-mixed at all scales. Under weak ergodic mixing characteristic of CS systems, the mix-norm $\Phi(c_n)$ decays algebraically with the number $n$ of iterations of the mixing process, whereas under strong mixing characteristic of SF systems $\Phi(c_n)$ decays exponentially with $n$. For SF systems that can be represented by a linear map such as $\mathcal{M}_a$, the mix-norm decays exponentially at a rate of half the Lyapunov exponent~\cite{Mathew:2005aa} as
\begin{equation}
\Phi(c_n)=\Phi(c_0)\exp\left(-\frac{\lambda_\infty n}{2}\right).\label{eqn:SF_mix_decay}
\end{equation}
Although many studies have considered mixing due to purely CS or SF actions, only a handful~\cite{Smith:2016aa,Smith:2019aa,Smith:2017aa,Smith:2017ab} have considered the impact of coupled CS and SF motions. \citet{Smith:2017aa,Smith:2017ab} show that the combination of CS and SF motions can either act to retard or enhance the rate of strong mixing given by SF actions alone, depending upon whether hyperbolic and pseudo-elliptic points form a complete set of building blocks for Lagrangian structures~\cite{Smith:2017ab}, similar to hyperbolic and elliptic points in SF-only systems. In this case CS can retard mixing and the Lyapunov exponent in (\ref{eqn:SF_mix_decay}) forms an upper bound for the decay rate of the mix-norm. For mixed CS and SF systems, a key question regarding the impact of discontinuous deformation upon SF mixing is the rate and extent to which the web of discontinuities becomes space-filling. To quantify this, we consider the growth rate of line elements in the pore geometries (i)-(iv) whose mixing dynamics are shown in Fig.~\ref{fig:web}.

\begin{figure}[h]
\begin{centering}
\begin{tabular}{c c}
\includegraphics[width=0.48\columnwidth]{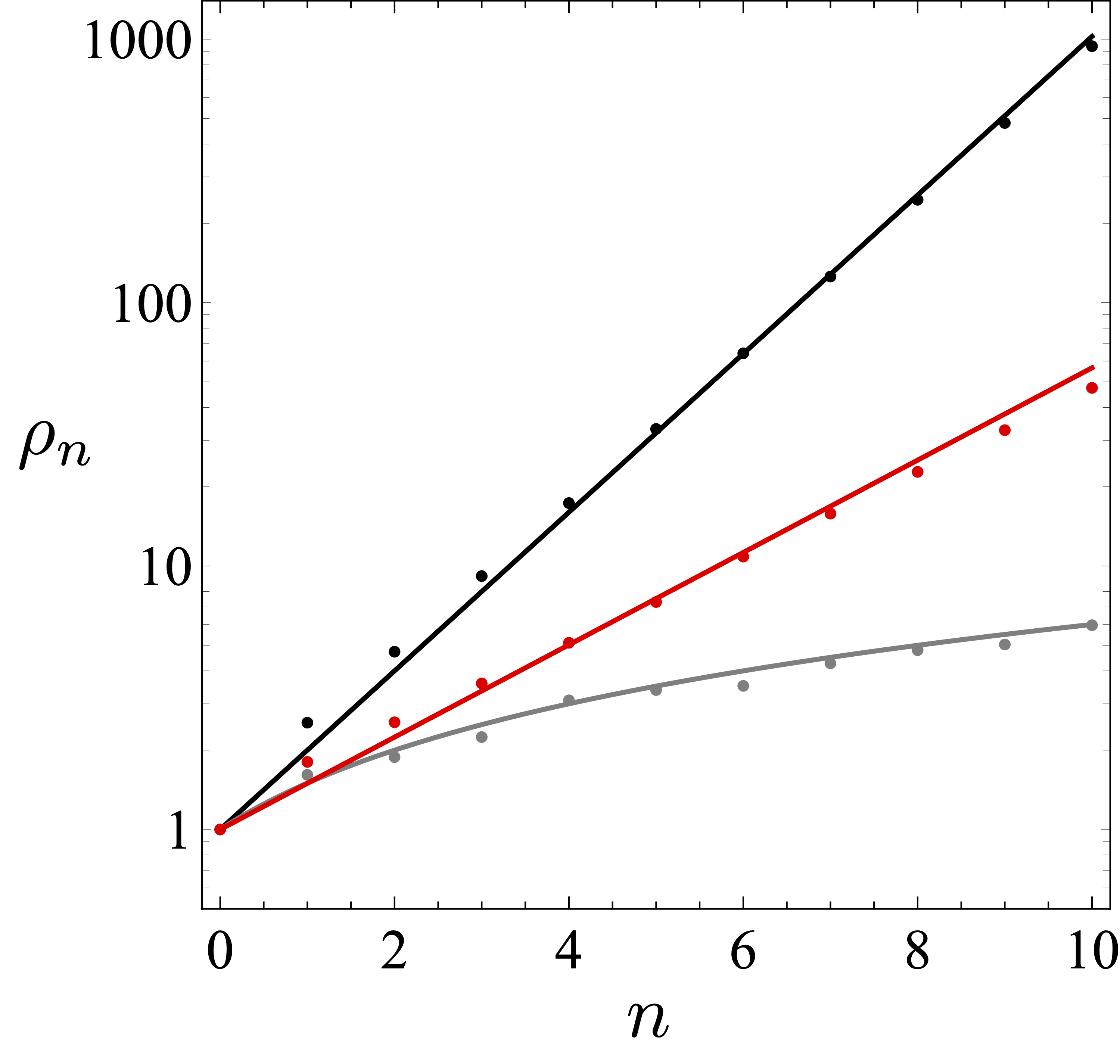}&
\includegraphics[width=0.48\columnwidth]{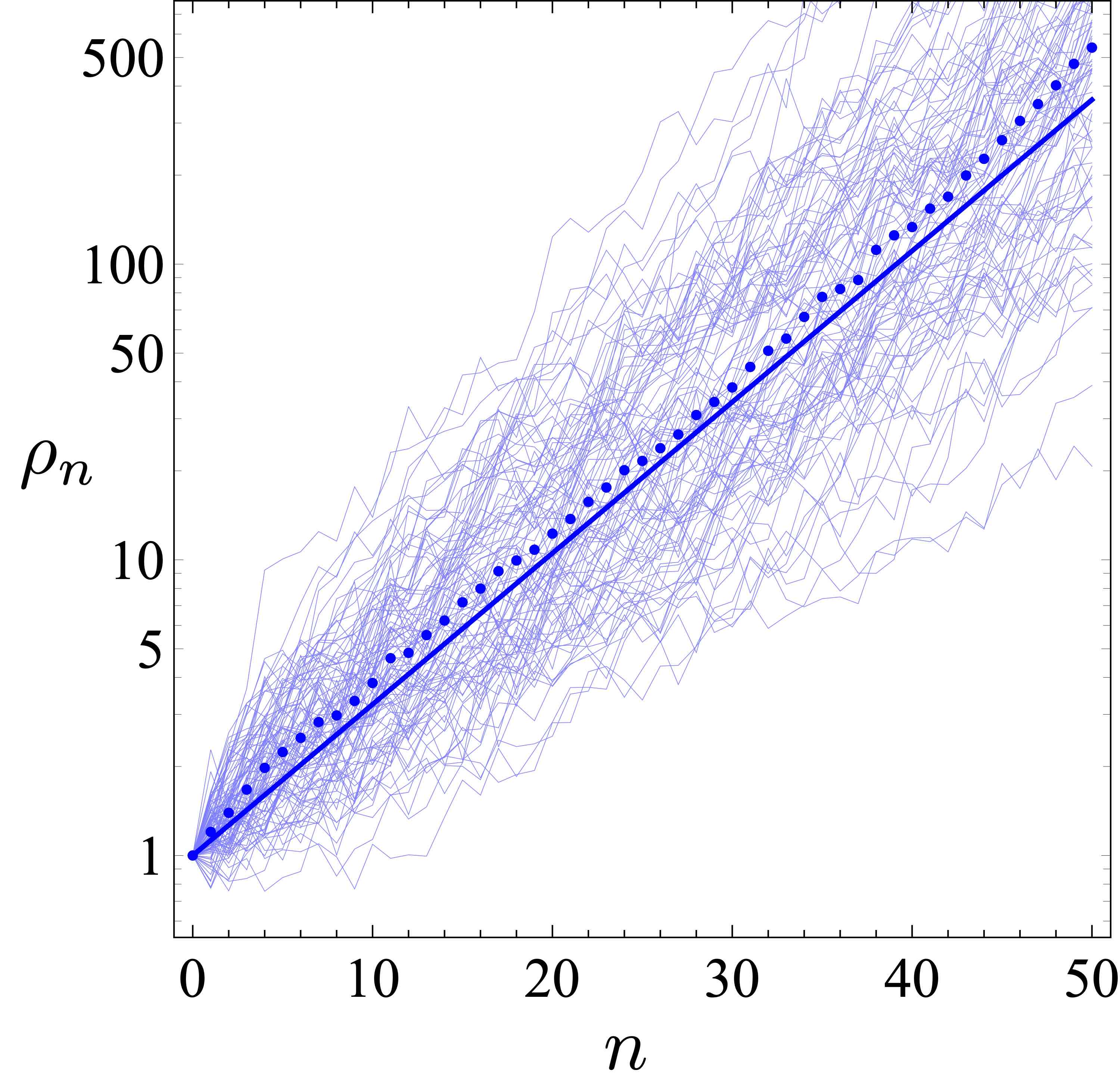}\\
(a) & (b)\\
\includegraphics[width=0.48\columnwidth]{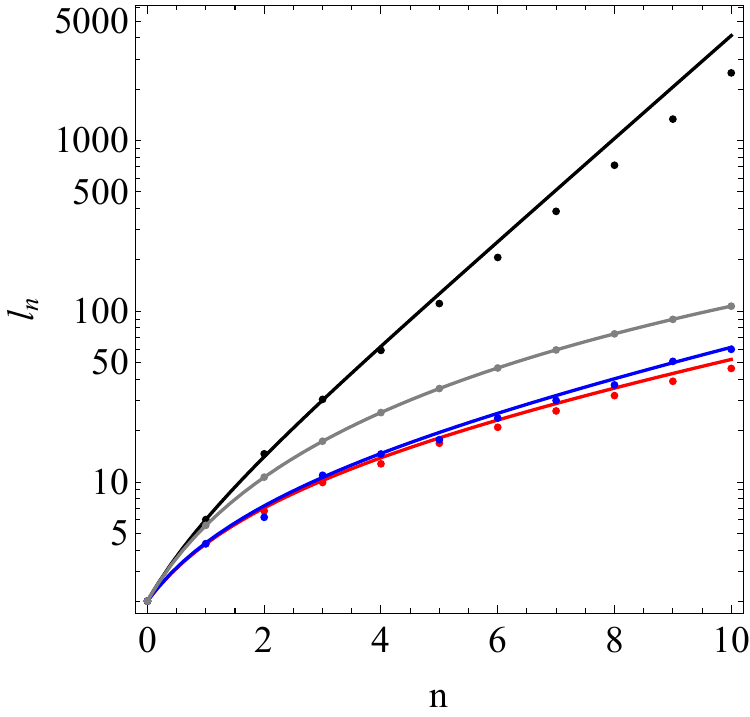}&
\includegraphics[width=0.48\columnwidth]{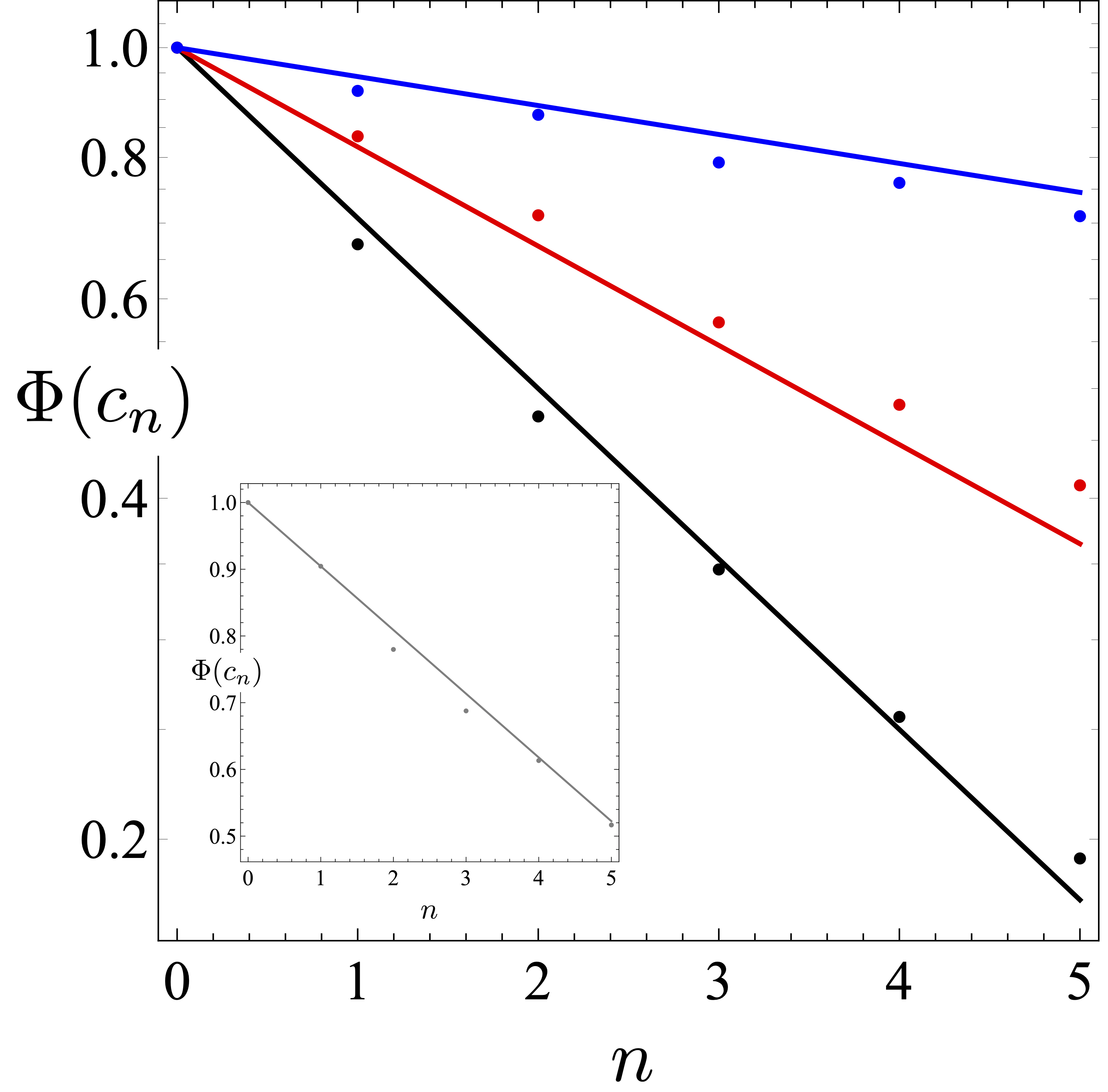}\\
(c) & (d)
\end{tabular}
\end{centering}
\caption{(a) Relative elongation $\rho$ of infinitesimal material lines in (a) ordered and (b) random pore networks where black, red, grey and blue lines and points respectively correspond to the baker's map (i), chaotic (ii), non-chaotic (iii), and 100 realisations of random (iv) pore geometries. Points represent numerical results from maps $\mathcal{S}_b$, $\mathcal{S}_m$, and lines represent stretching rates based upon the Lyapunov exponent for ordered (\ref{eqn:Lyapunov_cts_ordered}) and random (\ref{eqn:Lyapunov_cts_random}) networks, except for the non-chaotic case where stretching evolves linearly as $\rho_n=\rho_0+\alpha n$. Thin blue lines in (b) represent stretching of 100 realisations of the random network and the thick blue line the ensemble average. (c) Growth of the total length $l_n$ of the web of discontinuities with number $n$ of pore branches and mergers with same colour code as for (a) and (b). Points represent numerical results from maps $\mathcal{S}_b$, $\mathcal{S}_m$, and lines represent analytic predictions based on stretching rates for chaotic (\ref{eqn:expsum}) and non-chaotic (\ref{eqn:powsum}) cases. (d) Evolution of mix-norm $\Phi(c_n)$ in pore networks with chaotic and (inset) non-chaotic mixing, where points represent numerical results from maps $\mathcal{S}_b$, $\mathcal{S}_m$ and (\ref{eqn:mixnorm})-(\ref{eqn:davg}), and lines represent the mix-norm estimate (\ref{eqn:SF_mix_decay}) based upon pure SF motions and the Lyapunov exponent given by (\ref{eqn:Lyapunov_cts_ordered}), (\ref{eqn:Lyapunov_cts_random}).}\label{fig:linestretch}
\end{figure}

Fig.~\ref{fig:linestretch}a shows that the mean growth of the normalized length $\rho_n\equiv\langle\delta l_n/\delta l_0\rangle$ of an ensemble of infinitesimal line elements in the ordered and random pore networks is well approximated by the theoretical estimate (\ref{eqn:Lyapunov_cts_ordered}) of the Lyapunov exponent $\lambda_\infty$, hence exponential fluid stretching is captured by linearizations of the spatial maps $\mathcal{S}_b$, $\mathcal{S}_m$.  For the non-chaotic case ($\Delta=\pi/4$, $\delta=\pi/4$), the Lyapunov exponent is zero and the relative length $\rho_n$ grows linearly as $\rho_n\approx\rho_0+n/2$.  As all of these pore geometries completely mix as $t\rightarrow\infty$, then the growth of the total length $l_n$ of the web of discontinuities follows the same stretching process as the fluid phase.

As an additional cut is added to the web of discontinuities at each iteration of the map $\Lambda_n$ (\ref{eqn:Lambda_map}), then evolution of $l_n$ is related to the growth of the length $L_n$ of \emph{finite} material lines with initial length $L_0=2$ (corresponding to the length of the cut along the chord $y_r=0$) as $l_n=\sum_{i=0}^n L_i$. As the exponential stretching rate in chaotic pore-scale flows is log-normally distributed~\cite{Lester:2016aa} with mean stretching rate $\lambda_\infty$ and variance $\sigma_\lambda^2$, for pore geometries that admit chaotic mixing these finite line elements grow exponentially as $L_n\approx2\exp\left(\Lambda_\infty n\right)$, where $\Lambda_\infty=\lambda_\infty+\sigma_\lambda^2/2$ and $\sigma_\lambda^2$ is found~\cite{Lester:2016aa} to be $\sigma_\lambda^2\approx0.1144$ for random networks, whereas for the ordered networks this variance is negligible, $\sigma_\lambda^2\approx 0$. Thus the length $l_n$ of the web of discontinuities for pore networks that exhibit chaotic mixing evolves with $n$ as
\begin{equation}
l_n\approx2\frac{\exp\left(\Lambda_\infty(n+1)\right)-1}{\exp\left(\Lambda_\infty\right)-1}.\label{eqn:expsum}
\end{equation}
Conversely, for the non-chaotic case $\Delta=\delta=\pi/4$ the length of line elements grows linearly as $L_n\approx 2+n$ and so the total length of the web of discontinuities grows algebraically as
\begin{equation}
l_n\approx\frac{(n+1)(n+4)}{2}.\label{eqn:powsum}
\end{equation}
Fig.~\ref{fig:linestretch}b shows that for all pore geometries (i)-(iv), the growth rate of the web of discontinuities is well approximated by these scalings, indicating that the stretching of line elements mediates cutting and shuffling.

\subsection{Interactions Between Continuous and Discontinuous Mixing}

To determine how CS actions of these maps impact the mixing dynamics, we compute the mix-norm $\Phi(n)$ for all four cases, and the results are summarised in Fig.~\ref{fig:linestretch}c. For the chaotic mixing cases, Fig.~\ref{fig:linestretch}c indicates that the mix-norm decays exponentially at a rate that is half the Lyapunov exponent in (\ref{eqn:SF_mix_decay}), suggesting that the CS actions do not contribute significantly to rate of mixing, but rather these are dominated by the exponential SF dynamics. For the non-chaotic case, the mix-norm decays linearly at a rate that is also half the growth rate $\alpha$, also suggesting that mixing is governed by fluid stretching as stretching also mediates the CS process. Although it is tempting to conclude that SF motions dominate the mixing process, it is important to note that the estimate (\ref{eqn:SF_mix_decay}) of the mix-norm only accounts for the amount of stretching experienced by fluid elements, and does not consider whether packing of stretched material lines into a finite-sized pore is achieved by either folding or cutting of this material.

Fig.~\ref{fig:web} clearly shows that material lines experience much more cutting (due to strong 3D folding) rather than weak folding (arising from the non-affine map $\mathcal{M}_n$), hence cutting of material lines is integral to this packing process. Therefore although CS plays a critical role in mixing by facilitating packing of elongated material elements into the pore-space, of the SF motions, only fluid stretching is required to achieve chaotic mixing as weak fluid folding plays a secondary role. As such, the theoretical estimates for the Lypaunov exponents in both ordered (\ref{eqn:Lyapunov_cts_ordered}) and random (\ref{eqn:Lyapunov_cts_random}) networks (which have been validated against line stretching simulations using the spatial maps (\ref{eqn:spatial_maps})) provide an accurate means of predicting the mixing rate in continuous porous media, and represent useful quantitative tools for the control and optimisation of mixing in engineered porous media. These insights into these CS and SF actions provide a complete description of the mechanisms that govern mixing in continuous porous media.

\section{Unified Prediction of Lyapunov Exponent in Continuous and Discrete Porous Media}
\label{sec:lyapunov}

While the prediction of the Lyapunov exponent for continuous porous media is well established (see Section~\ref{subsec:lyapunov_cts}), current models for discrete porous media rely on experimental and numerical observations. Here we derive a general formulation for the Lyapunov exponent that encompasses continuous and discrete porous media.

\subsection{Lyapunov Model in Discrete Porous Media}
\label{subsec:lyapunov_discrete}

In the deformation model for continuous media presented in section \ref{subsec:lyapunov_cts}, fluid stretching and compression transverse to longitudinal flow direction is decoupled over the pore branch and merger. In a branch, fluid stretching (by a factor 2) in the transverse direction parallel to the branching is fully compensated by decceleration in the longitudinal direction, while the other transverse component is neutral. The reverse situation occurs symmetrically in a merger.

\begin{figure}[t]
\begin{centering}
\begin{tabular}{c c}
\includegraphics[width=0.9\columnwidth]{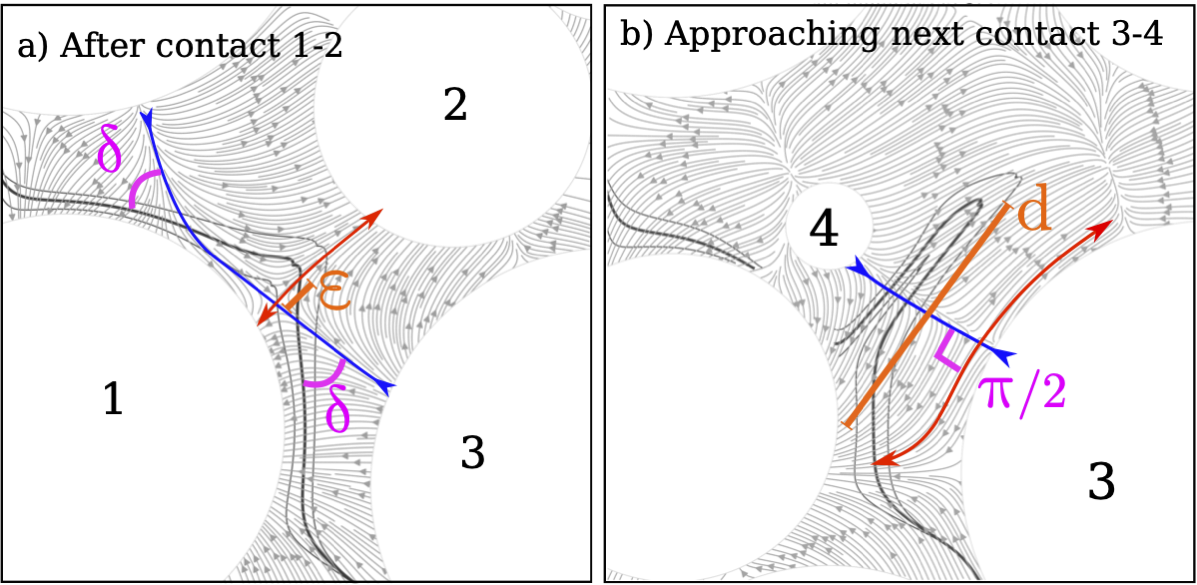}
\end{tabular}
\end{centering}
\caption{Illustration of the mechanism motivating the stretching model for discrete porous media, adapted from Fig. \ref{fig:3Dmanifold}. (a) Immediately after a contact, the material line (black) is oriented at a angle $\delta$ from the stable manifold (blue line) due to asymmetry of the bead pack. This leads to a cusp of initial size $\epsilon$ that elongates along the unstable manifold (red line). (b) When approaching the next contact, the cusp has elongated to a length approximately equal to the grain diameter $d$.}\label{fig:sketch}
\end{figure}

In discrete porous media, fluid stretching and compression acts in both transverse  directions (Fig. \ref{fig:3Dmanifold}). To formulate a Lyapunov model for discrete media, we thus assume that compression and stretching occur simultaneously and perpendicularly in the transverse plane along the stable and unstable manifolds respectively. Fluid deformation from one contact point to the next in the 2D plane transverse to the mean flow direction is approximated by the deformation gradient tensor
\begin{equation}
F=
\begin{pmatrix}
\alpha & 0\\
0 & 1/\alpha 
\end{pmatrix},
\label{deformation::discrete}
\end{equation}
where $\alpha\geqslant 1$ quantifies stretching along the unstable manifold. Note that the deformation represented by $F$ is equivalent to the deformation $\mathcal{D}(\Delta)$ over a couplet in the pore network model (\ref{eqn:S}) with $\Delta=\pi/2$ and $\alpha=2$. After passing the contact the un/stable manifolds are inverted (Fig.~\ref{fig:3Dmanifold}.3), which is reflected by a reorientation of $F$ of angle $\pi/2$ (transition from Fig. \ref{fig:3Dmanifold}.2 to Fig. \ref{fig:3Dmanifold}.4). However, the presence of other beads generates an asymmetry in the orientation of the hyperbolic manifolds upstream and downstream of the contact point, inducing a perturbation of angle $\delta$ (Fig. \ref{fig:3Dmanifold}.4). Therefore the downstream (un)stable manifold is oriented at an angle $\varphi=\pi/2 +\delta$ from the upstream (un)stable manifold. 
Since the material line at the contact point is aligned with the upstream unstable manifold, it becomes oriented at an angle $\delta$ from the compression direction downstream of the contact point as per Fig.~\ref{fig:sketch}.a and Fig. \ref{fig:3Dmanifold}.4.

\begin{figure}
    \includegraphics[width=\linewidth]{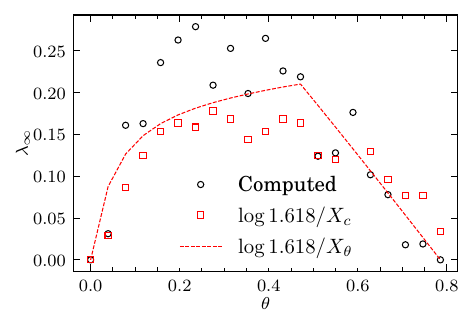}
    \caption{Comparison of predicted and computed~\cite{Turuban:2019aa} Lyapunov exponent $\lambda_\infty$ in BCC sphere lattices with various flow orientations $\theta$ with respect to the lattice symmetries. The numerically computed of Lyapunov exponents are shown as black circles. The predictions of equation \eqref{eq:log2_Xc} using the numerically measured distance $X_c$ and the analytical approximation $X_c(\theta)$ are shown as red squares and red dashed lines respectively.}
    \label{fig:comparison:BCC}
\end{figure}

The asymmetry that generates the perturbation angle $\delta$ also induces a small cusp of initial size $\epsilon$ (Fig.~\ref{fig:sketch}.a) that is amplified via compression and stretching to a fold on the order of the grain size $d$ at the following contact (Fig.s~\ref{fig:sketch}.b and \ref{fig:3Dmanifold}.5). Hence the deformation between the two successive contact points is given by $\alpha \approx d/\epsilon$. 
The net deformation $D_{2n}$ arising from subsequent stretching events between $2n+1$ contact points may then be represented as
\begin{equation}
    D_{2n}=\prod_{j=1}^n\left(F\cdot[R(\varphi)\cdot F\cdot R(\varphi)^\top]\right),
\end{equation}
where $R(\varphi)$ is the rotation matrix. The eigenvalues $r_i$ of $D_{2n}$ are
\begin{equation}
r_i=\{ \xi \pm \sqrt{\xi^2-1} \}^{n/2},\quad\xi =\cos^2{\delta}+\frac{\sin^2{\delta}}{2 \alpha^2}(1 + \alpha^4).
\end{equation}
The perturbation angle $\delta$ can be approximated from the initial cusp size $\epsilon$ as $\tan \delta \approx \epsilon /d$, hence for small perturbation angle $\delta$, $\delta \approx \tan \delta$ and  $\delta \approx \epsilon /d = 1/\alpha$. Under these assumptions and in the limit $\delta\rightarrow 0$, the eigenvalues $r_i$ are
\begin{equation}
r_i\approx \left(\frac{3\pm \sqrt{5}}{2}\right)^{n/2}
\label{eq:r:discontinuous}
\end{equation}
Hence the effective stretching from one contact point to the next ($n=1$) is given by the largest eigenvalue $r\approx 1.618$. If $X_c$ is the average length of the critical line connecting contact points \cite{Heyman:2020aa}, then the elongation $\rho$ over distance $X$
\begin{equation}
\rho=\exp \left( \frac{X}{X_c} \ln(r) \right),
\end{equation}
and so the dimensionless Lyapunov exponent is
\begin{equation}
\lambda_{\infty}=\frac{d}{X_c}\ln(r).
\label{eq:log2_Xc}
\end{equation}
This model (\ref{eq:log2_Xc}) can be tested against simulations of Stokes flow through bcc sphere lattices~\cite{Turuban:2018aa, Turuban:2019aa}. In this system, the Lagrangian kinematics and distance between contact points is varied by changing the mean flow orientation $\theta$ with respect to the lattices symmetries~\cite{Turuban:2019aa}. 
The latter was estimated by measuring the distance of critical lines between successive contacts, described by an analytical model using simple geometrical principles. The prediction of Eq. (\ref{eq:log2_Xc}) is in relatively good agreement with the measured Lyapunov exponents $\lambda_{\infty}$ for all angles $\theta$  using either numerical estimates or the analytical expression for $X_c$ (Fig.~\ref{fig:comparison:BCC}). 
It captures the evolution of the absolute value of the Lyapunov as a function of $\theta$, while previous predictions for this system were only relative ~\cite{Turuban:2019aa}.

Note that Heyman et al. \cite{Heyman:2020aa}, who measured experimentally the Lyapunov exponent $\lambda_{\infty},$ in random loose granular packings, linked it to a folding frequency assuming $r=ln(2)$. 
The slightly different value of 
$r$ derived here should therefore introduce a minor adjustment to the estimated folding frequency in random packings

\subsection{General Formulation for the Lyapunov Exponent}
\label{subsec:lyapunov_discrete_model}

\begin{figure}[ht]
\centering\includegraphics[width=\linewidth]{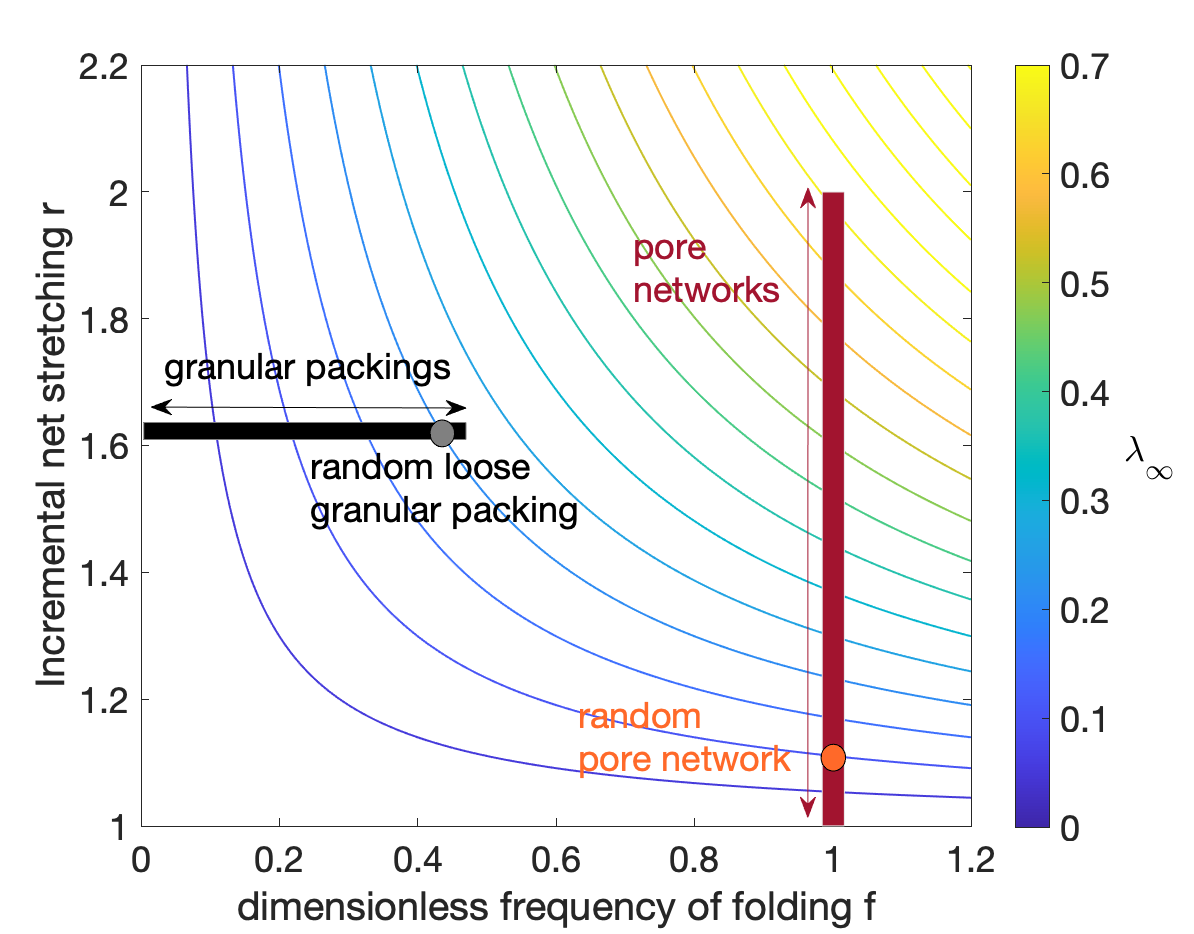}
\caption{Distribution of Lypapunov exponents $\lambda_{\infty}$ predicted from unified theory (\ref{eq:lambda:generic}) as a function of dimensionless folding frequency $f$ and net incremental stretching $r$. The Lyapunov exponent for granular packings and pore networks are located on the black and red lines respectively, and random loose granular packing and the random pore network are indicated by a grey and orange dots respectively.}\label{fig:synthesis}
\end{figure}

The models of equations \eqref{eqn:Lyapunov_cts_ordered} and \eqref{eq:log2_Xc} for the Lyapunov exponent in continuous and discrete porous media respectively, can be described by the general expression 
\begin{equation}
\lambda_{\infty}= f \ln(r) 
\label{eq:lambda:generic}
\end{equation}
with $f$ the dimensionless frequency of folding events and $r$ the average incremental stretch at each event,
\begin{equation}
r= \lim_{n\to\infty} \left\langle \frac{\rho_{n+1}}{\rho_{n}} \right\rangle.
\label{eq:r}
\end{equation}
For continuous media the frequency $f$ of folding events is unity, as folding occurs at every pore branching or merging, but the magnitude of stretching $r|\xi+\sqrt{\xi^2-1}|\in[1,2]$ varies with the pore network parameters $\delta$, $\Delta$ as per \eqref{eqn:Lyapunov_cts_ordered}~\cite{Lester:2013aa}. Conversely, for discrete media the stretching increment is fixed at $r=1.618$, but the frequency of contacts varies with the average length of critical lines between contact points as $f=d/X_c$ as per (\ref{eq:log2_Xc}) which can be controlled by e.g. altering the flow orientation in crystalline lattices~\cite{Turuban:2018aa}.

These distinct controls on the Lyapunov exponent $\lambda_\infty$ are illustrated in Fig. 
\ref{fig:synthesis} over the phase space $\{f,r \}$. For continuous media, the Lyapunov exponent can vary from zero to the theoretical maximum $\ln 2\approx 0.6931$ for continuous 3D systems given by the baker's map. For discrete media $\lambda_\infty$ ranges from zero to around 0.218 based on an upper bound of $d/X_c\approx 0.45$. While the considered granular packings and pore networks shown in Fig. 
\ref{fig:synthesis} cover limited regions of the phase space $\{f,r \}$, more complex porous media, such as rocks or hierarchical media, may have mixed continuous/discrete porous media attributes and thus lie in between these extremes.

\section{A Unified Description of Mixing in Porous Media}
\label{sec:conclusions}

Despite fundamental differences in pore-scale geometry and topology, discrete and continuous porous media share the same basic mechanism for the generation of chaotic mixing. The topological complexity inherent to all porous media generates exponential fluid deformation at critical points in the fluid/solid boundary that propagates into the fluid interior. This stretching manifests as 2D hyperbolic un/stable manifolds that form a transverse heteroclinic connection in all but highly symmetric cases, generating fluid stretching and folding and chaotic mixing. Previous studies have considered the mechanisms that lead to chaotic mixing in both continuous~\cite{Lester:2013aa,Lester:2016aa} and discrete porous media~\cite{Turuban:2018aa,Turuban:2019aa}, but there remain several open questions (i-vi) outlined in Section~\ref{sec:intro}. This study addresses these questions, leading to a unified description of pore-scale mixing in continuous and discrete porous media.

In Section~\ref{sec:topology}, continuous and discrete porous media were shown to be topologically equivalent by considering continuous porous media as discrete porous media with finite-sized connections between grains. Although this topological equivalence involves inversion of the fluid and solid phases, this does not impact chaotic mixing because this is generated by the total negative Gaussian curvature $K$ of the fluid-solid boundary $\delta\Omega$ in both classes of media. These media do however differ in that $\delta\Omega$ in continuous media has mostly negative curvature and so admits saddle points, whereas discrete media has mostly positive curvature with node points, except at cusp-shaped contacts which have divergent negative curvature and contact points. This difference has significant implications for generation of chaotic mixing as saddle points generate chaotic mixing, whereas node and contact points do not.

In Section~\ref{sec:stretching}, both classes of media were shown to exhibit similar fluid stretching dynamics, interior stable and unstable 2D manifolds that intersect transversely along 1D critical lines that connect with boundary saddle points (continuous media) or contact points (discrete media). It was also established, contrary to prior understanding~\cite{Turuban:2019aa}, that while contact points between grains in discrete media are \emph{degenerate} (in that the local velocity gradient tensor has a zero eigenvalue), in concert with node points, they can still generate hyperbolic 2D manifolds and chaotic mixing. This establishes that chaotic mixing is ubiquitous in discrete porous media as contact points are inherent to these materials.

The mechanisms that drive fluid folding in both discrete and continuous porous media were uncovered in Section~\ref{sec:folding}. For discrete porous media, hold-up of fluid at contact points generates strong folding during advection, which manifests as weak folding transverse to the mean flow direction due to symmetry breaking. This weak folding is then amplified by fluid stretching and compression, leading to the formation of highly striated material distributions. For continuous porous media, strong folding is generated by fluid hold-up at critical lines in pore branches. Although fluid mixing is continuous in the 3D pore space, when projected to downstream 2D planes transverse to the mean flow, this hold-up and strong folding manifests as discontinuous mixing, which is explored in Section~\ref{sec:disco}. Weak folding also occurs in continuous media due to advection over pore branches and mergers, which is also amplified by stretching and compression to form striated fluid distributions.

The impacts of discontinuous mixing due to fluid hold-up at pore branches in continuous media are considered in Section~\ref{sec:disco}. Although similar hold-up occurs in discrete porous media at contact points, their infinitesimal size means that they does not generate finite-sized discontinuities. Fluid elements in continuous media undergo both continuous (SF) and discontinuous (CS) actions , which may be summarised as "cut-stretch-fold-rotate" in pore branches and "compress-fold-glue-rotate" in pore mergers, leading to generation of a \emph{web of discontinuities} along which cutting and shuffling occur. For systems that undergo CS only, complete mixing occurs if this web becomes space filling. We show that growth of the web of discontinuities can be accurately predicted by fluid stretching models, and highly stretched fluid elements are predominantly packed into finite-sized pores by CS rather than folding. As CS is mediated by SF, evolution of the mix-norm measure~\cite{Mathew:2005aa} in continuous media are accurately predicted by estimates (\ref{eqn:SF_mix_decay}) based solely on the Lyapunov exponent.

In Section~\ref{sec:lyapunov}, a generalised model for the dimensionless Lyapunov exponent in discrete porous media is presented. This model (\ref{eq:log2_Xc}) encodes the fluid stretching mechanisms uncovered in Section~\ref{sec:stretching} and agrees well with numerical and experimental observations. A unified model for the dimensionless Lyapunov exponent in both discrete and continuous porous media is also developed which is comprised of the product of the magnitude of stretching increments $r$ and their dimensionless frequency $f$. For continuous media, the frequency is unity, but the stretching increment varies due to the orientation of the pore branches and mergers in the network. Conversely, for discrete media the stretching increment over a contact is fixed but the frequency varies with the length of critical lines that connect contact points. These specific models cover a limited range of the parameter space $\{f,r\}$, whereas more complex media such as heterogeneous materials may span this space.

As such, the Lyapunov exponent can be used to quantify mixing in all porous media (Fig. \ref{fig:synthesis}) and  provide a basis for optimisation of mixing and transport in engineered porous materials. Thus predictive theories~\cite{Lester:2013aa,Heyman:2020aa} for the Lyapunov exponent play an important role in understanding and quantifying mixing in porous media, and recently developed characterisation techniques~\cite{Heyman:2021aa} facilitate measurement of the Lyapunov exponent for a broad range of porous materials. In conjunction with previous studies~\cite{Lester:2013aa,Lester:2016aa,Kree:2017aa,Turuban:2018aa,Turuban:2019aa,Souzy:2020aa,Heyman:2020aa}, these insights provide a complete description of the mechanisms that govern mixing in porous media, and so facilitate the development of improved theories of a wide range of physical, chemical and biological fluid-borne phenomena in porous media, as well as design tools for novel porous architectures with tuneable mixing and transport properties.

\bibliography{biblio_chaos-2}

\begin{thebibliography}{54}%
\makeatletter
\providecommand \@ifxundefined [1]{%
 \@ifx{#1\undefined}
}%
\providecommand \@ifnum [1]{%
 \ifnum #1\expandafter \@firstoftwo
 \else \expandafter \@secondoftwo
 \fi
}%
\providecommand \@ifx [1]{%
 \ifx #1\expandafter \@firstoftwo
 \else \expandafter \@secondoftwo
 \fi
}%
\providecommand \natexlab [1]{#1}%
\providecommand \enquote  [1]{``#1''}%
\providecommand \bibnamefont  [1]{#1}%
\providecommand \bibfnamefont [1]{#1}%
\providecommand \citenamefont [1]{#1}%
\providecommand \href@noop [0]{\@secondoftwo}%
\providecommand \href [0]{\begingroup \@sanitize@url \@href}%
\providecommand \@href[1]{\@@startlink{#1}\@@href}%
\providecommand \@@href[1]{\endgroup#1\@@endlink}%
\providecommand \@sanitize@url [0]{\catcode `\\12\catcode `\$12\catcode
  `\&12\catcode `\#12\catcode `\^12\catcode `\_12\catcode `\%12\relax}%
\providecommand \@@startlink[1]{}%
\providecommand \@@endlink[0]{}%
\providecommand \url  [0]{\begingroup\@sanitize@url \@url }%
\providecommand \@url [1]{\endgroup\@href {#1}{\urlprefix }}%
\providecommand \urlprefix  [0]{URL }%
\providecommand \Eprint [0]{\href }%
\providecommand \doibase [0]{https://doi.org/}%
\providecommand \selectlanguage [0]{\@gobble}%
\providecommand \bibinfo  [0]{\@secondoftwo}%
\providecommand \bibfield  [0]{\@secondoftwo}%
\providecommand \translation [1]{[#1]}%
\providecommand \BibitemOpen [0]{}%
\providecommand \bibitemStop [0]{}%
\providecommand \bibitemNoStop [0]{.\EOS\space}%
\providecommand \EOS [0]{\spacefactor3000\relax}%
\providecommand \BibitemShut  [1]{\csname bibitem#1\endcsname}%
\let\auto@bib@innerbib\@empty
\bibitem [{\citenamefont {Dentz}\ \emph {et~al.}(2011)\citenamefont {Dentz},
  \citenamefont {Le~Borgne}, \citenamefont {Englert},\ and\ \citenamefont
  {Bijeljic}}]{Dentz:2011aa}%
  \BibitemOpen
  \bibfield  {author} {\bibinfo {author} {\bibfnamefont {M.}~\bibnamefont
  {Dentz}}, \bibinfo {author} {\bibfnamefont {T.}~\bibnamefont {Le~Borgne}},
  \bibinfo {author} {\bibfnamefont {A.}~\bibnamefont {Englert}},\ and\ \bibinfo
  {author} {\bibfnamefont {B.}~\bibnamefont {Bijeljic}},\ }\bibfield  {title}
  {\bibinfo {title} {Mixing, spreading and reaction in heterogeneous media: A
  brief review},\ }\href {https://doi.org/10.1016/j.jconhyd.2010.05.002}
  {\bibfield  {journal} {\bibinfo  {journal} {J. Cont. Hydrol.}\ }\textbf
  {\bibinfo {volume} {120-121}},\ \bibinfo {pages} {1} (\bibinfo {year}
  {2011})}\BibitemShut {NoStop}%
\bibitem [{\citenamefont {Rolle}\ and\ \citenamefont
  {Le~Borgne}(2019)}]{Rolle:2019aa}%
  \BibitemOpen
  \bibfield  {author} {\bibinfo {author} {\bibfnamefont {M.}~\bibnamefont
  {Rolle}}\ and\ \bibinfo {author} {\bibfnamefont {T.}~\bibnamefont
  {Le~Borgne}},\ }\bibfield  {title} {\bibinfo {title} {Mixing and reactive
  fronts in the subsurface},\ }\href@noop {} {\bibfield  {journal} {\bibinfo
  {journal} {Reviews in Mineralogy and Geochemistry}\ }\textbf {\bibinfo
  {volume} {85}},\ \bibinfo {pages} {111} (\bibinfo {year} {2019})}\BibitemShut
  {NoStop}%
\bibitem [{\citenamefont {Valocchi}\ \emph {et~al.}(2019)\citenamefont
  {Valocchi}, \citenamefont {Bolster},\ and\ \citenamefont
  {Werth}}]{Valocchi:2019aa}%
  \BibitemOpen
  \bibfield  {author} {\bibinfo {author} {\bibfnamefont {A.~J.}\ \bibnamefont
  {Valocchi}}, \bibinfo {author} {\bibfnamefont {D.}~\bibnamefont {Bolster}},\
  and\ \bibinfo {author} {\bibfnamefont {C.~J.}\ \bibnamefont {Werth}},\
  }\bibfield  {title} {\bibinfo {title} {Mixing-limited reactions in porous
  media},\ }\href@noop {} {\bibfield  {journal} {\bibinfo  {journal} {Transport
  in Porous Media}\ }\textbf {\bibinfo {volume} {130}},\ \bibinfo {pages} {157}
  (\bibinfo {year} {2019})}\BibitemShut {NoStop}%
\bibitem [{\citenamefont {Gramling}\ \emph {et~al.}(2002)\citenamefont
  {Gramling}, \citenamefont {Harvey},\ and\ \citenamefont
  {Meigs}}]{Gramling:2002aa}%
  \BibitemOpen
  \bibfield  {author} {\bibinfo {author} {\bibfnamefont {C.~M.}\ \bibnamefont
  {Gramling}}, \bibinfo {author} {\bibfnamefont {C.~F.}\ \bibnamefont
  {Harvey}},\ and\ \bibinfo {author} {\bibfnamefont {L.~C.}\ \bibnamefont
  {Meigs}},\ }\bibfield  {title} {\bibinfo {title} {Reactive transport in
  porous media: A comparison of model prediction with laboratory
  visualization},\ }\href@noop {} {\bibfield  {journal} {\bibinfo  {journal}
  {Environ. Sci. Technol.}\ }\textbf {\bibinfo {volume} {36}},\ \bibinfo
  {pages} {2508} (\bibinfo {year} {2002})}\BibitemShut {NoStop}%
\bibitem [{\citenamefont {Berkowitz}\ \emph {et~al.}(2016)\citenamefont
  {Berkowitz}, \citenamefont {Dror}, \citenamefont {Hansen},\ and\
  \citenamefont {Scher}}]{Berkowitz:2016aa}%
  \BibitemOpen
  \bibfield  {author} {\bibinfo {author} {\bibfnamefont {B.}~\bibnamefont
  {Berkowitz}}, \bibinfo {author} {\bibfnamefont {I.}~\bibnamefont {Dror}},
  \bibinfo {author} {\bibfnamefont {S.~K.}\ \bibnamefont {Hansen}},\ and\
  \bibinfo {author} {\bibfnamefont {H.}~\bibnamefont {Scher}},\ }\bibfield
  {title} {\bibinfo {title} {Measurements and models of reactive transport in
  geological media},\ }\href@noop {} {\bibfield  {journal} {\bibinfo  {journal}
  {Reviews of Geophysics}\ }\textbf {\bibinfo {volume} {54}},\ \bibinfo {pages}
  {930} (\bibinfo {year} {2016})}\BibitemShut {NoStop}%
\bibitem [{\citenamefont {Wright}\ \emph {et~al.}(2017)\citenamefont {Wright},
  \citenamefont {Zadrazil},\ and\ \citenamefont {Markides}}]{Wright:2017aa}%
  \BibitemOpen
  \bibfield  {author} {\bibinfo {author} {\bibfnamefont {S.~F.}\ \bibnamefont
  {Wright}}, \bibinfo {author} {\bibfnamefont {I.}~\bibnamefont {Zadrazil}},\
  and\ \bibinfo {author} {\bibfnamefont {C.~N.}\ \bibnamefont {Markides}},\
  }\bibfield  {title} {\bibinfo {title} {A review of solid-fluid selection
  options for optical-based measurements in single-phase liquid, two-phase
  liquid-liquid and multiphase solid-liquid flows},\ }\href@noop {} {\bibfield
  {journal} {\bibinfo  {journal} {Exp. Fluids}\ ,\ \bibinfo {pages} {58:108}}
  (\bibinfo {year} {2017})}\BibitemShut {NoStop}%
\bibitem [{\citenamefont {Lester}\ \emph {et~al.}(2013)\citenamefont {Lester},
  \citenamefont {Metcalfe},\ and\ \citenamefont {Trefry}}]{Lester:2013aa}%
  \BibitemOpen
  \bibfield  {author} {\bibinfo {author} {\bibfnamefont {D.~R.}\ \bibnamefont
  {Lester}}, \bibinfo {author} {\bibfnamefont {G.}~\bibnamefont {Metcalfe}},\
  and\ \bibinfo {author} {\bibfnamefont {M.~G.}\ \bibnamefont {Trefry}},\
  }\bibfield  {title} {\bibinfo {title} {Is chaotic advection inherent to
  porous media flow?},\ }\href {https://doi.org/10.1103/PhysRevLett.111.174101}
  {\bibfield  {journal} {\bibinfo  {journal} {Phys. Rev. Lett.}\ }\textbf
  {\bibinfo {volume} {111}},\ \bibinfo {pages} {174101} (\bibinfo {year}
  {2013})}\BibitemShut {NoStop}%
\bibitem [{\citenamefont {Lester}\ \emph
  {et~al.}(2016{\natexlab{a}})\citenamefont {Lester}, \citenamefont {Dentz},\
  and\ \citenamefont {Le~Borgne}}]{Lester:2016aa}%
  \BibitemOpen
  \bibfield  {author} {\bibinfo {author} {\bibfnamefont {D.~R.}\ \bibnamefont
  {Lester}}, \bibinfo {author} {\bibfnamefont {M.}~\bibnamefont {Dentz}},\ and\
  \bibinfo {author} {\bibfnamefont {T.}~\bibnamefont {Le~Borgne}},\ }\bibfield
  {title} {\bibinfo {title} {Chaotic mixing in three-dimensional porous
  media},\ }\href@noop {} {\bibfield  {journal} {\bibinfo  {journal} {J. Fluid
  Mech.}\ }\textbf {\bibinfo {volume} {803}},\ \bibinfo {pages} {144} (\bibinfo
  {year} {2016}{\natexlab{a}})}\BibitemShut {NoStop}%
\bibitem [{\citenamefont {Kree}\ and\ \citenamefont
  {Villermaux}(2017)}]{Kree:2017aa}%
  \BibitemOpen
  \bibfield  {author} {\bibinfo {author} {\bibfnamefont {M.}~\bibnamefont
  {Kree}}\ and\ \bibinfo {author} {\bibfnamefont {E.}~\bibnamefont
  {Villermaux}},\ }\bibfield  {title} {\bibinfo {title} {Scalar mixtures in
  porous media},\ }\href {https://doi.org/10.1103/PhysRevFluids.2.104502}
  {\bibfield  {journal} {\bibinfo  {journal} {Phys. Rev. Fluids}\ }\textbf
  {\bibinfo {volume} {2}},\ \bibinfo {pages} {104502} (\bibinfo {year}
  {2017})}\BibitemShut {NoStop}%
\bibitem [{\citenamefont {Turuban}\ \emph {et~al.}(2018)\citenamefont
  {Turuban}, \citenamefont {Lester}, \citenamefont {Le~Borgne},\ and\
  \citenamefont {M\'eheust}}]{Turuban:2018aa}%
  \BibitemOpen
  \bibfield  {author} {\bibinfo {author} {\bibfnamefont {R.}~\bibnamefont
  {Turuban}}, \bibinfo {author} {\bibfnamefont {D.~R.}\ \bibnamefont {Lester}},
  \bibinfo {author} {\bibfnamefont {T.}~\bibnamefont {Le~Borgne}},\ and\
  \bibinfo {author} {\bibfnamefont {Y.}~\bibnamefont {M\'eheust}},\ }\bibfield
  {title} {\bibinfo {title} {Space-group symmetries generate chaotic fluid
  advection in crystalline granular media},\ }\href
  {https://doi.org/10.1103/PhysRevLett.120.024501} {\bibfield  {journal}
  {\bibinfo  {journal} {Phys. Rev. Lett.}\ }\textbf {\bibinfo {volume} {120}},\
  \bibinfo {pages} {024501} (\bibinfo {year} {2018})}\BibitemShut {NoStop}%
\bibitem [{\citenamefont {Turuban}\ \emph {et~al.}(2019)\citenamefont
  {Turuban}, \citenamefont {Lester}, \citenamefont {Heyman}, \citenamefont
  {Borgne},\ and\ \citenamefont {Meheust}}]{Turuban:2019aa}%
  \BibitemOpen
  \bibfield  {author} {\bibinfo {author} {\bibfnamefont {R.}~\bibnamefont
  {Turuban}}, \bibinfo {author} {\bibfnamefont {D.~R.}\ \bibnamefont {Lester}},
  \bibinfo {author} {\bibfnamefont {J.}~\bibnamefont {Heyman}}, \bibinfo
  {author} {\bibfnamefont {T.~L.}\ \bibnamefont {Borgne}},\ and\ \bibinfo
  {author} {\bibfnamefont {Y.}~\bibnamefont {Meheust}},\ }\bibfield  {title}
  {\bibinfo {title} {Chaotic mixing in crystalline granular media},\ }\href
  {https://doi.org/10.1017/jfm.2019.245} {\bibfield  {journal} {\bibinfo
  {journal} {J. Fluid Mech.}\ }\textbf {\bibinfo {volume} {871}},\ \bibinfo
  {pages} {562} (\bibinfo {year} {2019})}\BibitemShut {NoStop}%
\bibitem [{\citenamefont {Souzy}\ \emph {et~al.}(2020)\citenamefont {Souzy},
  \citenamefont {Lhuissier}, \citenamefont {M{\'e}heust}, \citenamefont
  {Le~Borgne},\ and\ \citenamefont {Metzger}}]{Souzy:2020aa}%
  \BibitemOpen
  \bibfield  {author} {\bibinfo {author} {\bibfnamefont {M.}~\bibnamefont
  {Souzy}}, \bibinfo {author} {\bibfnamefont {H.}~\bibnamefont {Lhuissier}},
  \bibinfo {author} {\bibfnamefont {Y.}~\bibnamefont {M{\'e}heust}}, \bibinfo
  {author} {\bibfnamefont {T.}~\bibnamefont {Le~Borgne}},\ and\ \bibinfo
  {author} {\bibfnamefont {B.}~\bibnamefont {Metzger}},\ }\bibfield  {title}
  {\bibinfo {title} {Velocity distributions, dispersion and stretching in
  three-dimensional porous media},\ }\href
  {https://doi.org/10.1017/jfm.2020.113} {\bibfield  {journal} {\bibinfo
  {journal} {Journal of Fluid Mechanics}\ }\textbf {\bibinfo {volume} {891}},\
  \bibinfo {pages} {A16} (\bibinfo {year} {2020})}\BibitemShut {NoStop}%
\bibitem [{\citenamefont {Heyman}\ \emph {et~al.}(2020)\citenamefont {Heyman},
  \citenamefont {Lester}, \citenamefont {Turuban}, \citenamefont
  {M{\'e}heust},\ and\ \citenamefont {Le~Borgne}}]{Heyman:2020aa}%
  \BibitemOpen
  \bibfield  {author} {\bibinfo {author} {\bibfnamefont {J.}~\bibnamefont
  {Heyman}}, \bibinfo {author} {\bibfnamefont {D.~R.}\ \bibnamefont {Lester}},
  \bibinfo {author} {\bibfnamefont {R.}~\bibnamefont {Turuban}}, \bibinfo
  {author} {\bibfnamefont {Y.}~\bibnamefont {M{\'e}heust}},\ and\ \bibinfo
  {author} {\bibfnamefont {T.}~\bibnamefont {Le~Borgne}},\ }\bibfield  {title}
  {\bibinfo {title} {Stretching and folding sustain microscale chemical
  gradients in porous media},\ }\bibfield  {journal} {\bibinfo  {journal}
  {Proceedings of the National Academy of Sciences}\ }\href
  {https://doi.org/10.1073/pnas.2002858117} {10.1073/pnas.2002858117} (\bibinfo
  {year} {2020})\BibitemShut {NoStop}%
\bibitem [{\citenamefont {Heyman}\ \emph {et~al.}(2021)\citenamefont {Heyman},
  \citenamefont {Lester},\ and\ \citenamefont {Le~Borgne}}]{Heyman:2021aa}%
  \BibitemOpen
  \bibfield  {author} {\bibinfo {author} {\bibfnamefont {J.}~\bibnamefont
  {Heyman}}, \bibinfo {author} {\bibfnamefont {D.~R.}\ \bibnamefont {Lester}},\
  and\ \bibinfo {author} {\bibfnamefont {T.}~\bibnamefont {Le~Borgne}},\
  }\bibfield  {title} {\bibinfo {title} {Scalar signatures of chaotic mixing in
  porous media},\ }\href {https://doi.org/10.1103/PhysRevLett.126.034505}
  {\bibfield  {journal} {\bibinfo  {journal} {Phys. Rev. Lett.}\ }\textbf
  {\bibinfo {volume} {126}},\ \bibinfo {pages} {034505} (\bibinfo {year}
  {2021})}\BibitemShut {NoStop}%
\bibitem [{\citenamefont {Aref}\ \emph {et~al.}(2017)\citenamefont {Aref},
  \citenamefont {Blake}, \citenamefont {Budi{\v{s}}i{\'c}}, \citenamefont
  {Cardoso}, \citenamefont {Cartwright}, \citenamefont {Clercx}, \citenamefont
  {El~Omari}, \citenamefont {Feudel}, \citenamefont {Golestanian},
  \citenamefont {Gouillart} \emph {et~al.}}]{Aref:2017aa}%
  \BibitemOpen
  \bibfield  {author} {\bibinfo {author} {\bibfnamefont {H.}~\bibnamefont
  {Aref}}, \bibinfo {author} {\bibfnamefont {J.~R.}\ \bibnamefont {Blake}},
  \bibinfo {author} {\bibfnamefont {M.}~\bibnamefont {Budi{\v{s}}i{\'c}}},
  \bibinfo {author} {\bibfnamefont {S.~S.}\ \bibnamefont {Cardoso}}, \bibinfo
  {author} {\bibfnamefont {J.~H.}\ \bibnamefont {Cartwright}}, \bibinfo
  {author} {\bibfnamefont {H.~J.}\ \bibnamefont {Clercx}}, \bibinfo {author}
  {\bibfnamefont {K.}~\bibnamefont {El~Omari}}, \bibinfo {author}
  {\bibfnamefont {U.}~\bibnamefont {Feudel}}, \bibinfo {author} {\bibfnamefont
  {R.}~\bibnamefont {Golestanian}}, \bibinfo {author} {\bibfnamefont
  {E.}~\bibnamefont {Gouillart}}, \emph {et~al.},\ }\bibfield  {title}
  {\bibinfo {title} {Frontiers of chaotic advection},\ }\href@noop {}
  {\bibfield  {journal} {\bibinfo  {journal} {Reviews of Modern Physics}\
  }\textbf {\bibinfo {volume} {89}},\ \bibinfo {pages} {025007} (\bibinfo
  {year} {2017})}\BibitemShut {NoStop}%
\bibitem [{\citenamefont {Lester}\ \emph
  {et~al.}(2014{\natexlab{a}})\citenamefont {Lester}, \citenamefont
  {Metcalfe},\ and\ \citenamefont {Rudman}}]{Lester:2014aa}%
  \BibitemOpen
  \bibfield  {author} {\bibinfo {author} {\bibfnamefont {D.}~\bibnamefont
  {Lester}}, \bibinfo {author} {\bibfnamefont {G.}~\bibnamefont {Metcalfe}},\
  and\ \bibinfo {author} {\bibfnamefont {M.}~\bibnamefont {Rudman}},\
  }\bibfield  {title} {\bibinfo {title} {Control mechanisms for the global
  structure of scalar dispersion in chaotic flows},\ }\href@noop {} {\bibfield
  {journal} {\bibinfo  {journal} {Phys. Rev. E}\ }\textbf {\bibinfo {volume}
  {90}},\ \bibinfo {pages} {022908} (\bibinfo {year}
  {2014}{\natexlab{a}})}\BibitemShut {NoStop}%
\bibitem [{\citenamefont {Lester}\ \emph
  {et~al.}(2014{\natexlab{b}})\citenamefont {Lester}, \citenamefont
  {Metcalfe},\ and\ \citenamefont {Trefry}}]{Lester:2014ab}%
  \BibitemOpen
  \bibfield  {author} {\bibinfo {author} {\bibfnamefont {D.~R.}\ \bibnamefont
  {Lester}}, \bibinfo {author} {\bibfnamefont {G.}~\bibnamefont {Metcalfe}},\
  and\ \bibinfo {author} {\bibfnamefont {M.~G.}\ \bibnamefont {Trefry}},\
  }\bibfield  {title} {\bibinfo {title} {Anomalous transport and chaotic
  advection in homogeneous porous media},\ }\href
  {https://doi.org/10.1103/PhysRevE.90.063012} {\bibfield  {journal} {\bibinfo
  {journal} {Phys. Rev. E}\ }\textbf {\bibinfo {volume} {90}},\ \bibinfo
  {pages} {063012} (\bibinfo {year} {2014}{\natexlab{b}})}\BibitemShut
  {NoStop}%
\bibitem [{\citenamefont {Sapsis}\ and\ \citenamefont
  {Haller}(2010)}]{Sapsis:2010aa}%
  \BibitemOpen
  \bibfield  {author} {\bibinfo {author} {\bibfnamefont {T.}~\bibnamefont
  {Sapsis}}\ and\ \bibinfo {author} {\bibfnamefont {G.}~\bibnamefont
  {Haller}},\ }\bibfield  {title} {\bibinfo {title} {Clustering criterion for
  inertial particles in two-dimensional time-periodic and three-dimensional
  steady flows},\ }\href@noop {} {\bibfield  {journal} {\bibinfo  {journal}
  {Chaos: An Interdisciplinary Journal of Nonlinear Science}\ }\textbf
  {\bibinfo {volume} {20}},\ \bibinfo {pages} {017515} (\bibinfo {year}
  {2010})}\BibitemShut {NoStop}%
\bibitem [{\citenamefont {Ouellette}\ \emph {et~al.}(2008)\citenamefont
  {Ouellette}, \citenamefont {O'Malley},\ and\ \citenamefont
  {Gollub}}]{Ouellette:2008aa}%
  \BibitemOpen
  \bibfield  {author} {\bibinfo {author} {\bibfnamefont {N.~T.}\ \bibnamefont
  {Ouellette}}, \bibinfo {author} {\bibfnamefont {P.}~\bibnamefont
  {O'Malley}},\ and\ \bibinfo {author} {\bibfnamefont {J.~P.}\ \bibnamefont
  {Gollub}},\ }\bibfield  {title} {\bibinfo {title} {Transport of finite-sized
  particles in chaotic flow},\ }\href@noop {} {\bibfield  {journal} {\bibinfo
  {journal} {Phys. Rev. Lett.}\ }\textbf {\bibinfo {volume} {101}},\ \bibinfo
  {pages} {174504} (\bibinfo {year} {2008})}\BibitemShut {NoStop}%
\bibitem [{\citenamefont {Tel}\ \emph {et~al.}(2005)\citenamefont {Tel},
  \citenamefont {de~Moura}, \citenamefont {Grebogi},\ and\ \citenamefont
  {K{\'a}rolyi}}]{Tel:2005aa}%
  \BibitemOpen
  \bibfield  {author} {\bibinfo {author} {\bibfnamefont {T.}~\bibnamefont
  {Tel}}, \bibinfo {author} {\bibfnamefont {A.}~\bibnamefont {de~Moura}},
  \bibinfo {author} {\bibfnamefont {C.}~\bibnamefont {Grebogi}},\ and\ \bibinfo
  {author} {\bibfnamefont {G.}~\bibnamefont {K{\'a}rolyi}},\ }\bibfield
  {title} {\bibinfo {title} {Chemical and biological activity in open flows: A
  dynamical system approach},\ }\href@noop {} {\bibfield  {journal} {\bibinfo
  {journal} {Phys. Rep.}\ }\textbf {\bibinfo {volume} {413}},\ \bibinfo {pages}
  {91} (\bibinfo {year} {2005})}\BibitemShut {NoStop}%
\bibitem [{\citenamefont {K{\'a}rolyi}\ \emph {et~al.}(2000)\citenamefont
  {K{\'a}rolyi}, \citenamefont {P{\'e}ntek}, \citenamefont {Scheuring},
  \citenamefont {T{\'e}l},\ and\ \citenamefont {Toroczkai}}]{Karolyi:2000aa}%
  \BibitemOpen
  \bibfield  {author} {\bibinfo {author} {\bibfnamefont {G.}~\bibnamefont
  {K{\'a}rolyi}}, \bibinfo {author} {\bibfnamefont {{\'A}.}~\bibnamefont
  {P{\'e}ntek}}, \bibinfo {author} {\bibfnamefont {I.}~\bibnamefont
  {Scheuring}}, \bibinfo {author} {\bibfnamefont {T.}~\bibnamefont {T{\'e}l}},\
  and\ \bibinfo {author} {\bibfnamefont {Z.}~\bibnamefont {Toroczkai}},\
  }\bibfield  {title} {\bibinfo {title} {Chaotic flow: The physics of species
  coexistence},\ }\href@noop {} {\bibfield  {journal} {\bibinfo  {journal}
  {Proceedings of the National Academy of Sciences}\ }\textbf {\bibinfo
  {volume} {97}},\ \bibinfo {pages} {13661} (\bibinfo {year}
  {2000})}\BibitemShut {NoStop}%
\bibitem [{\citenamefont {Stocker}(2012)}]{Stocker:2012aa}%
  \BibitemOpen
  \bibfield  {author} {\bibinfo {author} {\bibfnamefont {R.}~\bibnamefont
  {Stocker}},\ }\bibfield  {title} {\bibinfo {title} {Marine microbes see a sea
  of gradients},\ }\href@noop {} {\bibfield  {journal} {\bibinfo  {journal}
  {Science}\ }\textbf {\bibinfo {volume} {338}},\ \bibinfo {pages} {6107}
  (\bibinfo {year} {2012})}\BibitemShut {NoStop}%
\bibitem [{\citenamefont {Neufeld}\ and\ \citenamefont
  {Hernandez-Garcia}(2009)}]{Neufeld:2009aa}%
  \BibitemOpen
  \bibfield  {author} {\bibinfo {author} {\bibfnamefont {Z.}~\bibnamefont
  {Neufeld}}\ and\ \bibinfo {author} {\bibfnamefont {E.}~\bibnamefont
  {Hernandez-Garcia}},\ }\href@noop {} {\emph {\bibinfo {title} {Chemical and
  biological processes in fluid flows: A dynamical systems approach}}}\
  (\bibinfo  {publisher} {Imperial College Press},\ \bibinfo {year}
  {2009})\BibitemShut {NoStop}%
\bibitem [{\citenamefont {K{\'a}rolyi}\ \emph {et~al.}(2002)\citenamefont
  {K{\'a}rolyi}, \citenamefont {Scheuring},\ and\ \citenamefont
  {Cz{\'a}r{\'a}n}}]{Karolyi:2002aa}%
  \BibitemOpen
  \bibfield  {author} {\bibinfo {author} {\bibfnamefont {G.}~\bibnamefont
  {K{\'a}rolyi}}, \bibinfo {author} {\bibfnamefont {I.}~\bibnamefont
  {Scheuring}},\ and\ \bibinfo {author} {\bibfnamefont {T.}~\bibnamefont
  {Cz{\'a}r{\'a}n}},\ }\bibfield  {title} {\bibinfo {title} {Metabolic network
  dynamics in open chaotic flow},\ }\href {https://doi.org/10.1063/1.1457468}
  {\bibfield  {journal} {\bibinfo  {journal} {Chaos: An Interdisciplinary
  Journal of Nonlinear Science}\ }\textbf {\bibinfo {volume} {12}},\ \bibinfo
  {pages} {460} (\bibinfo {year} {2002})},\ \Eprint
  {https://arxiv.org/abs/https://doi.org/10.1063/1.1457468}
  {https://doi.org/10.1063/1.1457468} \BibitemShut {NoStop}%
\bibitem [{\citenamefont {John}\ and\ \citenamefont
  {Mezi{\'c}}(2007)}]{John:2007aa}%
  \BibitemOpen
  \bibfield  {author} {\bibinfo {author} {\bibfnamefont {T.}~\bibnamefont
  {John}}\ and\ \bibinfo {author} {\bibfnamefont {I.}~\bibnamefont
  {Mezi{\'c}}},\ }\bibfield  {title} {\bibinfo {title} {Maximizing mixing and
  alignment of orientable particles for reaction enhancement},\ }\href
  {https://doi.org/10.1063/1.2819343} {\bibfield  {journal} {\bibinfo
  {journal} {Physics of Fluids}\ }\textbf {\bibinfo {volume} {19}},\ \bibinfo
  {pages} {123602} (\bibinfo {year} {2007})},\ \Eprint
  {https://arxiv.org/abs/https://doi.org/10.1063/1.2819343}
  {https://doi.org/10.1063/1.2819343} \BibitemShut {NoStop}%
\bibitem [{\citenamefont {Huang}\ \emph {et~al.}(2018)\citenamefont {Huang},
  \citenamefont {Huang}, \citenamefont {You}, \citenamefont {Liu},
  \citenamefont {Hollett}, \citenamefont {Kang}, \citenamefont {Gu},\ and\
  \citenamefont {Wu}}]{Huang:2018aa}%
  \BibitemOpen
  \bibfield  {author} {\bibinfo {author} {\bibfnamefont {J.}~\bibnamefont
  {Huang}}, \bibinfo {author} {\bibfnamefont {K.}~\bibnamefont {Huang}},
  \bibinfo {author} {\bibfnamefont {X.}~\bibnamefont {You}}, \bibinfo {author}
  {\bibfnamefont {G.}~\bibnamefont {Liu}}, \bibinfo {author} {\bibfnamefont
  {G.}~\bibnamefont {Hollett}}, \bibinfo {author} {\bibfnamefont
  {Y.}~\bibnamefont {Kang}}, \bibinfo {author} {\bibfnamefont {Z.}~\bibnamefont
  {Gu}},\ and\ \bibinfo {author} {\bibfnamefont {J.}~\bibnamefont {Wu}},\
  }\bibfield  {title} {\bibinfo {title} {Evaluation of tofu as a potential
  tissue engineering scaffold},\ }\href {https://doi.org/10.1039/C7TB02852K}
  {\bibfield  {journal} {\bibinfo  {journal} {J. Mater. Chem. B}\ }\textbf
  {\bibinfo {volume} {6}},\ \bibinfo {pages} {1328} (\bibinfo {year}
  {2018})}\BibitemShut {NoStop}%
\bibitem [{\citenamefont {Melchels}\ \emph {et~al.}(2009)\citenamefont
  {Melchels}, \citenamefont {Feijen},\ and\ \citenamefont
  {Grijpma}}]{Melchels:2009aa}%
  \BibitemOpen
  \bibfield  {author} {\bibinfo {author} {\bibfnamefont {F.~P.}\ \bibnamefont
  {Melchels}}, \bibinfo {author} {\bibfnamefont {J.}~\bibnamefont {Feijen}},\
  and\ \bibinfo {author} {\bibfnamefont {D.~W.}\ \bibnamefont {Grijpma}},\
  }\bibfield  {title} {\bibinfo {title} {A poly(d,l-lactide) resin for the
  preparation of tissue engineering scaffolds by stereolithography},\ }\href
  {https://doi.org/https://doi.org/10.1016/j.biomaterials.2009.03.055}
  {\bibfield  {journal} {\bibinfo  {journal} {Biomaterials}\ }\textbf {\bibinfo
  {volume} {30}},\ \bibinfo {pages} {3801} (\bibinfo {year}
  {2009})}\BibitemShut {NoStop}%
\bibitem [{\citenamefont {Huang}\ \emph {et~al.}(2009)\citenamefont {Huang},
  \citenamefont {Kim}, \citenamefont {Agrawal}, \citenamefont {Sudarsan},
  \citenamefont {Maxim}, \citenamefont {Jayaraman},\ and\ \citenamefont
  {Ugaz}}]{Huang:2009aa}%
  \BibitemOpen
  \bibfield  {author} {\bibinfo {author} {\bibfnamefont {J.-H.}\ \bibnamefont
  {Huang}}, \bibinfo {author} {\bibfnamefont {J.}~\bibnamefont {Kim}}, \bibinfo
  {author} {\bibfnamefont {N.}~\bibnamefont {Agrawal}}, \bibinfo {author}
  {\bibfnamefont {A.~P.}\ \bibnamefont {Sudarsan}}, \bibinfo {author}
  {\bibfnamefont {J.~E.}\ \bibnamefont {Maxim}}, \bibinfo {author}
  {\bibfnamefont {A.}~\bibnamefont {Jayaraman}},\ and\ \bibinfo {author}
  {\bibfnamefont {V.~M.}\ \bibnamefont {Ugaz}},\ }\bibfield  {title} {\bibinfo
  {title} {Rapid fabrication of bio-inspired 3d microfluidic vascular
  networks},\ }\href@noop {} {\bibfield  {journal} {\bibinfo  {journal}
  {Advanced Materials}\ }\textbf {\bibinfo {volume} {21}},\ \bibinfo {pages}
  {3567} (\bibinfo {year} {2009})}\BibitemShut {NoStop}%
\bibitem [{\citenamefont {Therriault}\ \emph {et~al.}(2003)\citenamefont
  {Therriault}, \citenamefont {White},\ and\ \citenamefont
  {Lewis}}]{Therriault:2003aa}%
  \BibitemOpen
  \bibfield  {author} {\bibinfo {author} {\bibfnamefont {D.}~\bibnamefont
  {Therriault}}, \bibinfo {author} {\bibfnamefont {S.~R.}\ \bibnamefont
  {White}},\ and\ \bibinfo {author} {\bibfnamefont {J.~A.}\ \bibnamefont
  {Lewis}},\ }\bibfield  {title} {\bibinfo {title} {Chaotic mixing in
  three-dimensional microvascular networks fabricated by direct-write
  assembly},\ }\href {https://doi.org/10.1038/nmat863} {\bibfield  {journal}
  {\bibinfo  {journal} {Nature Materials}\ }\textbf {\bibinfo {volume} {2}},\
  \bibinfo {pages} {265} (\bibinfo {year} {2003})}\BibitemShut {NoStop}%
\bibitem [{\citenamefont {{El Bied}}\ \emph {et~al.}(2002)\citenamefont {{El
  Bied}}, \citenamefont {Sulem},\ and\ \citenamefont
  {Martineau}}]{El-Bied:2002aa}%
  \BibitemOpen
  \bibfield  {author} {\bibinfo {author} {\bibfnamefont {A.}~\bibnamefont {{El
  Bied}}}, \bibinfo {author} {\bibfnamefont {J.}~\bibnamefont {Sulem}},\ and\
  \bibinfo {author} {\bibfnamefont {F.}~\bibnamefont {Martineau}},\ }\bibfield
  {title} {\bibinfo {title} {Microstructure of shear zones in fontainebleau
  sandstone},\ }\href
  {https://doi.org/https://doi.org/10.1016/S1365-1609(02)00068-0} {\bibfield
  {journal} {\bibinfo  {journal} {International Journal of Rock Mechanics and
  Mining Sciences}\ }\textbf {\bibinfo {volume} {39}},\ \bibinfo {pages} {917}
  (\bibinfo {year} {2002})}\BibitemShut {NoStop}%
\bibitem [{\citenamefont {Lester}\ and\ \citenamefont
  {Chryss}(2019)}]{Lester:2019aa}%
  \BibitemOpen
  \bibfield  {author} {\bibinfo {author} {\bibfnamefont {D.~R.}\ \bibnamefont
  {Lester}}\ and\ \bibinfo {author} {\bibfnamefont {A.}~\bibnamefont
  {Chryss}},\ }\bibfield  {title} {\bibinfo {title} {Topological mixing of
  yield stress materials},\ }\href
  {https://doi.org/10.1103/PhysRevFluids.4.064502} {\bibfield  {journal}
  {\bibinfo  {journal} {Phys. Rev. Fluids}\ }\textbf {\bibinfo {volume} {4}},\
  \bibinfo {pages} {064502} (\bibinfo {year} {2019})}\BibitemShut {NoStop}%
\bibitem [{\citenamefont {Thiffeault}(2004)}]{Thiffeault:2004aa}%
  \BibitemOpen
  \bibfield  {author} {\bibinfo {author} {\bibfnamefont {J.-L.}\ \bibnamefont
  {Thiffeault}},\ }\bibfield  {title} {\bibinfo {title} {Stretching and
  curvature of material lines in chaotic flows},\ }\href
  {https://doi.org/https://doi.org/10.1016/j.physd.2004.04.009} {\bibfield
  {journal} {\bibinfo  {journal} {Physica D: Nonlinear Phenomena}\ }\textbf
  {\bibinfo {volume} {198}},\ \bibinfo {pages} {169 } (\bibinfo {year}
  {2004})}\BibitemShut {NoStop}%
\bibitem [{\citenamefont {MacKay}(1994)}]{Mackay:1994aa}%
  \BibitemOpen
  \bibfield  {author} {\bibinfo {author} {\bibfnamefont {R.~S.}\ \bibnamefont
  {MacKay}},\ }\bibfield  {title} {\bibinfo {title} {Transport in 3d
  volume-preserving flows},\ }\href {https://doi.org/10.1007/BF02430637}
  {\bibfield  {journal} {\bibinfo  {journal} {Journal of Nonlinear Science}\
  }\textbf {\bibinfo {volume} {4}},\ \bibinfo {pages} {329} (\bibinfo {year}
  {1994})}\BibitemShut {NoStop}%
\bibitem [{\citenamefont {Surana}\ \emph {et~al.}(2006)\citenamefont {Surana},
  \citenamefont {Grunberg},\ and\ \citenamefont {Haller}}]{Surana:2006aa}%
  \BibitemOpen
  \bibfield  {author} {\bibinfo {author} {\bibfnamefont {A.}~\bibnamefont
  {Surana}}, \bibinfo {author} {\bibfnamefont {O.}~\bibnamefont {Grunberg}},\
  and\ \bibinfo {author} {\bibfnamefont {G.}~\bibnamefont {Haller}},\
  }\bibfield  {title} {\bibinfo {title} {Exact theory of three-dimensional flow
  separation. part 1. steady separation.},\ }\href@noop {} {\bibfield
  {journal} {\bibinfo  {journal} {Journal of Fluid Mechanics}\ }\textbf
  {\bibinfo {volume} {564}} (\bibinfo {year} {2006})}\BibitemShut {NoStop}%
\bibitem [{\citenamefont {Haller}\ and\ \citenamefont
  {Mezic}(1998)}]{Haller:1998ab}%
  \BibitemOpen
  \bibfield  {author} {\bibinfo {author} {\bibfnamefont {G.}~\bibnamefont
  {Haller}}\ and\ \bibinfo {author} {\bibfnamefont {I.}~\bibnamefont {Mezic}},\
  }\bibfield  {title} {\bibinfo {title} {Reduction of three-dimensional,
  volume-preserving flows with symmetry},\ }\href@noop {} {\bibfield  {journal}
  {\bibinfo  {journal} {Nonlinearity}\ }\textbf {\bibinfo {volume} {11}},\
  \bibinfo {pages} {319} (\bibinfo {year} {1998})}\BibitemShut {NoStop}%
\bibitem [{\citenamefont {Ott}(2002)}]{Ott:2002aa}%
  \BibitemOpen
  \bibfield  {author} {\bibinfo {author} {\bibfnamefont {E.}~\bibnamefont
  {Ott}},\ }\href {https://doi.org/10.1017/CBO9780511803260} {\emph {\bibinfo
  {title} {Chaos in Dynamical Systems}}},\ \bibinfo {edition} {2nd}\ ed.\
  (\bibinfo  {publisher} {Cambridge University Press},\ \bibinfo {year}
  {2002})\BibitemShut {NoStop}%
\bibitem [{\citenamefont {Ottino}(1989)}]{Ottino:1989aa}%
  \BibitemOpen
  \bibfield  {author} {\bibinfo {author} {\bibfnamefont {J.}~\bibnamefont
  {Ottino}},\ }\href@noop {} {\emph {\bibinfo {title} {The Kinematics of
  Mixing: Stretching, Chaos, and Transport}}}\ (\bibinfo  {publisher}
  {Cambridge University Press},\ \bibinfo {year} {1989})\BibitemShut {NoStop}%
\bibitem [{\citenamefont {Vogel}(2002)}]{Vogel:2002aa}%
  \BibitemOpen
  \bibfield  {author} {\bibinfo {author} {\bibfnamefont {H.-J.}\ \bibnamefont
  {Vogel}},\ }\bibinfo {title} {Topological characterization of porous media},\
  in\ \href {https://doi.org/10.1007/3-540-45782-8_3} {\emph {\bibinfo
  {booktitle} {Morphology of Condensed Matter: Physics and Geometry of
  Spatially Complex Systems}}},\ \bibinfo {editor} {edited by\ \bibinfo
  {editor} {\bibfnamefont {K.}~\bibnamefont {Mecke}}\ and\ \bibinfo {editor}
  {\bibfnamefont {D.}~\bibnamefont {Stoyan}}}\ (\bibinfo  {publisher} {Springer
  Berlin Heidelberg},\ \bibinfo {address} {Berlin, Heidelberg},\ \bibinfo
  {year} {2002})\ pp.\ \bibinfo {pages} {75--92}\BibitemShut {NoStop}%
\bibitem [{\citenamefont {Scholz}\ \emph {et~al.}(2012)\citenamefont {Scholz},
  \citenamefont {Wirner}, \citenamefont {G\''{o}tz}, \citenamefont {R\''{u}de},
  \citenamefont {Schr\''{o}der-Turk}, \citenamefont {Mecke},\ and\
  \citenamefont {Bechinger}}]{Scholz:2012aa}%
  \BibitemOpen
  \bibfield  {author} {\bibinfo {author} {\bibfnamefont {C.}~\bibnamefont
  {Scholz}}, \bibinfo {author} {\bibfnamefont {F.}~\bibnamefont {Wirner}},
  \bibinfo {author} {\bibfnamefont {J.}~\bibnamefont {G\''{o}tz}}, \bibinfo
  {author} {\bibfnamefont {U.}~\bibnamefont {R\''{u}de}}, \bibinfo {author}
  {\bibfnamefont {G.~E.}\ \bibnamefont {Schr\''{o}der-Turk}}, \bibinfo {author}
  {\bibfnamefont {K.}~\bibnamefont {Mecke}},\ and\ \bibinfo {author}
  {\bibfnamefont {C.}~\bibnamefont {Bechinger}},\ }\bibfield  {title} {\bibinfo
  {title} {Permeability of porous materials determined from the euler
  characteristic},\ }\href {https://doi.org/10.1103/PhysRevLett.109.264504}
  {\bibfield  {journal} {\bibinfo  {journal} {Physical Review Letters}\
  }\textbf {\bibinfo {volume} {109}},\ \bibinfo {pages} {264504} (\bibinfo
  {year} {2012})}\BibitemShut {NoStop}%
\bibitem [{\citenamefont {Armstrong}\ \emph {et~al.}(2019)\citenamefont
  {Armstrong}, \citenamefont {McClure}, \citenamefont {Robins}, \citenamefont
  {Liu}, \citenamefont {Arns}, \citenamefont {Schl{\"u}ter},\ and\
  \citenamefont {Berg}}]{Armstrong:2019aa}%
  \BibitemOpen
  \bibfield  {author} {\bibinfo {author} {\bibfnamefont {R.~T.}\ \bibnamefont
  {Armstrong}}, \bibinfo {author} {\bibfnamefont {J.~E.}\ \bibnamefont
  {McClure}}, \bibinfo {author} {\bibfnamefont {V.}~\bibnamefont {Robins}},
  \bibinfo {author} {\bibfnamefont {Z.}~\bibnamefont {Liu}}, \bibinfo {author}
  {\bibfnamefont {C.~H.}\ \bibnamefont {Arns}}, \bibinfo {author}
  {\bibfnamefont {S.}~\bibnamefont {Schl{\"u}ter}},\ and\ \bibinfo {author}
  {\bibfnamefont {S.}~\bibnamefont {Berg}},\ }\bibfield  {title} {\bibinfo
  {title} {Porous media characterization using minkowski functionals: Theories,
  applications and future directions},\ }\href
  {https://doi.org/10.1007/s11242-018-1201-4} {\bibfield  {journal} {\bibinfo
  {journal} {Transport in Porous Media}\ }\textbf {\bibinfo {volume} {130}},\
  \bibinfo {pages} {305} (\bibinfo {year} {2019})}\BibitemShut {NoStop}%
\bibitem [{\citenamefont {Carri\`{e}re}(2007)}]{Carriere:2007aa}%
  \BibitemOpen
  \bibfield  {author} {\bibinfo {author} {\bibfnamefont {P.}~\bibnamefont
  {Carri\`{e}re}},\ }\bibfield  {title} {\bibinfo {title} {On a
  three-dimensional implementation of the baker's transformation},\ }\href
  {https://doi.org/10.1063/1.2804959} {\bibfield  {journal} {\bibinfo
  {journal} {Physics of Fluids}\ }\textbf {\bibinfo {volume} {19}},\ \bibinfo
  {eid} {118110} (\bibinfo {year} {2007})}\BibitemShut {NoStop}%
\bibitem [{\citenamefont {Lester}\ \emph
  {et~al.}(2016{\natexlab{b}})\citenamefont {Lester}, \citenamefont {Trefry},\
  and\ \citenamefont {Metcalfe}}]{Lester:2016ab}%
  \BibitemOpen
  \bibfield  {author} {\bibinfo {author} {\bibfnamefont {D.}~\bibnamefont
  {Lester}}, \bibinfo {author} {\bibfnamefont {M.}~\bibnamefont {Trefry}},\
  and\ \bibinfo {author} {\bibfnamefont {G.}~\bibnamefont {Metcalfe}},\
  }\bibfield  {title} {\bibinfo {title} {Chaotic advection at the pore scale:
  Mechanisms, upscaling and implications for macroscopic transport},\
  }\href@noop {} {\bibfield  {journal} {\bibinfo  {journal} {Adv. Water
  Resour.}\ }\textbf {\bibinfo {volume} {97}},\ \bibinfo {pages} {175}
  (\bibinfo {year} {2016}{\natexlab{b}})}\BibitemShut {NoStop}%
\bibitem [{\citenamefont {Grinfeld}(2013)}]{Grinfeld:2013aa}%
  \BibitemOpen
  \bibfield  {author} {\bibinfo {author} {\bibfnamefont {P.}~\bibnamefont
  {Grinfeld}},\ }\href@noop {} {\emph {\bibinfo {title} {Introduction to Tensor
  Analysis and the Calculus of Moving Surfaces}}}\ (\bibinfo  {publisher}
  {Springer, New York, NY},\ \bibinfo {year} {2013})\BibitemShut {NoStop}%
\bibitem [{\citenamefont {Br\o{}ns}\ and\ \citenamefont
  {Hartnack}(1999)}]{Brons:1999aa}%
  \BibitemOpen
  \bibfield  {author} {\bibinfo {author} {\bibfnamefont {M.}~\bibnamefont
  {Br\o{}ns}}\ and\ \bibinfo {author} {\bibfnamefont {J.~N.}\ \bibnamefont
  {Hartnack}},\ }\bibfield  {title} {\bibinfo {title} {Streamline topologies
  near simple degenerate critical points in two-dimensional flow away from
  boundaries},\ }\href@noop {} {\bibfield  {journal} {\bibinfo  {journal}
  {Physics of Fluids}\ }\textbf {\bibinfo {volume} {11}},\ \bibinfo {pages}
  {314} (\bibinfo {year} {1999})}\BibitemShut {NoStop}%
\bibitem [{\citenamefont {Bakker}(1991)}]{Bakker:1991aa}%
  \BibitemOpen
  \bibfield  {author} {\bibinfo {author} {\bibfnamefont {P.~G.}\ \bibnamefont
  {Bakker}},\ }\href@noop {} {\emph {\bibinfo {title} {Bifurcations in flow
  patterns}}},\ Vol.~\bibinfo {volume} {2}\ (\bibinfo  {publisher} {Springer},\
  \bibinfo {address} {Netherlands},\ \bibinfo {year} {1991})\BibitemShut
  {NoStop}%
\bibitem [{\citenamefont {Smith}\ \emph {et~al.}(2016)\citenamefont {Smith},
  \citenamefont {Rudman}, \citenamefont {Lester},\ and\ \citenamefont
  {Metcalfe}}]{Smith:2016aa}%
  \BibitemOpen
  \bibfield  {author} {\bibinfo {author} {\bibfnamefont {L.~D.}\ \bibnamefont
  {Smith}}, \bibinfo {author} {\bibfnamefont {M.}~\bibnamefont {Rudman}},
  \bibinfo {author} {\bibfnamefont {D.~R.}\ \bibnamefont {Lester}},\ and\
  \bibinfo {author} {\bibfnamefont {G.}~\bibnamefont {Metcalfe}},\ }\bibfield
  {title} {\bibinfo {title} {Mixing of discontinuously deforming media},\
  }\href@noop {} {\bibfield  {journal} {\bibinfo  {journal} {Chaos: An
  Interdisciplinary Journal of Nonlinear Science}\ }\textbf {\bibinfo {volume}
  {26}},\ \bibinfo {pages} {023113} (\bibinfo {year} {2016})}\BibitemShut
  {NoStop}%
\bibitem [{Smi(2019)}]{Smith:2019aa}%
  \BibitemOpen
  \bibfield  {title} {\bibinfo {title} {The geometry of cutting and shuffling:
  An outline of possibilities for piecewise isometries},\ }\href
  {https://doi.org/https://doi.org/10.1016/j.physrep.2019.01.003} {\bibfield
  {journal} {\bibinfo  {journal} {Physics Reports}\ }\textbf {\bibinfo {volume}
  {802}},\ \bibinfo {pages} {1 } (\bibinfo {year} {2019})}\BibitemShut
  {NoStop}%
\bibitem [{\citenamefont {Smith}\ \emph
  {et~al.}(2017{\natexlab{a}})\citenamefont {Smith}, \citenamefont {Rudman},
  \citenamefont {Lester},\ and\ \citenamefont {Metcalfe}}]{Smith:2017aa}%
  \BibitemOpen
  \bibfield  {author} {\bibinfo {author} {\bibfnamefont {L.~D.}\ \bibnamefont
  {Smith}}, \bibinfo {author} {\bibfnamefont {M.}~\bibnamefont {Rudman}},
  \bibinfo {author} {\bibfnamefont {D.~R.}\ \bibnamefont {Lester}},\ and\
  \bibinfo {author} {\bibfnamefont {G.}~\bibnamefont {Metcalfe}},\ }\bibfield
  {title} {\bibinfo {title} {Impact of discontinuous deformation upon the rate
  of chaotic mixing},\ }\href {https://doi.org/10.1103/PhysRevE.95.022213}
  {\bibfield  {journal} {\bibinfo  {journal} {Phys. Rev. E}\ }\textbf {\bibinfo
  {volume} {95}},\ \bibinfo {pages} {022213} (\bibinfo {year}
  {2017}{\natexlab{a}})}\BibitemShut {NoStop}%
\bibitem [{\citenamefont {Smith}\ \emph
  {et~al.}(2017{\natexlab{b}})\citenamefont {Smith}, \citenamefont
  {Umbanhowar}, \citenamefont {Ottino},\ and\ \citenamefont
  {Lueptow}}]{Smith:2017ab}%
  \BibitemOpen
  \bibfield  {author} {\bibinfo {author} {\bibfnamefont {L.~D.}\ \bibnamefont
  {Smith}}, \bibinfo {author} {\bibfnamefont {P.~B.}\ \bibnamefont
  {Umbanhowar}}, \bibinfo {author} {\bibfnamefont {J.~M.}\ \bibnamefont
  {Ottino}},\ and\ \bibinfo {author} {\bibfnamefont {R.~M.}\ \bibnamefont
  {Lueptow}},\ }\bibfield  {title} {\bibinfo {title} {Mixing and transport from
  combined stretching-and-folding and cutting-and-shuffling},\ }\href
  {https://doi.org/10.1103/PhysRevE.96.042213} {\bibfield  {journal} {\bibinfo
  {journal} {Phys. Rev. E}\ }\textbf {\bibinfo {volume} {96}},\ \bibinfo
  {pages} {042213} (\bibinfo {year} {2017}{\natexlab{b}})}\BibitemShut
  {NoStop}%
\bibitem [{\citenamefont {Sturman}(2012)}]{Sturman:2012aa}%
  \BibitemOpen
  \bibfield  {author} {\bibinfo {author} {\bibfnamefont {R.}~\bibnamefont
  {Sturman}},\ }\bibfield  {title} {\bibinfo {title} {The role of
  discontinuities in mixing}\ }(\bibinfo  {publisher} {Elsevier},\ \bibinfo
  {year} {2012})\ pp.\ \bibinfo {pages} {51 -- 90}\BibitemShut {NoStop}%
\bibitem [{\citenamefont {Christov}\ \emph {et~al.}(2011)\citenamefont
  {Christov}, \citenamefont {Lueptow},\ and\ \citenamefont
  {Ottino}}]{Christov:2011aa}%
  \BibitemOpen
  \bibfield  {author} {\bibinfo {author} {\bibfnamefont {I.~C.}\ \bibnamefont
  {Christov}}, \bibinfo {author} {\bibfnamefont {R.~M.}\ \bibnamefont
  {Lueptow}},\ and\ \bibinfo {author} {\bibfnamefont {J.~M.}\ \bibnamefont
  {Ottino}},\ }\bibfield  {title} {\bibinfo {title} {Stretching and folding
  versus cutting and shuffling: An illustrated perspective on mixing and
  deformations of continua},\ }\href@noop {} {\bibfield  {journal} {\bibinfo
  {journal} {American Journal of Physics}\ }\textbf {\bibinfo {volume} {79}},\
  \bibinfo {pages} {359} (\bibinfo {year} {2011})}\BibitemShut {NoStop}%
\bibitem [{\citenamefont {Juarez}\ \emph {et~al.}(2010)\citenamefont {Juarez},
  \citenamefont {Lueptow}, \citenamefont {Ottino}, \citenamefont {Sturman},\
  and\ \citenamefont {Wiggins}}]{Juarez:2010aa}%
  \BibitemOpen
  \bibfield  {author} {\bibinfo {author} {\bibfnamefont {G.}~\bibnamefont
  {Juarez}}, \bibinfo {author} {\bibfnamefont {R.~M.}\ \bibnamefont {Lueptow}},
  \bibinfo {author} {\bibfnamefont {J.~M.}\ \bibnamefont {Ottino}}, \bibinfo
  {author} {\bibfnamefont {R.}~\bibnamefont {Sturman}},\ and\ \bibinfo {author}
  {\bibfnamefont {S.}~\bibnamefont {Wiggins}},\ }\bibfield  {title} {\bibinfo
  {title} {Mixing by cutting and shuffling},\ }\href
  {https://doi.org/10.1209/0295-5075/91/20003} {\bibfield  {journal} {\bibinfo
  {journal} {{EPL} (Europhysics Letters)}\ }\textbf {\bibinfo {volume} {91}},\
  \bibinfo {pages} {20003} (\bibinfo {year} {2010})}\BibitemShut {NoStop}%
\bibitem [{\citenamefont {Sudarsan}\ and\ \citenamefont
  {Ugaz}(2006)}]{Sudarsan:2006aa}%
  \BibitemOpen
  \bibfield  {author} {\bibinfo {author} {\bibfnamefont {A.~P.}\ \bibnamefont
  {Sudarsan}}\ and\ \bibinfo {author} {\bibfnamefont {V.~M.}\ \bibnamefont
  {Ugaz}},\ }\bibfield  {title} {\bibinfo {title} {Multivortex micromixing},\
  }\href@noop {} {\bibfield  {journal} {\bibinfo  {journal} {Proceedings of the
  National Academy of Sciences}\ }\textbf {\bibinfo {volume} {103}},\ \bibinfo
  {pages} {7228} (\bibinfo {year} {2006})}\BibitemShut {NoStop}%
\bibitem [{\citenamefont {Mathew}\ \emph {et~al.}(2005)\citenamefont {Mathew},
  \citenamefont {Mezi{\'c}},\ and\ \citenamefont {Petzold}}]{Mathew:2005aa}%
  \BibitemOpen
  \bibfield  {author} {\bibinfo {author} {\bibfnamefont {G.}~\bibnamefont
  {Mathew}}, \bibinfo {author} {\bibfnamefont {I.}~\bibnamefont {Mezi{\'c}}},\
  and\ \bibinfo {author} {\bibfnamefont {L.}~\bibnamefont {Petzold}},\
  }\bibfield  {title} {\bibinfo {title} {A multiscale measure for mixing},\
  }\href {https://doi.org/https://doi.org/10.1016/j.physd.2005.07.017}
  {\bibfield  {journal} {\bibinfo  {journal} {Physica D: Nonlinear Phenomena}\
  }\textbf {\bibinfo {volume} {211}},\ \bibinfo {pages} {23 } (\bibinfo {year}
  {2005})}\BibitemShut {NoStop}%
\end{thebibliography}%

\end{document}